\documentclass[preprint2]{aastex63}
\usepackage{listings}
\usepackage{color}
\usepackage{float}
\extrafloats{100}
\usepackage{pifont}
\usepackage{bbding}
\usepackage{longtable}

\usepackage{hyperref}
\usepackage{natbib}

\setcitestyle{citesep={,}}

\shorttitle{Validating PSZ2 clusters}
\shortauthors{Banerjee et al.}

\begin{document}

\title{Validating Planck SZ2 Clusters with Optical Counterparts}

\author{P. Banerjee}
\affiliation{Department of Physics and Astronomy, University of Southern
California, Los Angeles, CA 90089, USA}
\email{panchajb@usc.edu}

\author{E. Pierpaoli}
\affiliation{Department of Physics and Astronomy, University of Southern
California, Los Angeles, CA 90089, USA}
\email{pierpaol@usc.edu}

\author{N. Mirzatuny}
\affiliation{Department of Physics and Astronomy, University of Southern
California, Los Angeles, CA 90089, USA}
\email{mirzatun@usc.edu}

\author{K. Maamari}
\affiliation{Department of Physics and Astronomy, University of Southern
California, Los Angeles, CA 90089, USA}
\email{maamari@usc.edu}

\author{P. Rim}
\affiliation{Troy High School, 2200 Dorothy Ln, Fullerton, CA 92831, USA}
\email{800018315@fjuhsd.org}

\begin{abstract}

We perform an extensive analysis of optical counterparts of Planck PSZ2
clusters, considering matches with three recent catalogs built from Sloan
Digital Sky Survey (SDSS) data: AMF DR9, redMaPPer (v6.3) and Wen et al (WHL).
We significantly extend the number of
optical counterparts of detected Planck clusters, and characterize the
optical properties when multiple identifications in different
catalogs exist. For Planck clusters which already possess an external
validation, we analyze the redshift assignment for both optical and
X--ray determinations. We then analyze the Planck Cosmology sample and
comment on redshift determination and potential mass
mis-determinations due to alignment issues. Finally, we inspect the reconstructed $y$ map from
Planck and reason on the detectability of optical clusters.
Overall, AMF DR9 main
(extended) finds 485 (511) optical matches, with 45 (55) previously
unmatched PSZ2 clusters, to be compared with the 374 optical matches
already present in PSZ2. 
29 of the 55 previously unmatched clusters do not yet have a followup in the literature. 18 of these are found in more than one SDSS catalog with consistent redshifts. We provide redshift and mass estimates for the newly matched clusters, and discuss the comparison with the follow-ups, when present. We find good agreement between the redMaPPer and AMF DR9 redshift determinations. 
AMF DR9 tends to predict lower redshifts for a few PSZ2 high--redshift clusters which were previously validated by an optical counterpart.
From the Planck Cosmology sample, optical matches are found for 204 of the 278 objects in the observed area. 
We find 14 clusters which merit further investigation, and discuss possible alignment issues for 9 of these clusters. 
After inspecting the $y$ map, we provide a list of 229 optical clusters not included in the Planck PSZ2 catalog but showing a prominent $y$ signal. We have further investigated the 86 clusters with Planck S/N$>4.5$, 73 of which are unmasked by a nearby point source. From these potential clusters, using the MMF technique (applied to the Planck HFI maps), we were able to detect 20 new cluster candidates with S/N $>$ 4.5 that are not included in the PSZ2 catalog. 12 of these clusters have S/N $>$ 6 with 5 being in the close vicinity of at least one point source.

\end{abstract}


\section{Introduction} \label{sec:intro}

Galaxy clusters are the most massive gravitationally bound systems  
in the universe. Their study has been pursued for various different cosmological and astrophysical reasons
\citep{ref:2,ref:3,ref:4,ref:5,ref:6,ref:7}. They comprise very distinct components, such as dark matter (which constitutes most of the mass), 
cold gas and dust in galaxies, as well as the intra--cluster medium (ICM) which occurs in between galaxies.
 Emissions from these varied components
of a galaxy cluster allow to observe them in several different bands. Galaxies emit in the optical (as well as Infrared), while
the ICM emits in X--rays via thermal brehmsstrahlung. 
The Sunyaev-Z'eldovich (SZ) effect is caused by a spectral distortion of the Cosmic Microwave Background spectrum 
due to the scattering of CMB photons off high energy electrons in the ICM. 
A comparison of the various cluster-detection methods will provide a better understanding of 
cluster physics and the selection functions and biases for each one. This will allow unbiased determination of fundamental cluster parameters such as
 cluster mass and redshift for further cosmology.  Galaxy clusters identified using optical methods consider the over-density of galaxies in a cluster while SZ methods consider the gas in the ICM. It is useful to be able to compare the properties of clusters identified using these varied methodologies.  
  
 The AMF DR9 \citep{ref:1} is a new optical catalog of galaxy clusters compiled from the ninth data release of the
 Sloan Digital Sky Survey. It is constructed using a matched filter technique \citep{ref:10,ref:11,ref:12}. In this work, we characterize the SZ-selected galaxy clusters from the Planck SZ2 catalog, comparing the SZ cluster catalog with the AMF DR9 optical catalog, as well as a cross-section of new optical catalogs, such as the AMF DR9 extended, the WHL \citep{ref:13} and the newest version of the redMaPPer catalog (v6.3) \citep{ref:9}. Previously, the Planck PSZ2 catalog \citep{ref:8} had only been compared with the older version of the redMaPPer optical catalog (v5.10). We look at Planck cluster redshifts and investigate whether these can be confirmed by comparing with redshifts of their optical cluster counterparts. We assign redshifts to Planck clusters without a previous external validation using the redshift of the AMF counterpart. We calculate the SZ-determined masses ($M_{SZ}$ values) for these Planck clusters using the assigned redshifts. We formulate a scaling relation between the Planck mass estimate $M_{SZ}$ and AMF DR9 optical richness $\Lambda_{200}$ in order to identify the best counterpart for PSZ2 clusters with multiple possible AMF matches. We provide a list of $M_{SZ}$ values (and a reliability estimate) for Planck SZ2 clusters which did not possess an external counterpart in the original Planck paper. 
 
 	The paper is structured as follows: Section \ref{sec:data} outlines the catalogs (SZ and optical) that we use for our comparisons. Section \ref{sec:2dmatch} describes the 2D matching process by which we assign a counterpart to a PSZ2 cluster. In Section \ref{sec:scal_ang}, we establish two parameters (in richness and angular separation) in order to help identify the best AMF DR9 counterpart for PSZ2 clusters with multiple AMF matches. In Section \ref{sec:comp_msz} we investigate the PSZ2 clusters with external validation that have an AMF DR9 match, and comment on the respective characterizations of the Planck clusters and their optical counterparts. We separately look at the sub-samples of PSZ2 clusters which have a counterpart in both AMF and redMaPPer, and those that only have an AMF DR9 counterpart. In Section \ref{sec:comp_nomsz} we investigate the Planck clusters without a previous external validation (as listed in the PSZ2 catalog). We provide redshift and $M_{SZ}$ estimates for these clusters, in addition to providing a reliability flag based on how many other (non-AMF) counterparts we find for each of these clusters. We also comment on any relevant follow-up redshift investigations of these clusters. In Section \ref{sec:cosmo} we investigate the PSZ2 clusters which are part of the Planck Cosmology sample, and comment on the optical counterparts for these clusters, comparing the redshift estimates for these clusters. As part of Section \ref{sec:ext}, we work towards extending the PSZ2 catalog by identifying signal sources from the Compton y-map (both above and below the S/N threshold of the clusters included in the PSZ2 catalog) which correspond to the locations of AMF DR9 cluster centers. Our results are summarized in Section \ref{sec:conc}. In Appendix \ref{app:A} we discuss the PSZ2 clusters with multiple possible optical counterparts, both for Planck clusters with and without external validation. In Appendix \ref{app:B} we present the list of Planck clusters without external validation for which we identify an AMF DR9 counterpart.  We include $M_{SZ}$ estimates, flags for reliability of the characterization of these clusters, along with information about their optical counterparts, and whether there have been any follow-up redshift estimates. In Appendix \ref{app:C}, we provide the list of optical clusters not included in the PSZ2 catalog but which display a prominent $y$ signal. 
	
\begin{deluxetable*}{lccccc} [t]
\tabletypesize{\footnotesize}
\tablewidth{0pt}
\caption{Planck SZ2 Catalog Matches (2D matches) \label{table:one}}
\tablehead{
\multicolumn{1}{c}{Optical Catalog} &&  \multicolumn{1}{c}{No. of Clusters}
&& \multicolumn{1}{c}{Matches (With/Without $M_{SZ}$)} & \multicolumn{1}{c}{Optical Counterparts [With]/Without $M_{SZ}$(1/2/3)} 
}
\startdata
AMF DR9 main &&  46479 && 485 (440/45) & [(385/50/5)] (34/10/1) \\ 
AMF DR9 extended &&  79368 && 511 (456/55) & [(322/107/19)] (38/14/2) \\
redMaPPer (v6.3) &&  26111 && 383 (363/20) & [(278/71/12)] (16/4)\\
WHL && 132684 && 593 (510/83) & [(197/168/92)] (44/28/7) \\
\enddata
\tablecomments{The table lists the 2D matches within 10 arcmin radius between the PSZ2 catalog  and  the AMF DR9 main, the AMF DR9 extended as well as the redMaPPer (v6.3) catalogs. 805 PSZ2 clusters lie in the SDSS footprint. Column 1 and 2  lists the optical catalogs considered and the  total number of clusters in each catalog. Column 3 indicates the number of  matched Planck clusters (with/without external validation). Column 4 lists the number of instances  in which one Planck cluster matches with (at most) one, two or three optical clusters in a given optical catalog. \textbf{Note: }The total matched cluster numbers might not agree since in some cases one PSZ2 cluster may potentially have had 4 possible optical counterparts.}
\end{deluxetable*}

\section{Data} \label{sec:data}

\textbf{Planck SZ2:} The Planck PSZ2 catalog (\citet{ref:8}, hereafter Planck XXVII) of SZ-selected galaxy clusters contains 1653 objects, of which 805 lie in the SDSS DR9 coverage area. The Planck collaboration identified an external counterpart for 1091 clusters in total, with 613 in the SDSS DR9 area. In the Planck PSZ2 paper, 374 optical counterparts are found via comparison with a non-public version of the redMaPPer (v5.10) catalog  containing a low richness cutoff sample of 400,000 objects. 447 PSZ2 clusters (of which 196 are in the SDSS DR9 area) are listed to have an external X--ray counterpart sourced from the MCXC catalog \citep{ref:14}.
When considering these matching clusters we will make use of the Planck XXVII deliverables which include 
position, redshift, signal-to-noise (S/N) and mass $M_{SZ}$ (the hydrostatic mass expected for a cluster consistent with the assumed scaling relation, as in Equation (9) of \citet{ref:29}). The PSZ2 clusters without an external counterpart have all attributes but a redshift and a $M_{SZ}$ estimate. The S/N cutoff of the catalog is 4.5, while for the Planck Cosmology sample, the minimum S/N of selected clusters is 6.

\textbf{AMF DR9 main:} The AMF DR9 main catalog \citep{ref:1} contains 46,479 optically detected galaxy clusters, with richness $\Lambda_{200} > 20$ in the redshift range 0.05 $\le z \le$ 0.64 in $\sim$11,500 $deg^{2}$ of the sky. Angular position, richness and redshift estimates for these clusters are provided. The catalog was constructed using a maximum likelihood  technique based on a matched filter approach and does not rely on the red sequence for cluster detection,  potentially allowing for the detection of galaxy clusters that  do not possess a central luminous red galaxy. The richness $\Lambda_{200}$ is the total luminosity within $R_{200}$ in terms of $L^{*}$,  where $L^{*}$ brightens with redshift \citep{ref:11}. 

\textbf{AMF DR9 extended:} In addition to the main version of the catalog, Banerjee et al. provide an extended catalog with a lower richness cut, containing 79,368 clusters. This version  also contains clusters with richness $10<\Lambda_{200}<20$ which have a match in the DR8 catalog developed by Wen et al (WHL) \citep{ref:13}. 
 
 \textbf{redMaPPer:}  redMaPPer is a red-sequence cluster finder \citep{ref:9}. The finder identifies 26,111 clusters in $\sim$ 10,000 $deg^2$ of data acquired from SDSS DR8 in the range 0.08 $\le$ z $\le$ 0.55. The richness $\Lambda$ is defined as the sum of the membership probabilities over all galaxies within a scale-radius $R_{\lambda}$ (See Equation (2) in \citet{ref:9}). In our paper, we utilize the newest version of the redMaPPer catalog (v6.3). While the AMF DR9 $\Lambda_{200}$ and the redMaPPer $\Lambda$ are both measures of richness, they fundamentally differ in the sense that while the AMF richness is a measure of the luminosity within $R_{200}$ of a cluster center, the redMaPPer richness is a measure of the galaxy counts in a cluster. 

\textbf{WHL:} The WHL cluster catalog \citep{ref:13}  is constructed with a friend--of--friend algorithm applied to  the Sloan Digital Sky Survey III (DR8). It contains  $\sim$ 130,000 clusters in the redshift range of 0.05 $\le$ z $\le$ 0.8. $\sim$ 116,000 of these clusters lie in the SDSS DR9 coverage area. 
The authors define  cluster richness as $R_{L_*} = L_{200}/L^*$, where $L^*$ is the evolved characteristic luminosity of galaxies in the r-band, and $L_{200}$ is the total r-band luminosity within $R_{200}$ (See Equation (2) in \citet{ref:13}). 

\section{2D Optical Matches} \label{sec:2dmatch}

Taking the listed center of the Planck clusters as a reference, we first  look for matching optical counterparts within a 10 arc-minute radius.

We report the results for the AMF DR9 main, AMF DR9 extended, redMaPPer (v6.3) and WHL catalogs in Table \ref{table:one}. In general, all optical catalogs are found to have a higher number of matches than the ones quoted in Planck XXVII, and in some instances more than one potential counterpart for a given Planck cluster. Using the redMaPPer catalog (v6.3), we find 383 optical counterparts for the Planck SZ clusters, including
295 of the 374 redMaPPer clusters listed as an optical counterpart in the PSZ2 catalog. AMF DR9 main (extended) matches overall 485 (511) Planck clusters, 102 (128) more than redMaPPer. 440 (456) of these 485 (511) PSZ2 clusters have an external validation in Planck XXVII and are discussed in Section \ref{sec:comp_msz}. Of the clusters with a $M_{SZ}$ estimate, 385 just have one AMF DR9 main counterpart for the corresponding Planck cluster. The remaining multiply matched Planck clusters (with an external validation) are discussed in Section \ref{subsec:msz_mult}. 

AMF DR9 main (extended) also matches 45 (55) clusters without previously determined counterparts as per the Planck XXVII paper. Of these 55 clusters, 17 clusters are also matched by the most recent version of the redMaPPer catalog (v6.3). Among the 55 matches with PSZ2 clusters without external validation, 44 have just one AMF DR9 counterpart for the corresponding Planck cluster. These and the other 11 multiply matched clusters are discussed in Section \ref{sec:comp_nomsz}. The WHL catalog matches 108 more clusters than the AMF DR9 main, of which 78 have an external counterpart in the PSZ2 catalog, while 30 do not.

\section{Establishing a scaling relation and an angular separation parameter} \label{sec:scal_ang}

In order to facilitate the down--selection of  multiple optical matches, we define here two parameters that characterize richness and angular separation.
For the richness, we establish a scaling relation between $\Lambda_{200}$ and  $M_{SZ}$ for a subset of 140 Planck clusters whose redshift  is sourced from an X--ray counterpart, and which possess matches in both the AMF DR9 and redMaPPer catalogs with the redshift difference between the PSZ2 cluster and its richest optical counterpart (in the respective catalogs) $\le$ 0.05.
The scaling relation for AMF DR9 $\Lambda_{200}$ versus Planck $M_{SZ}$ is similar to the one used in Equation 8 of Planck XXVII \citep{ref:8} and reads: 
\begin{equation} \label{eq:1}
ln <\Lambda_{200}|M_{SZ} > = B + A ln(\frac{M_{SZ}}{C})
\end{equation}
where A = 0.7174 $\pm$ 0.075, B = 27.5354 $\pm$ 2.419, and C is a normalizing constant fixed at $5\times10^{14} M_{\odot}$. The scatter $\sigma_{ln \Lambda_{200}}$ for a given Planck $M_{SZ}$ is 0.299 $\pm$ 0.025. The deviation of the richness from the expected value, according to the above scaling relation, is measured by the quantity $\Delta_{\Lambda_{200}}$, 
which is formulated as $\Delta_{\Lambda_{200}} = (ln(\Lambda_{200} - <\ln \Lambda_{200}|M_{SZ}>))/ \sigma_{ln \Lambda_{200}}$. 

For the angular separation, we calculate an analogous parameter for every Planck-AMF cluster pair, $\Delta_{\theta} = \frac{\theta}{\theta_{err}}$, where $\theta$ is the angular separation (in arc-minutes) between the clusters and $\theta_{err}$ is the positional error provided for that Planck cluster in the PSZ2 catalog. 

\section{Optical counterparts of Planck clusters with external validation} \label{sec:comp_msz}

In this section, we focus on the Planck clusters with an external validation in the PSZ2 catalog, along with their matched optical counterparts (Section \ref{sec:2dmatch}) in order to validate the assigned redshifts and potentially identify PSZ2 clusters which could have a better redshift validation and line-of-sight alignment. The main characterizations are done using the AMF DR9 and redMaPPer catalogs. In Section \ref{subsec:msz_mult}, we analyze the multiply matched PSZ2 clusters using richness and angular separation elimination criteria that we have defined in Section \ref{sec:scal_ang}. Through this analysis, we aim to identify the best possible counterparts for the Planck clusters with more than one optical AMF match. The best characterized Planck clusters are likely be the ones which have optical counterparts in both the AMF DR9 and redMaPPer catalogs, so it is important to investigate how the properties of these clusters correlate. In Section \ref{subsec:msz_redoptcoun} we discuss the PSZ2 clusters which possess a counterpart in both the AMF DR9 and the redMaPPer (v6.3) catalog. We compare the redshifts between the PSZ2 clusters and those of their AMF DR9 counterparts, specifically considering the three Planck subsets whose redshifts are sourced from redMaPPer clusters, X--ray clusters or from other optical sources respectively. In addition, we also investigate the relationship between the Planck-provided mass and the optical richness estimates. In Section \ref{subsec:msz_new_mat} we analyze the PSZ2 Planck clusters that possess optical counterparts in the AMF DR9 catalog, but do not possess a counterpart in the redMaPPer (v6.3) catalog. In this process, we identify potential new optical counterparts for clusters in the PSZ2 catalog. While analyzing individual clusters in this section, we also consider whether they possess a match in the WHL catalog, and compare the properties of the AMF and WHL matches. 

\subsection{Analyzing multiple matches (with AMF DR9) for PSZ2 clusters with external validation} \label{subsec:msz_mult} 

In an effort to assign new AMF optical counterparts to the Planck clusters, we consider the richness, redshift and angular separation of the AMF clusters which might be possible matches for a given PSZ2 cluster. When there are more than one possible AMF counterpart for one PSZ2 cluster, we attempt to identify the better optical match. If one AMF counterpart had a richness exceeding at  least 10 greater than the others, had a redshift difference $|z_{AMF} - z_{Planck}| \le 0.05$ and had the lowest values of  both $\Delta_{\Lambda_{200}}$ and $\Delta_{\theta}$ (See Section \ref{sec:scal_ang}), then the richer cluster is deemed to be the best match. This process eliminates multiple optical counterparts for 35 of these 55 PSZ2 clusters that had multiple counterparts in the AMF (main and extended combined) catalogs. The optical counterparts so determined for these 35 PSZ2 clusters show a smaller median value in  $\Delta_{\Lambda_{200}}$ (1.41, as opposed to 3.96 for the lower richness clusters) and $\Delta_{\theta}$ (0.59, as opposed to 3.19 for the lower richness ones), as well as a lower $|z_{AMF} - z_{Planck}|$  (0.0008 versus 0.1135).

Among the remaining 20 Planck clusters, we identify 13 (\emph{99, 145, 237, 318, 621, 661, 709, 716, 791, 922,  959, 989, 1111}) where the Planck catalog might be characterizing the signal coming from two sources as one source. This is done by investigating the MILCA y-map \citep{ref:33} provided by the Planck mission. We reason that a Planck cluster may in fact be a blended cluster if the multiple optical counterparts lie on either side of the Planck cluster signal. In this context, a blended cluster refers to two distinct optical clusters at the same redshift which have been characterized as one cluster in the Planck sample. This is also supported if the optical counterparts lie below the $\Lambda_{200}$ - $M_{SZ}$ scaling relation, thus implying that the $M_{SZ}$ of the corresponding PSZ2 cluster may have been over-estimated as compared to the optical richness. 5 of these 12 clusters (\emph{99, 145, 237, 318, 922}) are a part of the Planck Cosmology sample (See Section \ref{sec:cosmo}). 

In addition, we find some clusters whose characterization may have been impacted by other physical factors, e.g. \emph{Cluster 122} (which is also a part of the Cosmology sample) might be undergoing a merger \citep{ref:18}, while \emph{Cluster 1299} is characterized as 'X-ray under-luminous' in the PSZ2 catalog. There are clusters (\emph{299, 850}) whose PSZ2 redshift is $>0.55$, which is typically beyond the redshift range of the AMF finder, while in the case of \emph{Planck Clusters 832, 868 and 945}, there is very little to distinguish between their respective multiple possible AMF DR9 counterparts. This analysis is presented in full in Appendix \ref{appA_sub1}. 

\subsection{Planck clusters with optical counterparts in AMF DR9 main and redMaPPer} \label{subsec:msz_redoptcoun}

\begin{figure}[h!]
\vspace*{2mm}
\centerline{\includegraphics[width=0.5\textwidth]{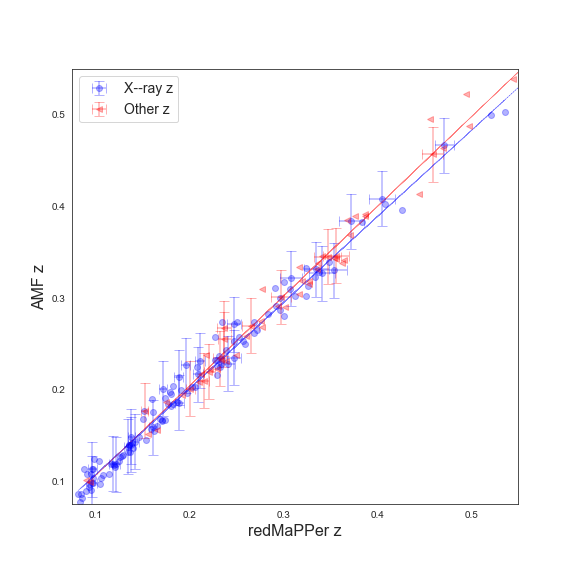}}
\caption{Plot for AMF z versus redMaPPer z for the 126 Planck clusters with an X--ray sourced redshift. These PSZ2 clusters possess an optical counterpart in both AMF DR9 and redMaPPer. We show the same plot for the 58 PSZ2 clusters which occur in both optical catalogs but have their redshifts sourced from a non-redMaPPer or X--ray source. The best-fit lines for $z_{AMF}$ vs $z_{redMaPPer}$ are shown in the figure. Error bars are shown for every fifth point.}
\label{fig:z_xotherscat}
\end{figure}

Overall, there are 295 PSZ2 clusters with a counterpart in both the AMF and redMaPPer optical catalogs, some of which possess multiple matches. In most cases, the criterium outlined in Section \ref{subsec:msz_mult} selects a best match, but for 7 clusters the AMF DR9 counterpart remains ambiguous (\emph{Planck clusters 99, 621, 709, 791, 850, 989, 1111}, see \ref{subsec:msz_mult} and \ref{appA_sub1}). Here we consider the remaining 288, and discuss their redshift determination and richnesses. The redshift of these clusters in the PSZ2 catalog are taken from  X--ray measurements (126 objects), redMaPPer (v5.10) estimates (104 objects), and other optical sources (58 objects). The redshifts of the same clusters in the two versions of the redMaPPer catalog (v5.10 and v6.3) are almost identical, with a mean difference of 0.0002 and a standard deviation of 0.003. 
 
Fig. \ref{fig:z_xotherscat} shows the redshift determination of redMaPPer and AMF finders for the PSZ2 clusters which occur in both catalogs and have their redshifts sourced from an X--ray cluster, as well as from a non-redMaPPer or X--ray source. We find that the redshift estimates for the optical counterparts are nearly identical at lower values of redshift ($z<0.35$) while the AMF redshift is typically lower than the quoted redMaPPer redshift for a Planck redshift $z\ge0.35$.This is particularly relevant because the quoted redshifts in PSZ2 in the redshift range 0.35-0.6 relies heavily on redMaPPer determinations. 
A linear fit to the PSZ2 clusters which occur in both optical catalogs but have their redshifts sourced from a non-redMaPPer or X--ray source provides nearly identical slopes for both $z<0.35$ and $z\ge0.35$, with the AMF redshifts being slightly lower as compared to the redMaPPer ones. 

\begin{figure}[H]
\vspace*{2mm}
\centerline{\includegraphics[width=0.5\textwidth]{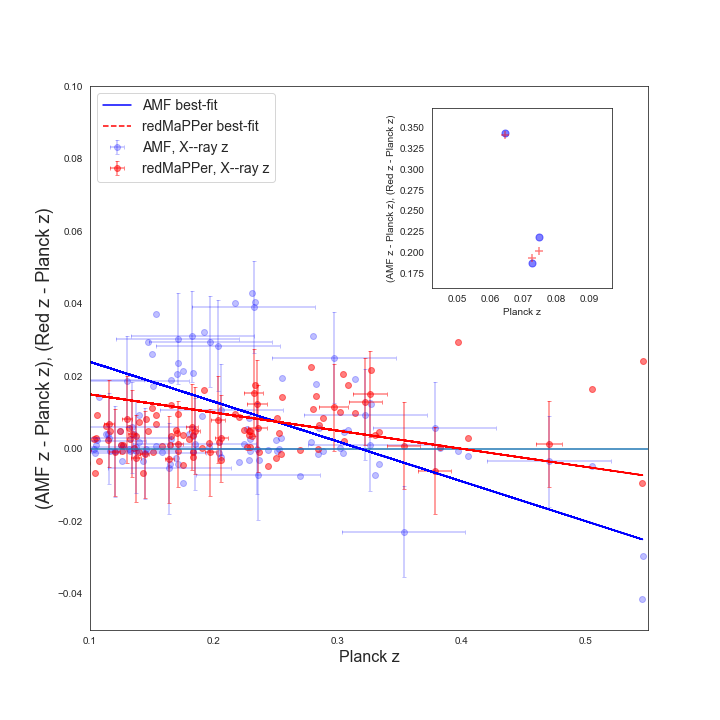}}
\caption{Plot for $z_{Planck}$ versus ($z_{AMF} - z_{Planck}$) and ($z_{redMaPPer}$ - $z_{Planck}$) for the 126 Planck clusters (with an X-ray validated redshift) with an optical counterpart in both AMF DR9 and redMaPPer. Inset image shows the outliers. Error bars are shown for every fifth point.}
\label{fig:zcomp_xz}
\end{figure}

Fig. \ref{fig:zcomp_xz} shows the scatter plot (with error bars) for $z_{Planck}$ vs ($z_{AMF}$ - $z_{Planck}$) and $z_{Planck}$ vs ($z_{redMaPPer}$ - $z_{Planck}$) for the clusters with a redshift validated from an X--ray source. Investigating the statistics of the clusters with X--ray sourced redshift values, we see that the mean and variance of the $z_{AMF} - z_{Planck}$ values for these clusters are 0.0135 and 0.0018 respectively, which are both lower than the corresponding values of 0.0143 and 0.0031 for the quantity $z_{redMaPPer} - z_{Planck}$ for these clusters. The lines of best fit (with the slopes being $-0.11$ and $-0.05$ for AMF and redMaPPer respectively) are also shown on the plot. 
However, a \emph{t-test} between the two samples yields no statistically significant difference. 
This is also consistent with our analysis of Fig. \ref{fig:z_xotherscat} above, where we inferred the redMaPPer z estimates to be slightly higher than the AMF z estimates for PSZ2 clusters with an X-ray sourced redshift, but, as mentioned, this difference does not appear to be statistically significant. The same trend holds when we consider the PSZ2 clusters with a redshift validated by neither an X-ray nor a redMaPPer cluster. 

The insert of Fig. \ref{fig:zcomp_xz} shows 3 clusters with X-ray redshift validation (\emph{Planck clusters 411, 1174 and 1522}) for which PSZ2 cites a redshift ($z<0.1$) which is much lower than the one determined from a matching cluster in redMaPPer, AMF or WHL catalogs. For these clusters, the redshift of the matching object in redMaPPer and AMF are nearly identical. For \emph{Planck clusters 411 and 1174}, the optical counterparts have very high richness ($\sim 100$). For these 2 clusters, the scatter ($\Delta_{\Lambda_{200}}$) decreases if we calculate the $M_{SZ}$ utilizing the redshift of the AMF counterpart. The angular separations of the optical matches (AMF and redMaPPer) from the Planck cluster center \emph{411} are 0.4 and 2.5 arc-minutes respectively while for \emph{cluster 1174} the corresponding separations are 2.5 and 1.6 arc-minutes respectively. WHL also possess a match for these two Planck clusters with identical properties in richness and redshift. Thus, it seems likely that the PSZ2 cluster might have chosen a nearby cluster as the optimal redshift source, while there could have been more viable richer clusters at a higher redshift. For \emph{Cluster 1522}, there are redMaPPer and WHL clusters at an equivalent distance (as the AMF cluster) from the Planck cluster center ($\sim$ 6 arc-minutes) and an identical redshift to the AMF counterpart ($\sim 0.4$). The X-ray luminosities of the X-ray clusters from which these PSZ2 cluster redshifts are sourced are all between 1-2 (in units of $10^{44} \, ergs \, s^{-1}$). All the optical matches are at a low richness, thus, it is possible that the PSZ2 cluster is actually a superposition of the low redshift X-ray cluster from which its catalog redshift is drawn, as well as a more distant lower richness cluster. For these clusters, there seems to be at least an alignment problem if not a mischaracterization. These 3 clusters are discussed in more detail in Appendix \ref{appA_sub2}.

\begin{figure}[t]
\vspace*{2mm}
\centerline{\includegraphics[width=0.5\textwidth]{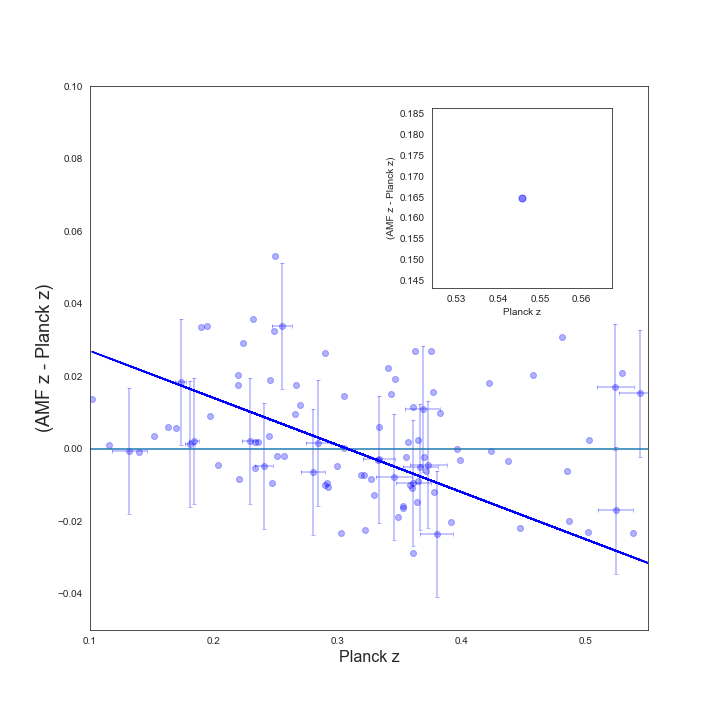}}
\caption{Plot for $z_{Planck}$ versus ($z_{AMF} - z_{Planck}$) for the 104 Planck clusters (with a redMaPPer sourced redshift) with an optical counterpart in both AMF DR9 and redMaPPer. Inset image shows the outlier. Error bars are shown for every fifth point.}
\label{fig:zcomp_redz}
\end{figure}

Fig. \ref{fig:zcomp_redz} shows the $z_{Planck}$ vs ($z_{AMF}$ - $z_{Planck}$) (with error bars) for PSZ2 clusters with a redMaPPer sourced redshift. From the statistics of these clusters, we see that the mean and variance of the $z_{AMF} - z_{Planck}$ values for these clusters are 0.0009 and 0.0004 respectively. The line of best-fit (shown on the plot) gives a slope of -0.13. Thus, in the case of these PSZ2 clusters as well, the AMF redshift values are lower than the Planck redshift values.  

\begin{figure}[t]
\vspace*{2mm}
\centerline{\includegraphics[width=0.5\textwidth]{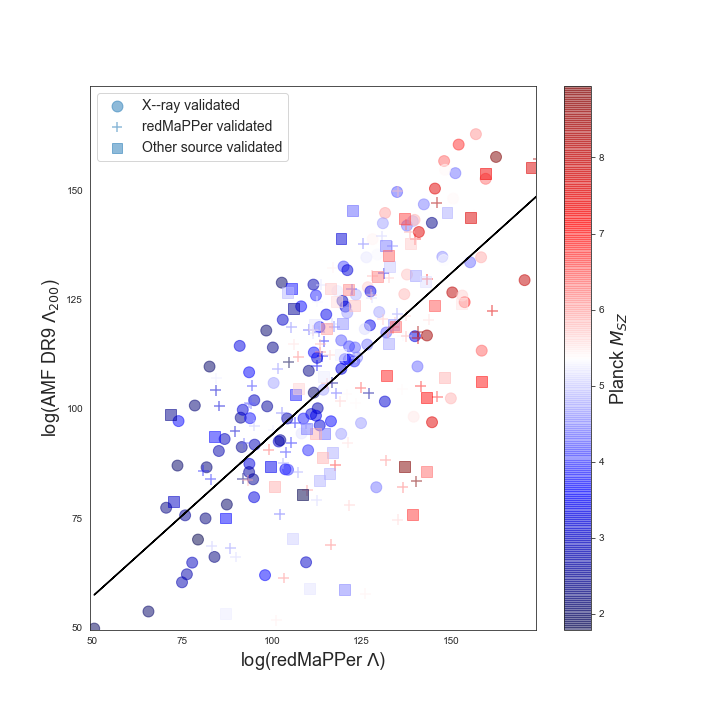}}
\caption{Plot for AMF $\Lambda_{200}$ versus redMaPPer $\Lambda$ for the 288 Planck clusters with an optical counterpart in both AMF DR9 and redMaPPer. The clusters are colored according to their Planck $M_{SZ}$. The plot shows the line of best--fit between the optical richnesses.}
\label{fig:rscat}
\end{figure}

The outlier from Fig. \ref{fig:zcomp_redz} (\emph{Planck cluster 637}) has a much lower value of redshift in AMF DR9 ($\sim0.16$) as opposed to the redMaPPer (or Planck) redshift ($\sim0.54$). The AMF counterpart is $\sim 9 $ arc-minutes away from the PSZ2 cluster center. It seems likely that the AMF counterpart does not correspond to the same cluster due to line-of-sight projection effects. Fig. \ref{fig:rscat} shows the scatter between the two optical richnesses, colored according to the Planck-quoted $M_{SZ}$ of the corresponding clusters. In general, the two richnesses are well correlated, although with considerable scatter, and the estimated $M_{SZ}$ correlates well with richness. The best--fit is $log_{10} ( \Lambda_{200}) = (0.74 \pm 0.05) \, log_{10} (\Lambda) + 0.43 \pm 0.1$. 

In this section, we have discussed the PSZ2 clusters which possess a counterpart in both the AMF DR9 and the redMaPPer (v6.3) catalogs. We compare the redshifts between the PSZ2 clusters and those of their AMF DR9 counterparts, specifically considering the three Planck subsets whose redshifts are sourced from redMaPPer clusters, X--ray clusters or from other optical sources respectively. For PSZ2 clusters with an X--ray sourced redshift, we find that the redshift estimates (for the AMF and redMaPPer counterparts) are nearly identical at lower values of redshift of the corresponding PSZ2 cluster ($z<0.35$) while the AMF redshift is typically lower than the quoted redMaPPer redshift for a cluster with Planck redshift $z\ge0.35$. There are 3 outliers (with X-ray redshift validation) for which PSZ2 cites a redshift much lower ($<0.1$) than the one determined from a matching high richness cluster in the redMaPPer, AMF or WHL catalogs. For two of these clusters, the optical properties are very similar, and it is likely that the PSZ2 redshift may have been better validated by a more distant cluster. We reason that the third cluster might be blended. We identify one PSZ2 cluster where the discrepancy with the AMF counterpart might be due to line-of-sight projection effects. We also investigate the relationship between the Planck $M_{SZ}$ and the optical richness estimates, and find good correlation between the optical richnesses, as well as between the $M_{SZ}$ and the AMF richness. For PSZ2 clusters with a counterpart in both AMF DR9 and redMaPPer, we find, on average the AMF redshift estimates to be lower than that of the corresponding redMaPPer clusters. However, this effect does not appear to be statistically significant. 

\begin{figure}[H]
\vspace*{1mm}
\centerline{\includegraphics[width=0.5\textwidth]{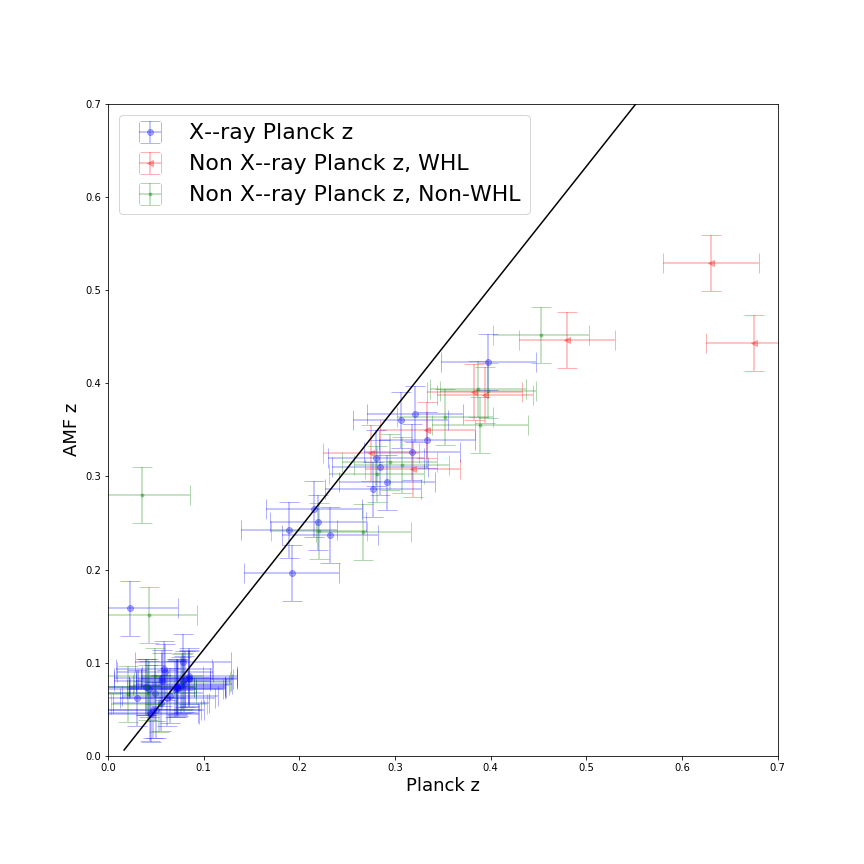}}
\caption{Plot for AMF z versus Planck z for the 85 Planck clusters with an optical counterpart in AMF DR9, but not in redMaPPer (v6.3). We plot separately the scatter for PSZ2 clusters which have a redshift sourced from an X--ray cluster (typically from the MCXC catalog), and PSZ2 clusters which take their redshift from a non X--ray source (such as Abell or WHL). The one-to-one line is also shown in the plot.}
\label{fig:zscat_msz}
\end{figure}

\subsection{AMF DR9 matches with PSZ2 clusters with external validation, without a redMaPPer (v6.3) counterpart} \label{subsec:msz_new_mat}

Overall, there are 92 PSZ2 clusters with an external validation that possess one or more matches in AMF but not in redMaPPer. However, seven of these are multiply matched clusters with an ambiguous AMF DR9 counterpart (\emph{Planck clusters 122, 318, 661, 716, 832, 868, 922}), and are discussed in Appendix \ref{appA_sub1}. The remaining 85 clusters are discussed here. Fig. \ref{fig:zscat_msz} shows the redshift relationship between the AMF and PSZ2 clusters. We plot separately the redshift for PSZ2 clusters which have a redshift sourced from an X--ray observation, and PSZ2 clusters which take their redshift from a non X--ray source (such as Abell or WHL). We investigate whether the AMF DR9 catalog can be used to distinguish between PSZ2 clusters with different redshift sources. Fitting the data for PSZ2 clusters with X--ray sourced redshifts (which are typically more abundant at lower values of $z$) we find good agreement between the AMF and the Planck-quoted redshifts. For the PSZ2 clusters with a redshift drawn from neither an X--ray nor a redMaPPer cluster, the AMF redshift tends to be lower than the redshift of the corresponding PSZ2 cluster. 

From Fig. \ref{fig:zscat_msz} we see that there is good agreement between the AMF and other source redshifts which are not validated by WHL. There are 2 clusters ($|z_{AMF} - z_{Planck}|>0.05$) which do not have an X--ray counterpart (\emph{Planck clusters 390 and 478}). Both of these PSZ2 clusters have their redshifts validated by WHL clusters and have redshift values $>$ 0.6. In general, optical cluster finders do not identify many high redshift clusters so there is cause for investigation here. These outliers are discussed in more detail in Appendix \ref{appA_sub3}. When the redshifts are drawn from X--ray sources, the scatter decreases from z$<0.1$ to z$\ge0.1$. For the clusters with redshifts drawn from other sources, the scatter shows the same trend. 

We assume that the best match is the one that brings better agreement with the $M_{SZ}$ that is expected from the 
e $\Lambda_{200} - M_{SZ}$ scaling relation  (See \ref{sec:scal_ang}) given the new z value. There are 3 PSZ2 clusters (\emph{cluster 102, 238 and 425}) where $|z_{AMF} - z_{Planck}|>0.05$. These clusters have Planck redshifts 0.04, 0.02 and 0.04 respectively, which lie below the lower bound of the redshift range of the AMF finder. We investigate these clusters for potential line-of-sight alignment issues. In all 3 cases, the AMF cluster lies at a z$>0.1$ and since both of the clusters \emph{102} and \emph{238} have very high S/N values (15 and 35 respectively), it is possible that they might be blended clusters. That is, the PSZ2 cluster signal takes into account the contribution of the signals of more than one cluster along the line-of-sight. 


In this section we have analyzed the 85 PSZ2 Planck clusters that possess optical counterparts in the AMF DR9 catalog, but do not possess a counterpart in the redMaPPer (v6.3) catalog. In this process, we identify potential new optical counterparts for clusters in the PSZ2 catalog. While analyzing individual clusters in this section, we also consider whether they possess a match in the WHL catalog, and compare the properties of the AMF and WHL matches. Fitting the data for PSZ2 clusters with X--ray sourced redshifts we find good agreement between the AMF and the Planck-quoted redshifts. For the PSZ2 clusters with a redshift drawn from neither an X--ray nor a redMaPPer cluster, the AMF redshift tends to be lower than the redshift of the corresponding PSZ2 cluster. We discuss the 4 outliers with $|z_{AMF} - z_{Planck}|>0.05$, 2 of which have a $z<0.04$ and the other 2 have a high z ($>0.6$). We reason that the two clusters at a lower redshift (which have a very high S/N) may be blended clusters. The two outliers at higher redshifts have their redshifts sourced from an optical catalog (WHL) and optical cluster finders typically do not perform well at such high redshift values. This discrepancy might be due to the fact that optical finders are less reliable at high redshift values. 

\section{Optical counterparts of Planck clusters without external validation} \label{sec:comp_nomsz}

\begin{deluxetable*}{lcccccc} [t]
\tablewidth{\linewidth}
\caption{Optical Counterparts for PSZ2 clusters without external validation \label{table:two}}
\tablehead{
\colhead{Planck Cluster}  & \colhead{$z_{AMF}$} & \colhead{$z_{AMF}$ - $z_{Follow-up}$}
& \colhead{$z_{AMF}$ - $z_{redMaPPer}$} & \colhead{$z_{AMF}$ - $z_{WHL}$} &\colhead{Reliability}
& \colhead{Calculated $M_{SZ}$}}
\startdata
161  & 0.38 & 0.0 & 0.02 & 0.04 & A & 6.10\\ 
176  & 0.26 & 0.06 & x & 0.04 & C & 4.56\\
194  & 0.37 & x & x & x & E & 5.85\\
243  & 0.37 & -0.04 & x & x & C & 5.09\\
\enddata
\tablecomments{The table displays the clusters (first 4 rows displayed) with a counterpart in the extended AMF catalog, along with information about whether they occur in the AMF main, WHL and redMaPPer catalogs, the redshift of their AMF counterpart, and whether they have a redshift estimate in a follow-up study. In addition the table also lists the redshift differences between the matched optical counterparts and whether the PSZ2 clusters have had other follow-up studies. If there are no counterparts to calculate the redshift difference from then the column entry is an 'x'. If a particular cluster does not occur in AMF DR9 main then the redshift difference is calculated from the corresponding entry in the AMF DR9 extended catalog. Column 6 lists the reliability estimate of the cluster, on a scale of most reliable (A) to least reliable (E). The details of the reliability flag are provided in Appendix \ref{app:B}. Column 7 lists the calculated $M_{SZ}$ values (in units of $10^{14}$ $M_{\odot}$) for these PSZ2 clusters, according to the redshift of their AMF DR9 counterpart. Multiple entries in the columns indicate multiple possible AMF counterparts for the given PSZ2 cluster. The full table is presented in Appendix \ref{app:B}.}
\end{deluxetable*}

In this section, we focus on the Planck clusters without an external validation in the PSZ2 catalog. We find AMF  counterparts to assign, for the first time, new redshifts to these clusters. Angular matches for each PSZ2 cluster are identified in accordance with the method outlined in Section \ref{sec:2dmatch}. PSZ2 clusters without any listed external counterparts do not have an $M_{SZ}$ estimate, as a redshift is needed for this purpose. We calculate the masses of these PSZ2 clusters according to a technique outlined in Section \ref{subsec:msz_calc}. We use these calculated $M_{SZ}$ values to determine the viability of the optical counterparts according to the scaling relation outlined in Section \ref{sec:scal_ang}. In Section \ref{subsec:no_msz}, we investigate the properties of the Planck clusters with matches in one or more optical catalogs, including the AMF DR9, WHL and the redMaPPer. We discuss whether any of these PSZ2 clusters have a follow-up redshift estimate and how that compares to the redshift of the matching AMF cluster. 

\subsection{Calculation of $M_{SZ}$ for PSZ2 clusters without external validation} \label{subsec:msz_calc}

The PSZ2 catalog is constructed by combining the detections made by three methods (MMF1, MMF3 and PwS) into a union catalog. The authors provide for all entries (for each individual detection method) an array of masses as a function of redshift ($M_{SZ}(z)$), which is obtained by intersecting the degeneracy curves with the expected function for different redshift values, from z = 0 to z = 1. 

Provided as part of the PSZ2 catalog, the third extension HDU (Header Data Unit) contains a three-dimensional image with the $M_{SZ}$ observable information per cluster as a function of assumed redshift. For each cluster, the PSZ2 catalog provides a flag (\emph{PIPE DET}), which refers to which of the three pipelines (MMF1, MMF3 or PwS) the cluster has the highest value of S/N in. We choose the $M_{SZ}(z)$ array from the corresponding pipeline according to the flag. We fit the array to a third degree polynomial, and interpolate the curve at the redshift of the AMF counterpart to estimate the $M_{SZ}$ value of the corresponding PSZ2 cluster.

\subsection{Characterizing PSZ2 clusters via their AMF counterparts} \label{subsec:no_msz}

Of the 562 PSZ2 clusters without redshift information, 57 occur in the SDSS DR9 coverage area. Overall, we find AMF DR9 (both main and extended) counterparts for 55 of these Planck clusters. Table \ref{table:two} lists a sub-sample of these clusters, along with information about whether they occur in the WHL and redMaPPer catalogs, and whether they have a redshift estimate in a follow-up study. In addition, the table also provides a $M_{SZ}$ estimate for these clusters, and a reliability flag, on a scale of most reliable (\textbf{A}) to least reliable (\textbf{E}). Further details of the reliability flag are provided, along with the full table, in Appendix \ref{app:B}. 

\subsubsection{Clusters with follow-up redshifts}

26 of these 55 clusters have had a follow-up redshift estimate in the literature since the PSZ2 catalog was published. 5 clusters (\emph{161, 327, 371, 667, 739}) have a counterpart in both the redMaPPer and WHL catalogs, as well as a follow-up redshift, and the redshift difference between the matched AMF cluster and these 3 counterparts are $\le0.05$ (These are designated to be the most reliable with the \textbf{A} reliability flag). 18 of these 26 clusters have a redshift difference $z_{AMF} - z_{Followup}<=0.05$. Of the 8 clusters where the $z_{AMF} - z_{Followup}>0.05$, 6 of them have redshifts (\emph{Planck clusters 303, 381, 394, 462, 483, 681}) which are at the upper-limit or beyond the range of the AMF finder ($z>0.6$). The other 2 clusters \emph{Planck clusters 295, 673} (with a discrepant AMF and Follow-up z citep{ref:26}) have counterparts in WHL with a redshift identical to the AMF counterpart. Thus we reason that the AMF matches for these 2 Planck clusters could be the genuine optical counterparts. 

\subsubsection{Clusters without follow-up redshift estimates}

Of the 29 clusters without a follow-up redshift estimate, 18 have a counterpart in either WHL or redMaPPer. In all of these cases, there exists an optical counterpart (either WHL or redMaPPer) which has a redshift difference of $<0.05$ with the AMF counterpart. 6 clusters occur in all 3 optical catalogs (\emph{279, 423, 668, 906, 921, 1070}). Aside from \emph{906} (where the AMF counterpart has a redshift of 0.27), in all the other 5 cases the AMF counterpart is at a $z>0.4$. The calculated $M_{SZ}$ values for these clusters are $>6$ (in units of $10^{14} M_{\odot}$). There is  good agreement with the redMaPPer counterpart z for 4 of these 6 clusters (\emph{423, 668, 906, 1070}). There are 11 clusters which possess a match only in the AMF catalog (and are flagged \textbf{E} in the reliability estimate). Of these 11 clusters, 4 are multiply matched clusters (\emph{410, 426, 468, 875}) and have been discussed in \ref{subsubsec:mult}. Clusters \emph{410} and \emph{468} are designated as ambiguous matches, while \emph{426} and \emph{875} are likely to be blended clusters. For the other 7 PSZ2 clusters which only possess a counterpart in AMF DR9, the scatter $\Delta_{\Lambda_{200}} \sim 4$, while the AMF richness $\Lambda_{200}<30$, with the angular separations between the AMF and Planck cluster centers being in the range $4-6$ arc-minutes. Thus, these PSZ2 clusters correspond to low richness AMF DR9 clusters which lie close to each other in two dimensions. These clusters, along with their reliability flags, are presented in the full table in Appendix \ref{app:B}. 

\subsubsection{Multiply Matched Clusters} \label{subsubsec:mult}

Among the 55 matches with PSZ2 clusters without external validation, 44 have just one AMF DR9 counterpart for the corresponding Planck cluster, while 11 PSZ2 clusters have multiple possible optical counterparts. From the PSZ2 clusters with multiple optical counterparts (\emph{Planck clusters 279, 284, 381, 410, 415, 426, 438, 468, 875, 906, 920}), we find 6 objects which might be blended clusters. For these clusters (\emph{279, 381, 426, 875, 906, 920}), the difference in redshifts of the matching AMF clusters are less than the error in redshift (0.03) of the AMF catalog. In addition, upon inspection of the signal map, the multiple AMF counterparts lie on either side of the PSZ2 cluster center. Upon calculating the $M_{SZ}$ values of these clusters (utilizing the redshift of the AMF counterparts), we estimate that in all these instances, the calculated $M_{SZ}$ value is higher than the cluster richness according to the $\Lambda_{200}$-$M_{SZ}$ scaling relation (\ref{sec:scal_ang}). We reason that this implies that the signal for the PSZ2 cluster might actually be coming from two sources, which correspond to the two optical counterparts. For the other 5 PSZ2 clusters \emph{284, 410, 415, 438, 468}, there is a redshift difference dz$>$0.05 between the possible AMF counterparts for each Planck cluster. In all these 5 cases, the higher richness match has a lower value of scatter with respect to the $\Lambda_{200}$-$M_{SZ}$ scaling relation. However, in 4 of the 5 cases (\emph{Clusters 410, 415, 438, 468}), the richer AMF cluster has a much higher value of angular separation from the PSZ2 cluster center as compared to the less rich cluster. Hence, it is difficult to determine a better match, or comment on possible line-of-sight projection effects for these clusters.

Thus, we assign 44 new optical counterparts for PSZ2 clusters without an external validation. Subsequently, we assign new redshift estimates (based on the redshift of the matching AMF cluster) to these cluster and calculate their $M_{SZ}$ values as well. This information is presented in Table \ref{table:two_full}. Among the remaining 11 clusters, we identify 6 cases which might be instances of blended clusters, with the PSZ2 catalog having characterized 2 clusters (which correspond to the possible optical counterparts) as one object. For the remaining 5 clusters, we cannot determine an unambiguous optical counterpart via our analysis, nor can we reason that these clusters might be blended. There has been a recent published cluster catalog which presents cluster candidates obtained from joint X-ray-SZ detections using observations from the Planck satellite and the ROSAT all-sky survey (RASS) \citep{ref:32}. However, none of the 55 clusters we have analyzed in this section have had a follow-up in the aforementioned paper. 

\section{Optical counterparts in the Planck Cosmology sample} \label{sec:cosmo}

\begin{deluxetable*}{lccccc} 
\tablewidth{\linewidth}
\caption{Possible clusters to investigate in the Planck Cosmology sample \label{table:three}}
\tablehead{
\colhead{Planck Cluster} & \colhead{AMF $\Lambda_{200}$} & \colhead{Ang. sep (arcmins)} & \colhead{$z_{AMF}$} 
& \colhead{$z_{Planck}$}}
\startdata
43 & 30 & 5.5 & 0.22 & 0.04\\ 
102 & 24 & 3.8 & 0.28 & 0.04\\
238  & 42 & 5.3 & 0.16 & 0.02\\
425  & 25 & 9.6 & 0.15 & 0.04\\
478  & 42 & 0.5 & 0.53 & 0.63\\
1101 & 33 & 4.0 & 0.36 & 0.04\\ 
99  & 59/36 & 3.0/6.4 & 0.39/0.09 & 0.09\\ 
122 & 29/21 & 1.7/9.4 & 0.21/0.47 & 0.04\\
145  & 79/57 & 7.5/3.0 & 0.48/0.08 & 0.06\\
237  & 68/53 & 9.7/1.2 & 0.21/0.07 & 0.07\\
318  & 133/29 & 2.8/9.7 & 0.34/0.26 & 0.25\\
922 & 95/90 & 3.5/1.0 & 0.14/0.15 & 0.12 \\ 
809 & 47/24 & 1.8/3.6 & 0.51/0.49 & 0.54 \\
987 & 99/75 & 3.4/5.4 & 0.23/0.23 & 0.23
\enddata
\tablecomments{The table shows the 14 PSZ2 clusters with ambiguous AMF counterparts, along with their redshift value. Column 1 lists the Planck cluster number. Columns 2, 3 and 4 list the AMF richness(es), angular separation(s) of the AMF cluster and Planck cluster center(s) and the redshift(s) of the matching cluster(s). These columns have multiple entries if the PSZ2 cluster has more than one possible AMF counterpart. Column 5 lists the Planck redshift.} 
\end{deluxetable*}

In the PSZ2 catalog presented by Planck XXVII, there are 507 clusters which are part of the Planck Cosmology sample, 278 of which lie in the SDSS DR9 coverage area. All 278 clusters possess a redshift estimate. The masses and redshifts of these clusters are used to determine cosmology so it is important to investigate whether they have been characterized properly. Overall we find 204 matches between the Planck Cosmology clusters and the optical clusters in AMF DR9 (both main and extended). 176 PSZ2 clusters have single matches with optical AMF clusters, while 28 Planck clusters have multiple possible AMF counterparts. We investigate the redshifts and counterparts of the Planck Cosmology clusters, and discuss potential issues in the characterization of the Planck clusters or their possible AMF DR9 counterparts. Table \ref{table:three} presents the clusters that we conclude might have alignment or mis-characterization issues, along with their properties (such as AMF redshift and $\Lambda_{200}$). 

170 of the 176 Planck Cosmology sample clusters with a single AMF DR9 counterpart have a redshift match ($|z_{AMF} - z_{Planck}|\le0.05$). 
The other 6 all lie at low z ($<0.05$) or high z ($>0.6$), beyond the redshift range of the AMF finder. In these 6 cases, we check if the Planck signal is perturbed by the presence of any of the more distant (or nearer) optical matches, due to line-of-sight alignment effects. 2 of the 6 clusters with $|z_{AMF} - z_{Planck}|>0.05$ (\emph{Planck clusters 43 and 1101}) possess counterparts in redMaPPer and in both these cases, $z_{AMF}- z_{Planck}$ and $z_{redMaPPer}-z_{Planck}$ values are $>0.1$. The Planck redshift (drawn from an X--ray cluster), in both these cases, is $<0.05$, while the optical catalogs place the cluster at a much farther redshift ($>0.2$). In particular, in the case of \emph{Cluster 1101}, both the AMF and redMaPPer counterparts possess identical optical richnesses $\sim 30$, redshifts ($\sim 0.36$) and lie at the same location, $\sim$ 4 arc-minutes away from the PSZ2 cluster center. From the $\Lambda_{200}$ - $M_{SZ}$ scaling relation, the value of $M_{SZ}$ (provided by Planck) for these clusters is larger with respect to the corresponding $\Lambda_{200}$. This is an indication that the signal could possibly accommodate another object along the line of sight. This could imply that the PSZ2 catalog chose a nearby X--ray source to characterize the cluster, while there could also have been contributions to the SZ signal from clusters at a higher redshift. Considering the remaining 4 clusters with $|z_{AMF} - z_{Planck}|>0.05$, \emph{102, 238, 425 and 478} do not possess counterparts in redMaPPer (v6.3) and could be subject to line-of-sight alignment effects. Of these, \emph{102, 238} and \emph{425} lie at low redshift values while \emph{478} has a WHL-sourced redshift of 0.63 (See \ref{subsec:msz_new_mat} and Appendix \ref{appA_sub3}). 

Of the 28 Planck Cosmology clusters with multiple AMF counterparts, 22 clusters have a clearly determined counterpart according to the criteria outlined in Section \ref{subsec:msz_mult}. There are 6 clusters with ambiguous AMF counterparts, 5 of which (\emph{99, 122, 237, 318, 922}) have their PSZ2 redshifts sourced from Abell clusters, while the other one (\emph{145}) has its redshift validated by an X--ray cluster. These clusters have already been discussed in detail in \ref{subsec:msz_mult} (Also see Appendix \ref{app:A}). In the case of clusters \emph{99,145,237, 318} and \emph{922}, we investigated the MILCA y-map and concluded that the Planck catalog might be characterizing the signal coming from two sources as one source. Cluster \emph{122} might be undergoing a merger \citep{ref:18}. Thus, we reason that these 6 PSZ2 clusters, which have been included as part of the Planck Cosmology sample, need to be investigated further. 

As part of our efforts to find the best possible optical counterpart for Planck cosmology clusters, we revisit the down-selection criteria for the 22 multiply-matched PSZ2 clusters with a clearly determined counterpart (according to the previous cutoff, See \ref{subsec:msz_mult}). Upon relaxing the richness cutoff threshold, we find 2 clusters (\emph{809, 987}) which merit further investigation. For both of these clusters, the AMF counterpart centers are nearly identical in angular position and redshift. In the case of PSZ2 cluster \emph{809}, the high value of the Planck quoted S/N (7.1) and the comparatively low value of the AMF richnesses (46.5 and 24.4) imply that the PSZ2 catalog might have characterized two smaller clusters as one larger (blended) cluster. In the case of cluster \emph{987}, there is a much richer AMF DR9 match, which also has a counterpart (similar optical richness, angular separation as well as redshift) in both the WHL and redMaPPer clusters. In this case as well, the Planck S/N is very high ($\sim 8$), which might imply the same blending phenomenon as for cluster \emph{809}. 

So, overall, we match 204 of 278 clusters from the Planck Cosmology sample. 190 of these 204 clusters possess one clearly determined counterpart in the AMF DR9 catalogs (both main and extended). There are 14 other matched clusters with ambiguous counterparts, which are shown in Table \ref{table:three}. Of these 14, there are 6 clusters with a single AMF match but a redshift discrepancy of $>0.05$ between the PSZ2 cluster and its AMF DR9 counterpart. These clusters all lie at low z or high z, beyond the redshift range of the AMF finder. In these 6 cases, we check if the Planck signal is perturbed by the presence of any of the more distant optical matches, due to line-of-sight alignment effects. In 2 of these cases, we find matching clusters in AMF and redMaPPer with nearly identical redshifts, but the PSZ2 cluster redshift is at a much closer value. These 2 clusters have X-ray sourced redshifts in the PSZ2 catalog, so the mis-match with the optical counterparts could be due to line-of-sight alignment effects. There are 8 other clusters with ambiguous multiple possible AMF counterparts and in 1 of these cases, an external physical process, such as a merger, might have interfered with the cluster identification. There could be potential alignment issues for the other 7 clusters, due to either Planck choosing the redshift of a nearer cluster, despite there being a richer cluster at a higher redshift, or considering the signal from two clusters along the line-of-sight as the signal for one cluster. 

\section{Towards extending the PSZ2 Catalog} \label{sec:ext}

As part of this section, we investigate whether the PSZ2 catalog could be expanded to include more clusters. We investigate the $y$-distortion signal detected by Planck in the direction of the AMF DR9 optical clusters. The scopes of this investigation are many. Firstly we can statistically inspect potential spurious detections of optical clusters, then we can see whether some likely massive clusters have remained undetected by Planck and address the reasons for such non-detections. After we establish a list of potential SZ-selected clusters, and proceed to use the multifrequency matched filtering (MMF) technique to estimate the signal-to-noise for these SZ galaxy cluster candidates using the Planck HFI maps.

\subsection{Investigating the $y$-map for new SZ clusters} \label{subsec:ext_1}

We utilized the y-map produced by the NILC component separation method, together with its standard deviation, both of which are available in the Planck Legacy Archive. In addition we also used the 61$\%$ CO map \citep{ref:31} as a galaxy foreground mask. We preprocessed the y-map following the procedure of Burenin \citep{ref:27}. Specifically, the signal to noise map was obtained from the map of the y-parameter and the smoothed standard deviation. The smoothing occurs via a 1 degree radius-median filter. The large scale anisotropy (which is estimated by further smoothing the S/N) is then subtracted from the S/N to obtain the final map where we investigate the galaxy cluster locations.

\begin{figure}[t]
\vspace*{2mm}
\centerline{\includegraphics[width=0.5\textwidth]{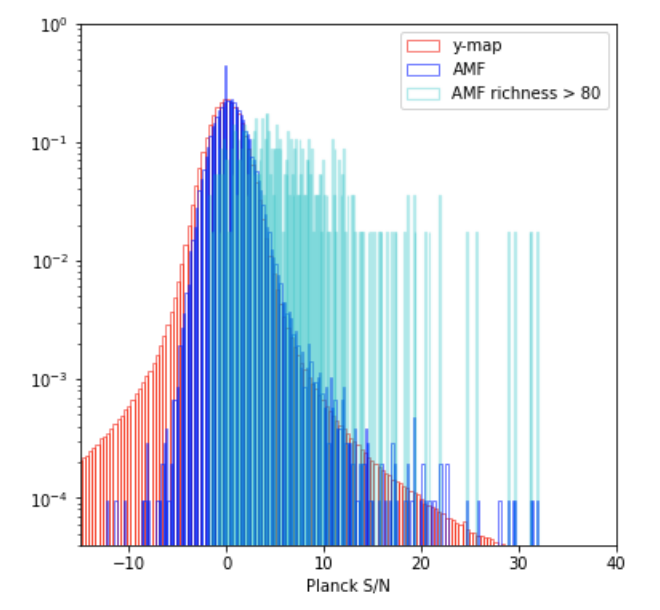}}
\caption{ S/N distribution obtained from the y-map using the Burenin procedure outlined above. We also overplot the S/N corresponding to the sources which align with AMF DR9 cluster centers as well as the S/N distribution for high richness ($\Lambda_{200}>80$) AMF DR9 clusters}
\label{fig:snhist}
\end{figure}

First, we compare the overall signal in the direction of the AMF clusters  with the one of the whole map. The histograms are displayed in Fig. \ref{fig:snhist}. From the plot, it is visible that the overall distribution of AMF clusters is skewed positively with respect to the one of the S/N signal in the map, while restricting the sample to the richest clusters cuts out the lower S/N tail and retains the highest S/N sample. Therefore, the location of potential clusters have an overall more prominent y-signal, and there is a trend in richness as expected. A low S/N, however, does not necessarily indicate a false optical cluster detection, so we cannot use the y-map to exclude optical clusters from the catalog. However, we can inspect which optically detected clusters  should have been detected and weren't, as well as which ones might have been close to detection. In order to do so, we need to also investigate lower S/N values than the threshold of 4.5 used in the PSZ2 paper. 

\begin{figure}[t]
\vspace*{2mm}
\centerline{\includegraphics[width=0.5\textwidth]{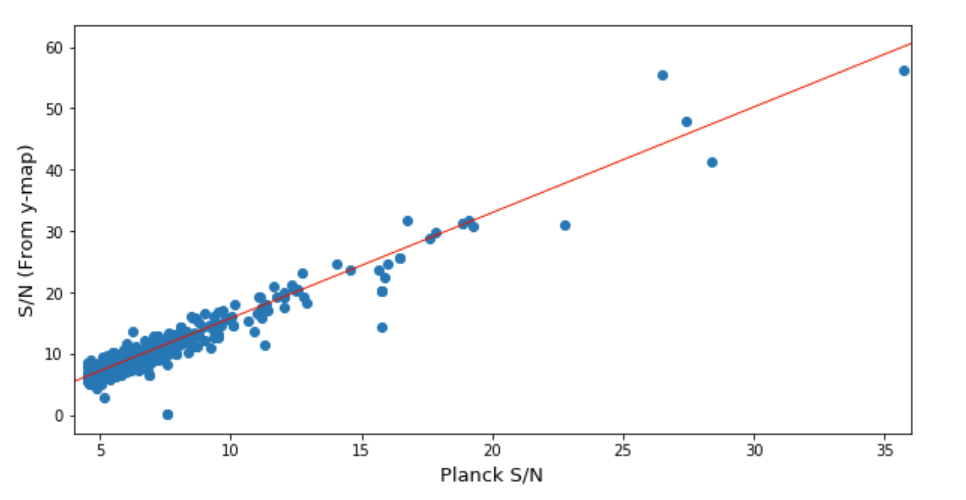}}
\caption{Plot for Planck S/N versus the S/N calculated from the y-maps using the procedure outlined in the text. The best-fit line is also shown.}
\label{fig:sn_scat}
\end{figure}

The S/N determination procedure outlined above differs from that used to determine the Planck S/N in the PSZ2 paper. Hence, the S/Ns of the sources corresponding to AMF cluster centers must be corrected to the scale of the PSZ2 catalog. In order to rescale the S/Ns to those of the PSZ2 catalog, we calibrate a best-fit relation between the S/Ns of the two separate measurements for the cluster catalogs (AMF and Planck) where we obtain a slope of 1.65 and an intercept of -1.15. Fig. \ref{fig:sn_scat} shows the relationship between the Planck provided S/N and the S/N generated from the y-map using the Burenin procedure. The scatter is calculated to be $\sim 22 \%$ for the entire Planck S/N range, as well as for the Planck S/N range between 5 and 10. From here onwards, S/N refers to the inferred S/N from the scaling, corresponding to the one in the PSZ2 paper for the individual cluster detection \citep{ref:8}. 

\begin{figure}[t]
\vspace*{2mm}
\centerline{\includegraphics[width=0.5\textwidth]{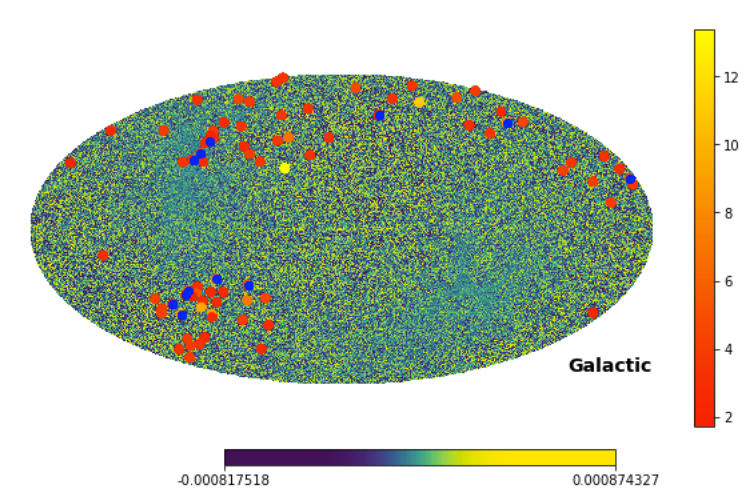}}
\caption{Map of signal sources corresponding to AMF DR9 cluster locations (with S/N $>4.5$), overlaid on the noise map. The sources are colored according to their signal to noise. The color-bar at the bottom of the figure has units in mK. The masked sources are shown in blue.}
\label{fig:noisegt45}
\end{figure}

\begin{figure}[t]
\vspace*{2mm}
\centerline{\includegraphics[width=0.5\textwidth]{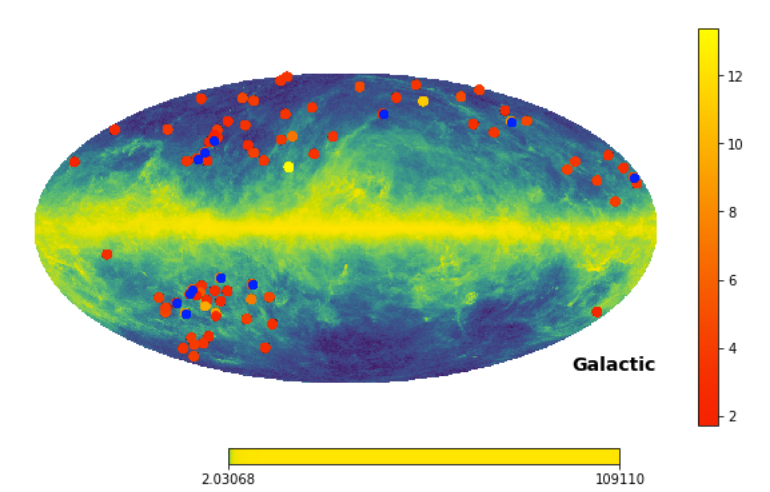}}
\caption{Map of signal sources corresponding to AMF DR9 cluster locations (with S/N $>4.5$), overlaid on the foreground map. The sources are colored according to their signal to noise. The color-bar at the bottom of the figure has units in mK. The masked sources are shown in blue.}
\label{fig:dustgt45}
\end{figure}

\begin{figure}[t]
\vspace*{2mm}
\centerline{\includegraphics[width=0.5\textwidth]{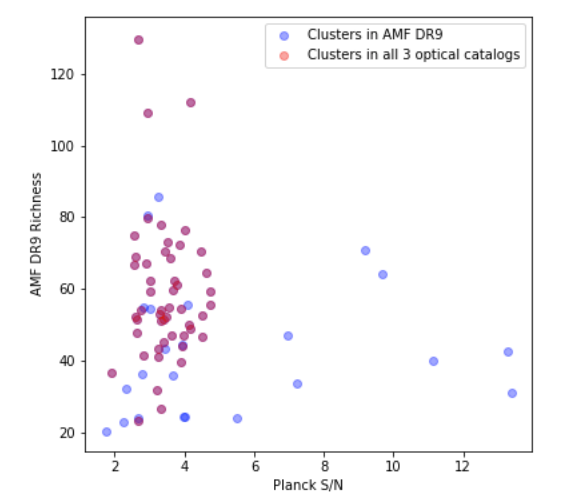}}
\caption{Scatter between the AMF DR9 $\Lambda_{200}$ and the Planck S/N for the AMF DR9 clusters with a corresponding Planck S/N$>4.5$}
\label{fig:sngt45}
\end{figure}

In order to inspect whether certain rich clusters have not been detected by Planck, we first consider the sources with S/N in the range that should have produced a detection according to the criteria outlined in the PSZ2 paper, but were not detected. There exist 86 AMF clusters with a signal-to-noise greater than 4.5 which were not included in the PSZ2 catalog. 73 of these clusters are not masked by a point source (See \ref{subsec:ext_2}). 50 of these clusters were detected by both WHL and redMaPPer, while 8 of these clusters occurred only in the AMF DR9 catalog. Thus there are 50 such sources that occur in all 3 optical catalogs which could be potential Planck clusters and were not included among the 1653 clusters of the PSZ2 catalog. 
We investigate the maps to see whether these clusters may have been missed by the PSZ2 catalog due to high galactic contamination (dust) or noise. Figs. \ref{fig:noisegt45} and \ref{fig:dustgt45} show the locations of the clusters (with S/N $>4.5$) overlaid on the noise map and the galactic dust map respectively \citep{ref:34,  ref:35}. From the distribution of the clusters it would seem that they do not preferentially lie in areas typically contaminated by high foregrounds or noise levels. Fig \ref{fig:sngt45} shows the scatter between the AMF DR9 $\Lambda_{200}$ and the Planck signal-to-noise for the clusters with a S/N$>4.5$. While an apparent trend does not exist in Fig. \ref{fig:sngt45}, we find that 3 of the clusters have a richness ($\Lambda_{200}$) greater than 100. We find 5 AMF clusters that correspond to sources with a S/N$>5.5$ in the y-map. 3 of these clusters also have a counterpart in both the redMaPPer and WHL optical catalogs. The details of these clusters are provided as a table (Table \ref{table:app}) in Appendix \ref{app:C}. 

Sources (corresponding to AMF cluster centers) with $4< $S/N $<4.5$ were checked against the redMaPPer and WHL catalogs to determine whether or not they were likely to be galaxy clusters. Within this S/N range, we found 143 possible cluster candidates not detected in the PSZ2 catalog, the spatial position of which are shown in Fig. \ref{fig:spat_sn4to45}. These clusters are shown in Table \ref{table:sn445} in Appendix \ref{app:C}. 122 of these clusters were detected by WHL and 103 were detected by redMaPPer while 102 of these clusters occurred in all 3 optical catalogs. 

\begin{figure}[t]
\vspace*{2mm}
\centerline{\includegraphics[width=0.5\textwidth]{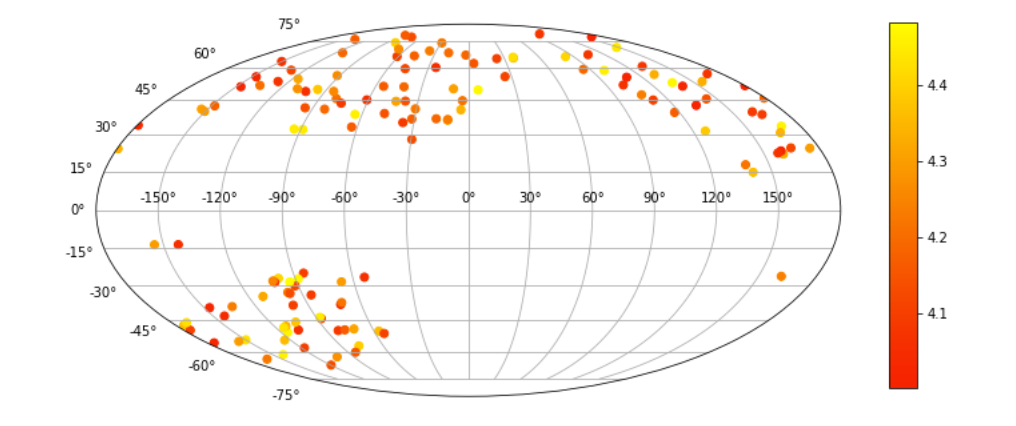}}
\caption{Spatial distribution of the possible PSZ2 cluster candidates in the S/N range 4 to 4.5. The cluster candidates are colored according to their S/N.}
\label{fig:spat_sn4to45}
\end{figure}

\subsection{Using a Multi-frequency Matched Filter to detect SZ clusters} \label{subsec:ext_2}

In this section we use the multifrequency matched filtering (MMF) technique \citep{ref:36, ref:37, ref:38, ref:39} to estimate the signal-to-noise for SZ galaxy cluster candidates that are not included in the Planck PSZ2 catalog \citep{ref:8} but are detected in one or more of the optical catalogs discussed above. The MMF code is only run on cluster candidates with a signal-to-noise $>$ 4.5. The analysis is based on the publicly available data from Planck mission. Namely, Planck HFI maps and the point source catalogues for each frequency channel have been used. We model the HFI maps as a linear combination of the cluster SZ signal plus astrophysical and instrumental noise:
\begin{equation}
    \textit{\textbf{M}}(\textbf{x}) = y_0 \textit{\textbf{j}}_{\nu}T_{\theta_c}(\textbf{x}) + 
    \textit{\textbf{N}}(\textbf{x})
\end{equation}
where $y_0$ is the central Comptonization parameter for a cluster with a core radius of $\theta_c$, $\textit{\textbf{M}}(x)$ is a column vector whose $i^{th}$ component is the HFI map at frequency $\nu_i$, $\textit{\textbf{j}}_{\nu}$ is a vector containing the SZ spectral function at each frequency and $\textit{\textbf{N}}(x)$ is a vector whose $i^{th}$ component is the noise map at $\nu_i$. The noise vector by design incorporates both astrophysical and instrumental components including but not limited to the primordial CMB anisotropy, extragalactic foregrounds and so on. $T_{\theta_c}(\textbf{x})$ is the projected spatial profile of the cluster. We use a spherical $\beta-$profile in this study defined as,
\begin{equation}
    T_{\theta_c}(\textbf{x}) = \bigg(1 + |\textbf{x}|^2/\theta_{c}^2\bigg)^{\frac{-(3\beta - 1)}{2}}
\end{equation}
with $\beta = 1$. Using a matched multifreqeuncy filter, $\mathbf{\Psi}_{\theta_c}$, we are able to recover an unbiased estimate of the central Comptonization parameter, $\hat{y}_0$:
\begin{equation}
    \hat{y}_0 = \int d^2 x\ \mathbf{\Psi}_{\theta_c}^T (\mathbf{x})\ .\  \textit{\textbf{M}}(\textbf{x}),
\end{equation}
where the superscript $"T"$ denotes the transpose of the matched multi-filter. Demanding a minimum variance estimate, the filter in Fourier space can be obtained as \citep{ref:37}:
\begin{equation}
    \mathbf{\Psi}_{\theta_c}(\textbf{\textit{k}}) = \sigma_{\theta_c}^2 \textit{\textbf{P}}^{-1}(\textbf{\textit{k}})\ .\ \textit{\textbf{F}}_{\theta_c}(\textbf{\textit{k}})
\end{equation}
where
\begin{equation}
     \textit{\textbf{F}}_{\theta_c}(\textbf{\textit{k}}) \equiv \textit{\textbf{j}}_{\nu} T_{\theta_c}(\textit{\textbf{k}})B_{\nu}(\textit{\textbf{k}})
\end{equation}
with $B_{\nu}(\textit{\textbf{k}})$ being the Fourier transform of the beam for each frequency channel and
\begin{equation}
    \sigma_{\theta_c} \equiv \bigg[\int d^2 k \ \textit{\textbf{F}}_{\theta_c}^T (\textbf{\textit{k}})\ .\ \textit{\textbf{P}}^{-1}(\textbf{\textit{k}})\ .\ \textit{\textbf{F}}_{\theta_c}(\textbf{\textit{k}}) \bigg]^{-1/2}.
\end{equation}
Here $\textit{\textbf{P}}(\textbf{\textit{k}})$ is the cross-channel noise power spectrum matrix defined as $P_{ij}(\textbf{\textit{k}})\delta(\textbf{\textit{k}} - \textbf{\textit{k}}^\prime) = \langle N_{\nu_i}(\textbf{\textit{k}})N^{\ast}_{\nu_j}(\textbf{\textit{k}}) \rangle$, where $^{\ast}$ denotes the complex conjugate. In Table \ref{table:nareg} we summarize the effective frequency and the full-width-at-half-maximum of the Gaussian beam for each channel. The signal-to-noise of a detection is defined as $\hat{y}_0/\sigma_{\theta_c}$. 

We have applied the filter to $14^{\circ}\times 14^{\circ}$ patches of the CMB sky where the potential cluster candidate is located at the central position. For viable candidates we have re-applied the filter to a smaller patch to obtain the best S/N value. When preparing the HFI maps for the analysis we have masked the point sources with S/N $>$ 10 from Planck's PCNT catalog in each channel with disks of radius $3\sigma_{beam}$ and replaced the disk-area containing the point source with the local average of the patch. We have used the filter with various $\theta_c$ values in the range of $[0.8, 32]$ in arc-minutes in order to find the maximum S/N of each detection.

\begin{table}[t] 
	\centering
	\caption{FWHM and the effective frequency for Planck HFI channels.\newline}
	\label{table:nareg}
	\begin{tabular}{lccccr} 
		\hline
		\hline
	    Channel & FWHM\ (arcmin) & $\nu_{eff}$\ (GHz) \\
		\hline
		\hline
		100 & 9.659 & 103.416 \\
		143 & 7.220 & 144.903 \\
		217 & 4.900 & 222.598 \\
		353 & 4.916 & 355.218 \\
		545 & 4.675 & 528.400 \\
		857 & 4.216 & 776.582 \\
		\hline
	\end{tabular}
\end{table}

Our results are summarized in Table \ref{table:app} of Appendix \ref{app:C}. The table also provides information on whether or not there are any point sources from Planck's PCCS catalog in the vicinity of the cluster candidate, where we have defined the vicinity to be a disk of radius $5\sigma_{\mathrm{beam}}$ with $\sigma_{\mathrm{beam}} = \mathrm{FWHM}/\sqrt{8\log(2)}$ and the cluster candidate positioned at the center of the disk.

Using the MMF filter we were able to detect 20 new cluster candidates with S/N $>$ 4.5 that are not included in PSZ2 cluster catalog. 8 of these candidates have close point source counterparts in the Planck point sources catalog. In particular, among the 20 clusters, 12 have S/N $>$ 6 with 5 being in the close vicinity of at least one point source. Considering the large S/N of these 12 candidates, they can potentially be used in cosmological parameter estimation with SZ cluster number counts. The redshift measurements for the twelve high S/N candidates are in the range [0.078, 0.41]. The richness parameter for these candidates lies within the [30, 71] range. In addition, 3 of the candidates with S/N $>$ 6 occur in all three optical catalogs (AMF, WHL and redMaPPer), flagged A in Table \ref{table:app}. 3 occur in AMF and any one other optical catalog, flagged B and the remaining 6 clusters occur only in the AMF DR9 catalog, indicated with flag C in table 6. We highlight the 20 candidates with S/N $>$ 4.5  with an $\ast$ next to their corresponding \emph{AMF Cluster} column entry in Table \ref{table:app} of Appendix \ref{app:C}. Finally, we report applying the MMF filter to candidates exclusive to the WHL or redMaPPer catalogs alone with no significant results.  

Thus, as part of our efforts to extend the PSZ2 catalog, we identify potential clusters both below and above the signal-to-noise threshold used to identify clusters in the Planck catalog. We find 86 such clusters (of which 73 are unmasked by a nearby point source) with Planck S/N$>$4.5, where the signal source corresponds to AMF DR9 cluster centers. 50 of these 73 clusters also possess matches in the redMaPPer and WHL catalogs. These cluster positions are not correlated with high galactic emission, or the instrumental noise, which could have been two possible reasons for their exclusion from the original PSZ2 catalog. Investigating sources below the S/N threshold of the Planck catalog($<4.5$), we find 143 possible cluster candidates not detected in the PSZ2 catalog, and 102 objects which occurred in all 3 considered optical catalogs (AMF DR9, redMaPPer and WHL). So overall, we provide a list of 229 optical clusters not included in the PSZ2 catalog but showing a prominent $y$ signal. We have further investigated the 86 clusters (with Planck S/N$>4.5$) and we have used the MMF technique (applied to the Planck HFI maps) to estimate the signal-to-noise for these SZ galaxy cluster candidates. We were able to detect 20 new cluster candidates with S/N $>$ 4.5 that were not included in the PSZ2 catalog. Among the 20, 12 have S/N $>$ 6 with 5 being in the close vicinity of at least one point source. The details of these clusters are provided as Table \ref{table:app} in Appendix \ref{app:C}.

\section{Conclusion} \label{sec:conc}

As part of this paper, we have performed an extensive analysis of optical counterparts of Planck clusters, considering matches with recent catalogs built from Sloan Digital Sky Survey (SDSS) data. These newer catalogs provide a larger set of optical matches, including matches for some PSZ2 clusters which previously did not posses any counterpart. AMF DR9 main (extended) finds 485 (511) optical matches, with 45 (55) previously unmatched PSZ2 clusters, to be compared with the 374 optical (redMaPPer v5.10) matches already present in PSZ2.

For clusters that are detected in both redMaPPer (v6.3) and AMF DR9, we find good agreement between both redMaPPer and AMF DR9 determinations with the PSZ2 clusters whose redshift is validated by an X-ray cluster. For the Planck clusters with a redMaPPer (v5.10) sourced redshift, we do not find a statistically significant difference between the quoted Planck redshift and the redshift of the AMF DR9 counterpart. In general, clusters AMF redshift determinations correlate well with redMaPPer ones, although they tend to provide a slightly lower redshift determination for clusters at $z\ge 0.35$. 
For PSZ2 clusters without a redMaPPer counterpart, we find that the AMF DR9 redshifts match very closely the X--ray determined ones, while the AMF tends to infer lower redshifts for the other clusters with redshifts typically determined by other optical sources such as Abell or WHL. 

We assign 44 new (uniquely matched) optical counterparts for PSZ2 clusters without a previously determined external counterpart. Subsequently, we assign new redshift estimates (based on the redshift of the matching AMF cluster) to these cluster and calculate their $M_{SZ}$ values as well. We use the calculated $M_{SZ}$ values to determine the validity of the AMF matches, as well as determining the best counterpart if one PSZ2 cluster has multiple possible AMF DR9 matches.  Among the 11 multiply-matched PSZ2 clusters (with no previous external validation), we identify 6 cases which might be instances of blended clusters, with the PSZ2 catalog having characterized 2 clusters (which correspond to the possible optical counterparts) as one object. 

Of the 55 matched clusters without an external validation, there are 26 clusters with a follow-up redshift estimate, and out of these AMF DR9 has a redshift match with 18 clusters. Of the 29 PSZ2 clusters without a follow-up redshift estimate that have a match with AMF DR9, 18 have another optical counterpart (either redMaPPer or WHL) which has a redshift identical to the matched AMF cluster. The AMF DR9 matches 11 new PSZ2 clusters with no other optical counterparts or follow-up redshift estimates.

Upon investigation of the PSZ2 clusters which are part of the Planck Cosmology sample, we match 204 of 278 such clusters in the SDSS DR9 coverage area. We find an unambiguous AMF DR9 match for 190 of these clusters. There could be potential alignment issues for 14 clusters with an ambiguous AMF DR9 counterpart, due to either Planck choosing the redshift of a nearer cluster despite there being a richer cluster at a higher redshift, or by considering the signal from two clusters along the line-of-sight as the signal for one cluster.

As part of our efforts to extend the PSZ2 catalog, we investigate the y-distortion signal detected by Planck in the direction of AMF optical clusters and find 86 (73 unmasked) and 143 cluster candidates respectively that could be included as part of the Planck catalog, both above and below the S/N threshold used in the PSZ2 paper. We have further investigated the 86 clusters (with Planck S/N$>4.5$) and have used the MMF technique (applied to the Planck HFI maps) to estimate signal-to-noise values for these SZ-selected galaxy cluster candidates. We were able to detect 20 new cluster candidates with S/N $>$ 4.5 that are not included in the PSZ2 catalog. In particular, among the twenty, twelve have S/N $>$ 6 with five being in a close vicinity of at least one point source. Considering the large S/N of these twelve candidates, they can potentially be used in cosmological parameter estimation with SZ cluster number counts.

\vspace{1cm}

\section{Acknowledgements}

EP and NM are supported by NASA grant 80NSSC18K0403. PB was partially supported by the WiSE major support for faculty, awarded
to EP. KM was partially supported by the SOAR grant from USC. Funding for the Sloan Digital Sky Survey (SDSS) has been provided by the Alfred P. Sloan Foundation, the Participating Institutions, the
National Aeronautics and Space Administration, the National Science Foundation, the U.S. Department of Energy, the Japanese Monbukagakusho,
and the Max Planck Society. The SDSS Web site is http://www.sdss.org/. The SDSS is managed by the Astrophysical Research Consortium (ARC)
for the Participating Institutions.

\appendix
\section{Appendix A: PSZ2 clusters with external validation} \label{app:A}
 
\subsection{Analysis of Multiple AMF Counterparts for PSZ2 clusters (with external validation)} \label{appA_sub1} 

In the following analyses, the AMF counterparts are ranked by richness, that is, if there are two possible AMF counterparts for a given Planck cluster, then the richer optical cluster is Ranked 1, and the AMF cluster of lower richness is Ranked 2. To look at the relative positions of the matching cluster centers in the SZ map, we utilize the MILCA full mission y maps (Compton parameter maps) as well as the standard deviation maps provided by the Planck collaboration. We obtained a signal-to-noise map from the y-map and the standard deviation map, and smoothed the result by 1$^\circ$. or the SZ maps, the plots span 50 arc-minutes width and height.

\vspace{1cm}

\begin{figure}[H]
\centering
\begin{minipage}{.75\textwidth}
  \centering
  \includegraphics[width=.5\textwidth, height=.5\textwidth]{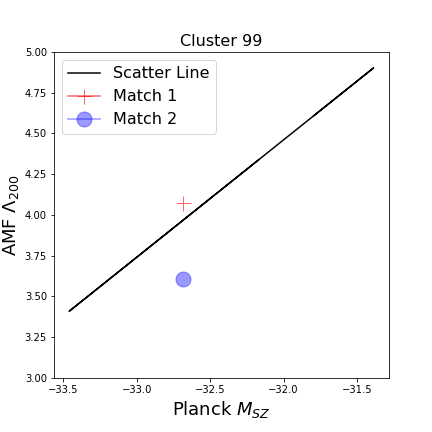}
  \caption{Cluster positions with respect to the scaling line for PSZ2 99}
  \label{fig:2}
\end{minipage}%
\begin{minipage}{.75\textwidth}
  \centering
  \includegraphics[width=.6\textwidth, height=.6\textwidth]{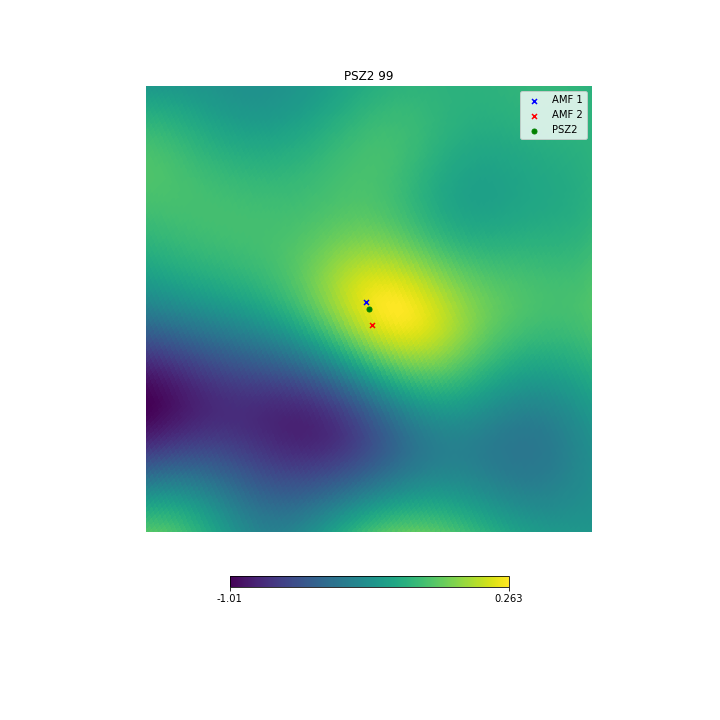}
  \caption{SZ map for PSZ2 99 and its AMF counterparts}
  \label{fig:3}
\end{minipage}
\end{figure}

\textbf{1. PSZ2 G028.63+50.15 (Cluster 99)}: Two AMF counterparts (of richness $\Lambda_{200}$) 59 and 37 and $\Delta z$ 0.29 and 0.0002 (from the PSZ2 redshift z=0.0916). The angular separations $\Delta_{\theta}$ of the 2 clusters are 1.19 and 2.55 respectively. The AMF clusters lie on either side of the scaling relation (Fig. \ref{fig:2}), while from Fig.\ref{fig:3} we see that the richer AMF cluster most likely contributes to the SZ signal. According to Hogan et al. \citep{ref:17} a brighter source nearby could have swamped catalog detections. The PSZ2 redshift is sourced from an Abell cluster (A2108). 

\begin{figure}[H]
\centering
\begin{minipage}{.8\textwidth}
  \centering
  \includegraphics[width=.5\textwidth, height=.5\textwidth]{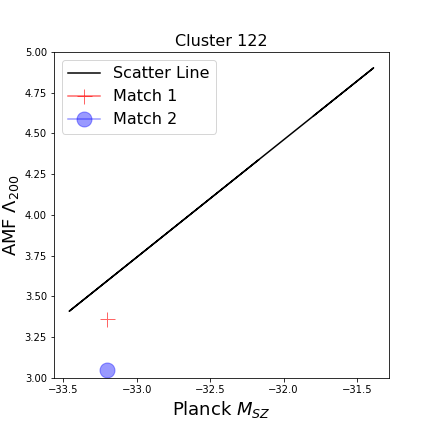}
  \caption{Cluster positions with respect to the scaling line for PSZ2 122}
  \label{fig:4}
\end{minipage}%
\begin{minipage}{.8\textwidth}
  \centering
  \includegraphics[width=.7\textwidth, height=.7\textwidth]{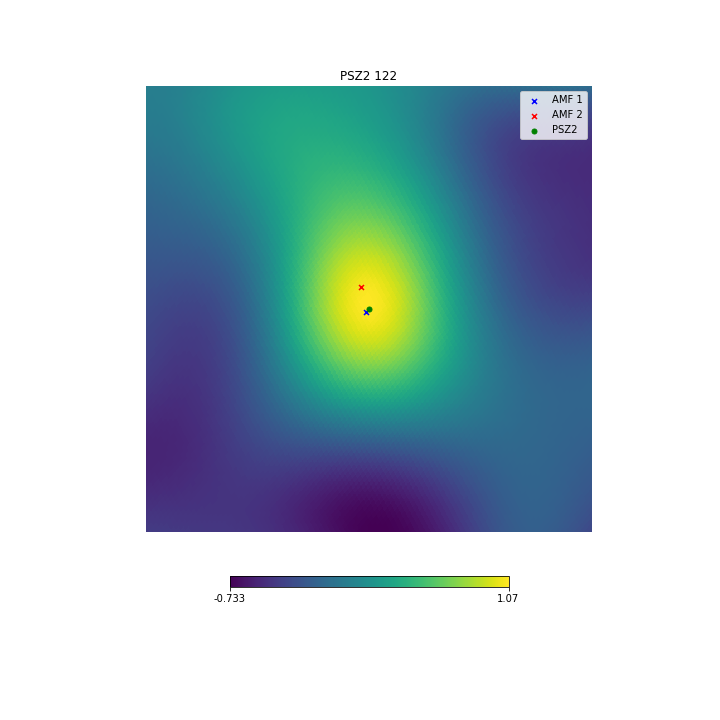}
  \caption{SZ map for PSZ2 122 and its AMF counterparts}
  \label{fig:5}
\end{minipage}
\end{figure}

\textbf{2. PSZ2 G034.38+51.57 (Cluster 122)}:  Two AMF counterparts (of richness $\Lambda_{200}$) 29 and 21 and $\Delta z$ 0.1647 and 0.4303 (from the PSZ2 redshift z=0.0411). The angular separations $\Delta_{\theta}$ of the 2 clusters are 0.29 and 1.55 respectively.. Both optical richnesses lie below the scaling relation (Fig. \ref{fig:4}), implying that the SZ signal could potentially have characterized two lower richness clusters as one cluster, while the SZ map (Fig. \ref{fig:5}) shows that the bulk of the SZ signal contribution comes from the region of the higher richness AMF cluster. According to Fujita et al \citep{ref:18} this cluster is undergoing a merger. The PSZ2 redshift is sourced from an Abell cluster (A2107). 

\textbf{3. PSZ2 G040.03+74.95 (Cluster 145)}: Two AMF counterparts (of richness $\Lambda_{200}$) 79 and 57 and $\Delta z$ 0.42 and 0.017 (from the PSZ2 redshift z=0.0612). The angular separations $\Delta_{\theta}$ of the 2 clusters are 3.15 and 1.29 respectively. From Fig. \ref{fig:4}, both optical richnesses lie above the scaling relation (which would imply that PSZ2 has not overestimated the $M_{SZ}$ value and characterized two clusters as one). The SZ map image shows that the SZ signal source is most likely at the location of the 2nd ranked AMF cluster (Fig. \ref{fig:5}). The PSZ2 redshift is sourced from a Reflex cluster (RXC J1359.2+2758). 

\begin{figure}[H]
\centering
\begin{minipage}{.8\textwidth}
  \centering
  \includegraphics[width=.5\textwidth, height=.5\textwidth]{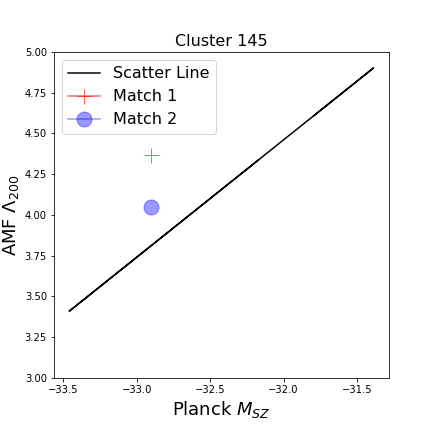}
  \caption{Cluster positions with respect to the scaling line for PSZ2 145}
  \label{fig:6}
\end{minipage}%
\begin{minipage}{.8\textwidth}
  \centering
  \includegraphics[width=.7\textwidth, height=.7\textwidth]{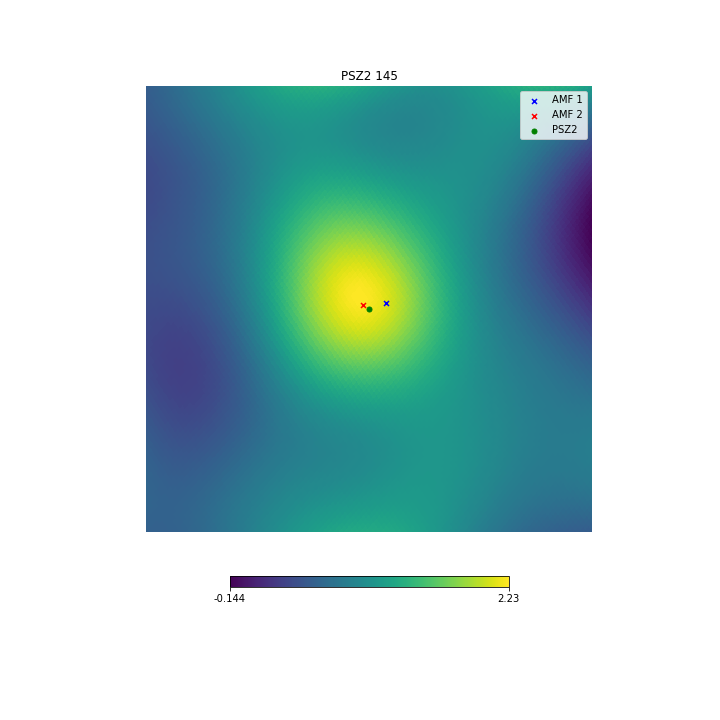}
  \caption{SZ map for PSZ2 145 and its AMF counterparts}
  \label{fig:7}
\end{minipage}
\end{figure}

\begin{figure}[H]
\centering
\begin{minipage}{.75\textwidth}
  \centering
  \includegraphics[width=.5\textwidth, height=.5\textwidth]{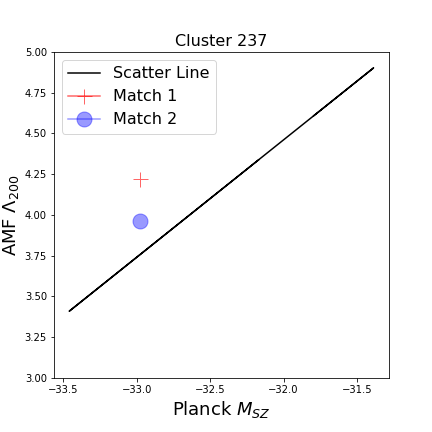}
  \caption{Cluster positions with respect to the scaling line for PSZ2 237}
  \label{fig:8}
\end{minipage}%
\begin{minipage}{.75\textwidth}
  \centering
  \includegraphics[width=.55\textwidth, height=.55\textwidth]{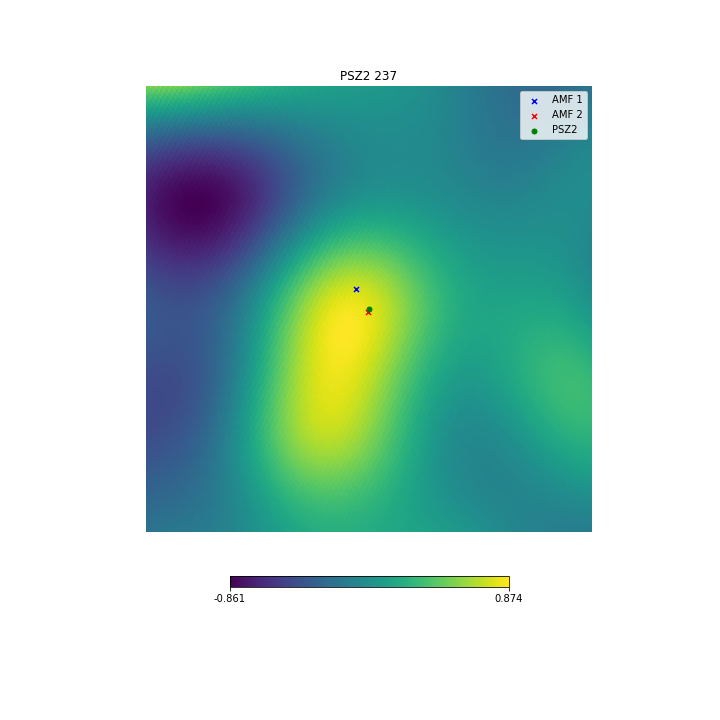}
  \caption{SZ map for PSZ2 237 and its AMF counterparts}
  \label{fig:9}
\end{minipage}
\end{figure}

\textbf{4. PSZ2 G057.78+52.32 (Cluster 237)}:  Two AMF counterparts (of richness $\Lambda_{200}$) 68 and 53 and $\Delta z$ 0.1422 and 0.0023 (from the PSZ2 redshift z=0.0654). The angular separations $\Delta_{\theta}$ of the 2 clusters are 3.99 and 0.54 respectively. From Fig. \ref{fig:8}, both optical richnesses lie above the scaling relation (which would imply that PSZ2 has not overestimated the $M_{SZ}$ value and characterized two clusters as one). However, from the SZ map (Fig \ref{fig:9}) we see that though the lower richness optical cluster center coincides with the SZ2 cluster center, the SZ signal could have had contributions from sources at the locations of both AMF clusters. The PSZ2 redshift is sourced from an Abell cluster (A2124). 

\begin{figure}[H]
\centering
\begin{minipage}{.75\textwidth}
  \centering
  \includegraphics[width=.5\textwidth, height=.5\textwidth]{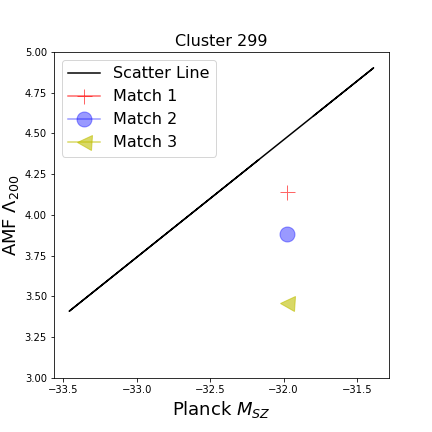}
  \caption{Cluster positions with respect to the scaling line for PSZ2 299}
  \label{fig:12}
\end{minipage}%
\begin{minipage}{.75\textwidth}
  \centering
  \includegraphics[width=.6\textwidth, height=.6\textwidth]{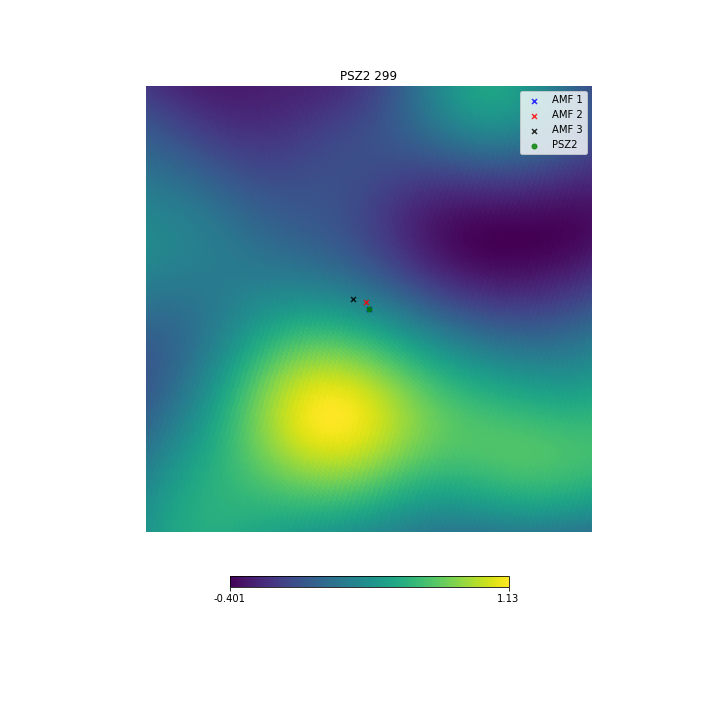}
  \caption{SZ map for PSZ2 299 and its AMF counterparts}
  \label{fig:13}
\end{minipage}
\end{figure}

\textbf{5. PSZ2 G070.89+49.26 (Cluster 299)}:  Three AMF counterparts (of richness $\Lambda_{200}$) 63, 49 and 32 and $\Delta z$ 0.0962, 0.1063, 0.1245 from the PSZ2 redshift). The angular separations $\Delta_{\theta}$ of the 3 clusters are 0.068, 1.29 and 3.1 respectively. All 3 optical clusters lie below the scaling relation (Fig \ref{fig:12}) while the SZ map shows possible contributions from the region of the 1st and 2nd ranked clusters, with the 1st ranked cluster corresponding to the peak of the SZ signal (Fig. \ref{fig:13}). The PSZ2 cluster in question does not have a redshift source listed in the PSZ2 paper \citep{ref:8}. In general, the high redshift of this Planck cluster (z=0.61) makes it unreliable for optical redshift characterization. 

\begin{figure}[H]
\centering
\begin{minipage}{.8\textwidth}
  \centering
  \includegraphics[width=.5\textwidth, height=.5\textwidth]{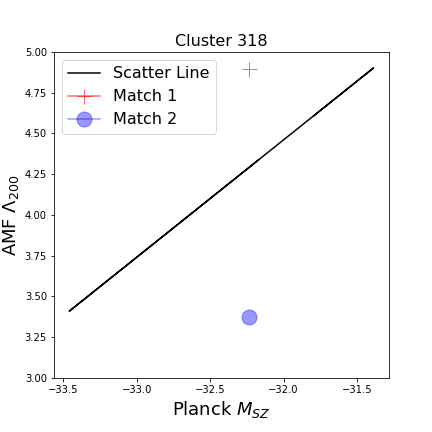}
  \caption{Cluster positions with respect to the scaling line for PSZ2 318}
  \label{fig:14}
\end{minipage}%
\begin{minipage}{.8\textwidth}
  \centering
  \includegraphics[width=.7\textwidth, height=.7\textwidth]{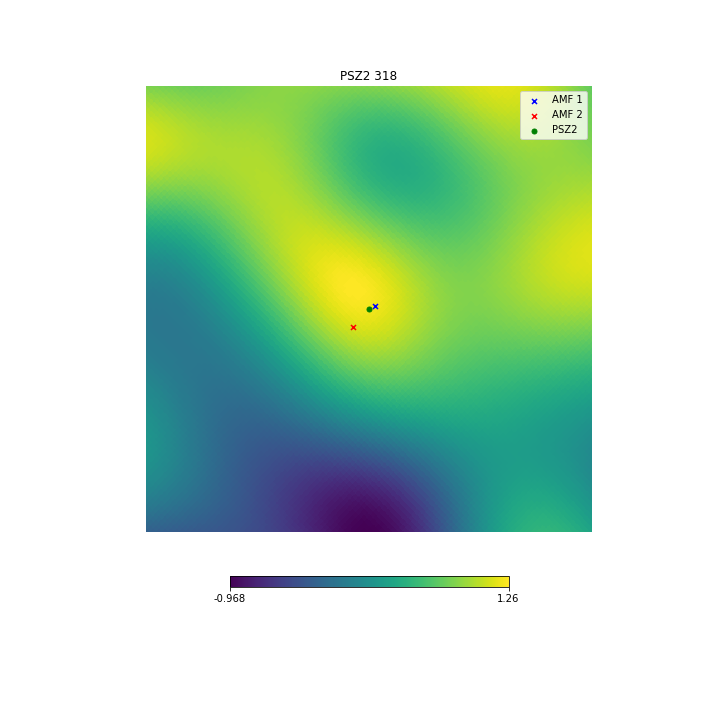}
  \caption{SZ map for PSZ2 318 and its AMF counterparts}
  \label{fig:15}
\end{minipage}
\end{figure}

\textbf{6. PSZ2 G074.75-24.59 (Cluster 318)}: Two AMF counterparts (of richness $\Lambda_{200}$) 133 and 29 and $\Delta z$ 0.0944 and 0.0162 (from the PSZ2 redshift z=0.25). The angular separations $\Delta_{\theta}$ of the 2 clusters are 1.61 and 3.38 respectively. The AMF clusters lie on either side of the scaling relation (Fig. \ref{fig:14}),  while from Fig.\ref{fig:15} we see that the most likely contribution to the SZ signal arises from the region of the richer AMF cluster center, which implies that the AMF redshift characterization might need to be corrected. WHL cluster identifies a cluster with the same z ($\sim 0.34$) in the same region of the sky. The PSZ2 cluster redshift is sourced from an Abell cluster with a redshift follow-up (ZwCl 2143.5+2014).

\begin{figure}[H]
\centering
\begin{minipage}{.75\textwidth}
  \centering
  \includegraphics[width=.5\textwidth, height=.5\textwidth]{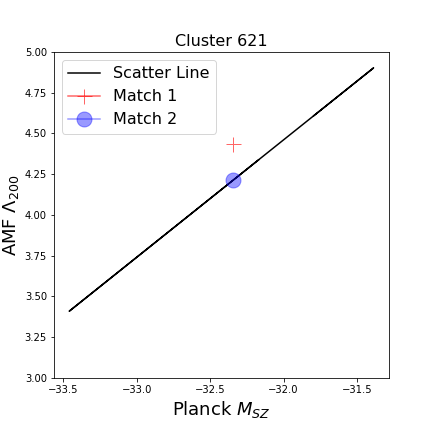}
  \caption{Cluster positions with respect to the scaling line for PSZ2 621}
  \label{fig:16}
\end{minipage}%
\begin{minipage}{.75\textwidth}
  \centering
  \includegraphics[width=.6\textwidth, height=.6\textwidth]{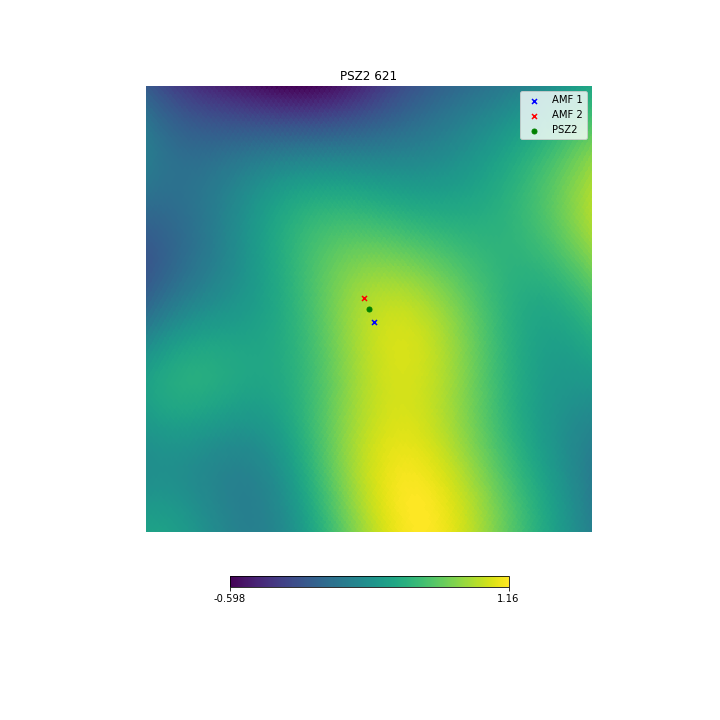}
  \caption{SZ map for PSZ2 621 and its AMF counterparts}
  \label{fig:17}
\end{minipage}
\end{figure}

\textbf{7. PSZ2 G126.07-49.55 (Cluster 621)}: Two AMF counterparts (of richness $\Lambda_{200}$) 85 and 68 and $\Delta z$ 0.020 and 0.048 (from the PSZ2 redshift z=0.49). The angular separations $\Delta_{\theta}$ of the 2 clusters are 0.81 and 0.71 respectively. The lower richness cluster lies on the regression line for the scaling relation while the 1st ranked match lies above it (Fig. \ref{fig:16}). The SZ map shows optical clusters lying on either side of the PSZ2 cluster center, thus implying that clusters at the locations of the optical clusters could have potentially contributed to the SZ signal (Fig. \ref{fig:17}). The CAMIRA catalog \citep{ref:15} identifies a cluster in the region at z $\sim$ 0.5.  

\begin{figure}[H]
\centering
\begin{minipage}{.75\textwidth}
  \centering
  \includegraphics[width=.5\textwidth, height=.5\textwidth]{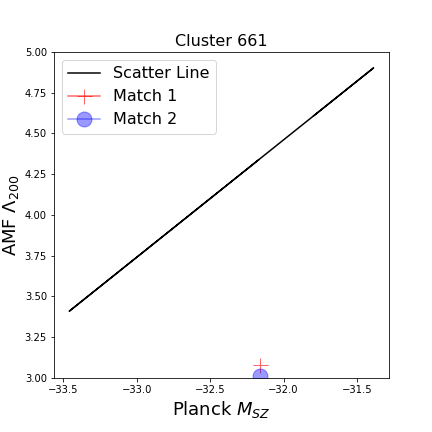}
  \caption{Cluster positions with respect to the scaling line for PSZ2 661}
  \label{fig:18}
\end{minipage}%
\begin{minipage}{.75\textwidth}
  \centering
  \includegraphics[width=.6\textwidth, height=.6\textwidth]{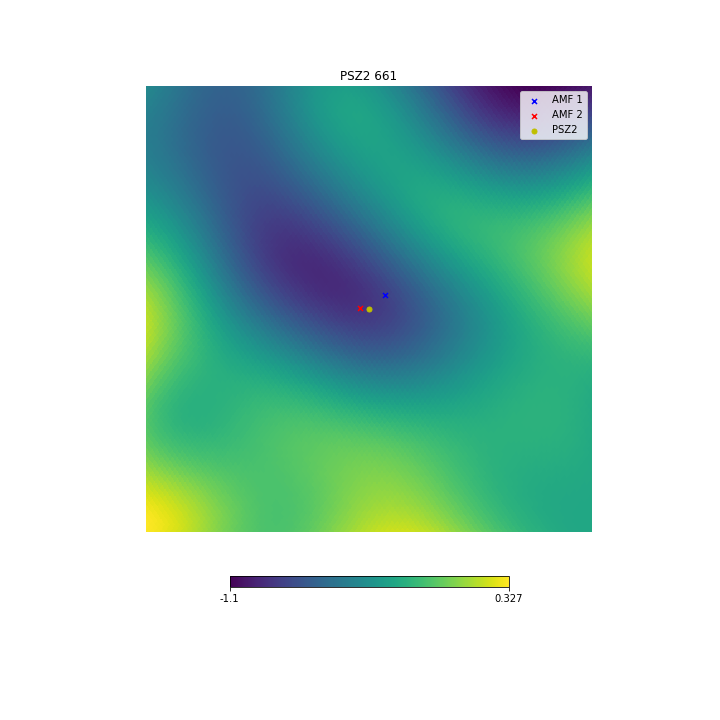}
  \caption{SZ map for PSZ2 661 and its AMF counterparts}
  \label{fig:19}
\end{minipage}
\end{figure}

\textbf{8. PSZ2 G135.06+54.39 (Cluster 661)}:  Two AMF counterparts (of richness $\Lambda_{200}$) 22 and 20 and $\Delta z$ 0.1494 and -0.0002 (from the PSZ2 redshift z=0.3169). The angular separations $\Delta_{\theta}$ of the 2 clusters are 4.25 and 1.82 respectively. From Fig. \ref{fig:18}, both optical richnesses lie below the scaling relation, implying that the SZ signal could potentially have characterized two lower richness clusters as one cluster. The SZ map (Fig \ref{fig:19}) shows that the optical cluster centers lie on either side of the PSZ2 center. 

\begin{figure}[H]
\centering
\begin{minipage}{.8\textwidth}
  \centering
  \includegraphics[width=.5\textwidth, height=.5\textwidth]{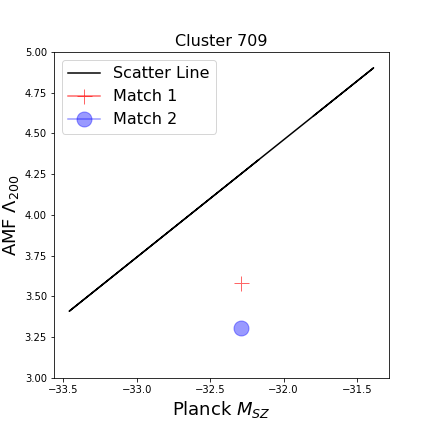}
  \caption{Cluster positions with respect to the scaling line for PSZ2 709}
  \label{fig:21}
\end{minipage}%
\begin{minipage}{.8\textwidth}
  \centering
  \includegraphics[width=.7\textwidth, height=.7\textwidth]{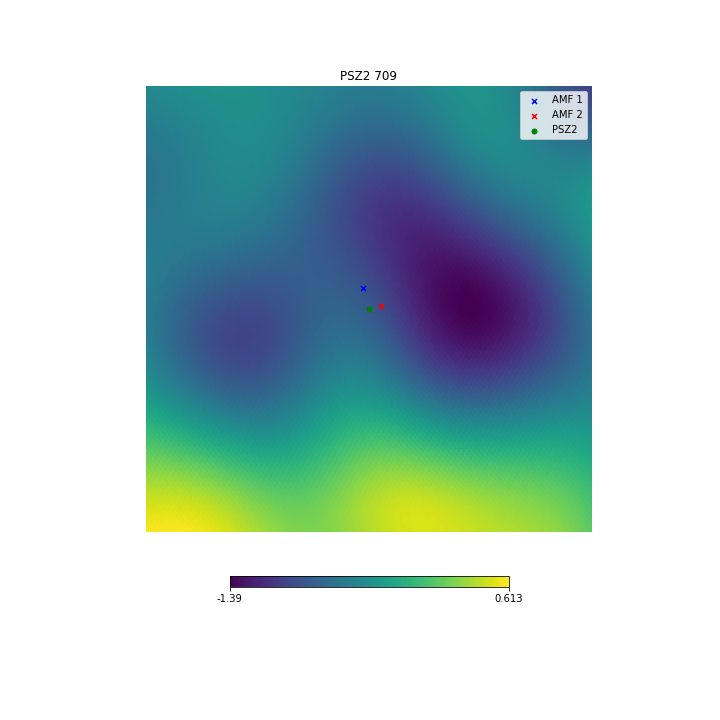}
  \caption{SZ map for PSZ2 709 and its AMF counterparts}
  \label{fig:22}
\end{minipage}
\end{figure}

\textbf{9. PSZ2 G145.65+59.30 (Cluster 709)}: Two AMF counterparts (of richness $\Lambda_{200}$) 36 and 27 and $\Delta z$ 0.007 and -0.0023 (from the PSZ2 redshift z=0.3475). The angular separations $\Delta_{\theta}$ of the 2 clusters are 3.69 and 2.1 respectively. From Fig. \ref{fig:21}, both optical richnesses lie below the scaling relation, implying that the SZ signal could potentially have characterized two lower richness clusters as one cluster. The SZ map (Fig \ref{fig:22}) shows that the optical cluster centers lie on either side of the PSZ2 center. In the same region of the sky, CAMIRA \citep{ref:15} and LRG \citep{ref:16} both identify a cluster with z$\sim$0.34. 

\begin{figure}[H]
\centering
\begin{minipage}{.8\textwidth}
  \centering
  \includegraphics[width=.5\textwidth, height=.5\textwidth]{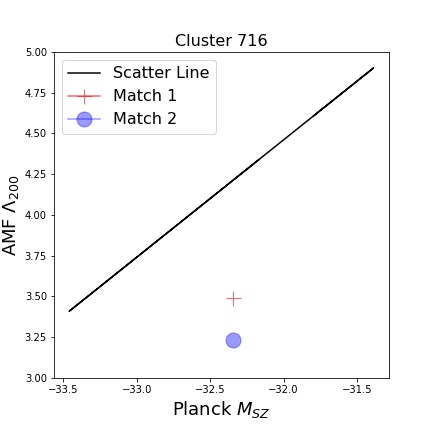}
  \caption{Cluster positions with respect to the scaling line for PSZ2 716}
  \label{fig:23}
\end{minipage}%
\begin{minipage}{.8\textwidth}
  \centering
  \includegraphics[width=.7\textwidth, height=.7\textwidth]{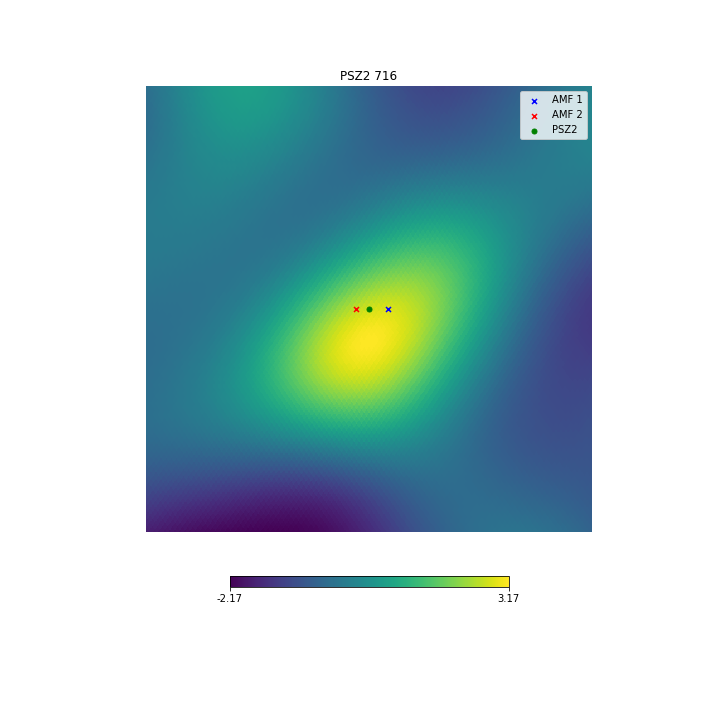}
  \caption{SZ map for PSZ2 716 and its AMF counterparts}
  \label{fig:24}
\end{minipage}
\end{figure}

\textbf{10. PSZ2 G146.82+40.97 (Cluster 716)}:  Two AMF counterparts (of richness $\Lambda_{200}$) 33 and 25 and $\Delta z$ 0.0729 and 0.0363 (from the PSZ2 redshift z=0.26). The angular separations $\Delta_{\theta}$ of the 2 clusters are 1.98 and 1.38 respectively. Both optical richnesses lie below the scaling relation, implying that the SZ signal could potentially have characterized two lower richness clusters as one cluster, while the SZ map shows optical clusters lying on either side of the PSZ2 cluster center. (Figs. \ref{fig:23} and \ref{fig:24}). The LRG catalog \citep{ref:16} locates a cluster in the same region of the sky at z $\sim$ 0.34, the same redshift  as the richer AMF cluster.  

\begin{figure}[H]
\centering
\begin{minipage}{.8\textwidth}
  \centering
  \includegraphics[width=.5\textwidth, height=.5\textwidth]{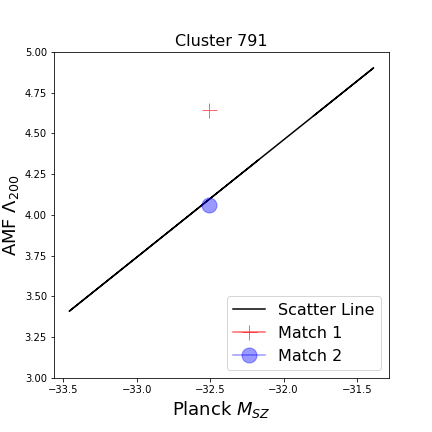}
  \caption{Cluster positions with respect to the scaling line for PSZ2 791}
  \label{fig:25}
\end{minipage}%
\begin{minipage}{.8\textwidth}
  \centering
  \includegraphics[width=.7\textwidth, height=.7\textwidth]{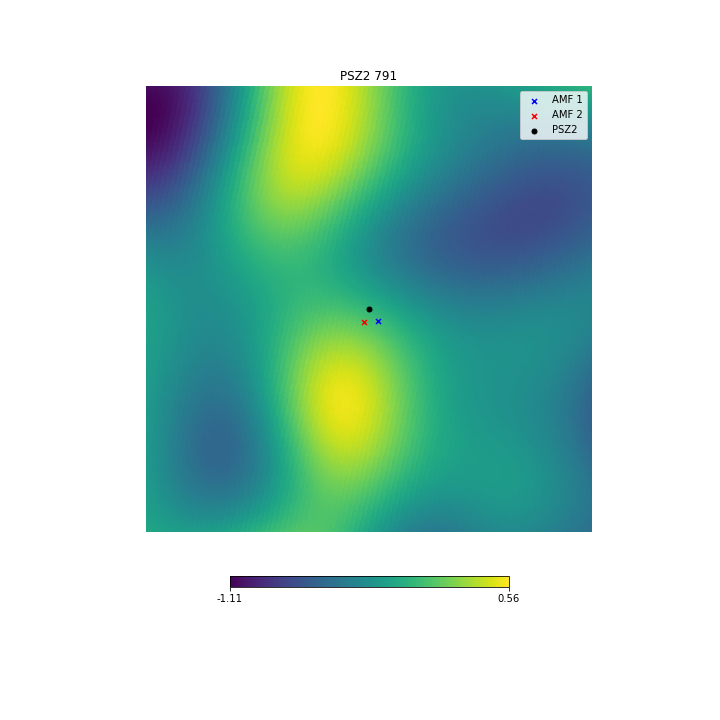}
  \caption{SZ map for PSZ2 791 and its AMF counterparts}
  \label{fig:26}
\end{minipage}
\end{figure}

\textbf{11. PSZ2 G165.68+44.01 (Cluster 791)}: Two AMF counterparts (of richness $\Lambda_{200}$) 104 and 32 and $\Delta z$ -0.0012 and -0.0015 (from the PSZ2 redshift z=0.2097). The angular separations $\Delta_{\theta}$ of the 2 clusters are 2.45 and 2.33 respectively. The lower ranked cluster lies on the regression line for the scaling relation, while the richer AMF cluster lies above it (Fig. \ref{fig:25}). From the SZ map, the 2 optical clusters lie on either side of the PSZ2 cluster center (Fig. \ref{fig:26}). Both the redMaPPer and WHL catalogs have rich optical clusters lying at about the same redshift as the PSZ2 cluster. 

\begin{figure}[H]
\centering
\begin{minipage}{.8\textwidth}
  \centering
  \includegraphics[width=.5\textwidth, height=.5\textwidth]{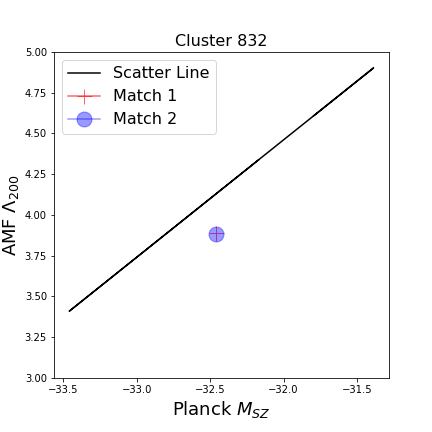}
  \caption{Cluster positions with respect to the scaling line for PSZ2 832}
  \label{fig:27}
\end{minipage}%
\begin{minipage}{.8\textwidth}
  \centering
  \includegraphics[width=.7\textwidth, height=.7\textwidth]{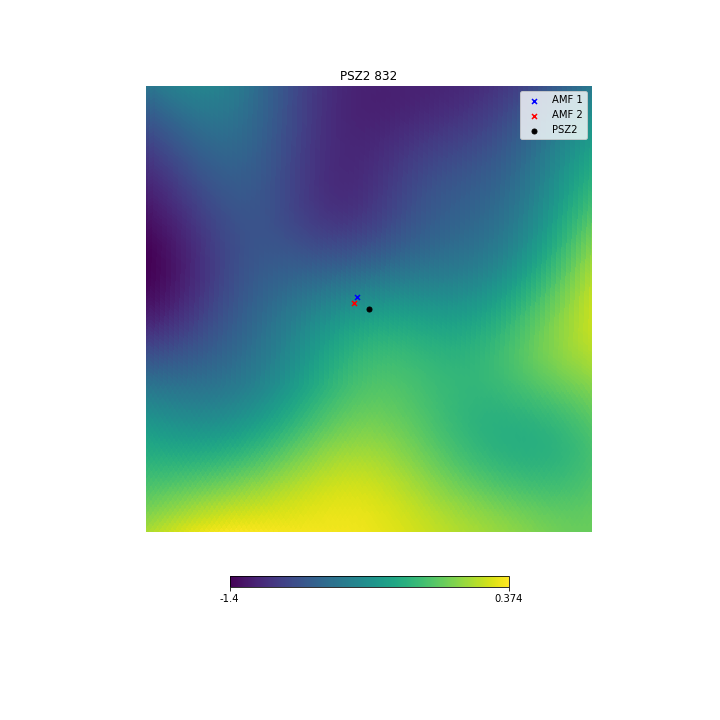}
  \caption{SZ map for PSZ2 832 and its AMF counterparts}
  \label{fig:28}
\end{minipage}
\end{figure}

\textbf{12. PSZ2 G178.00+42.32 (Cluster 832)}:  Two AMF counterparts (of richness $\Lambda_{200}$) 48.9 and 48.7 and $\Delta z$ 0.1499 and 0.1281 (from the PSZ2 redshift z=0.2368). The angular separations $\Delta_{\theta}$ of the 2 clusters are 2.84 and 2.64 respectively. Both AMF clusters lie below the regression line in the scaling relation (Fig. \ref{fig:27}). The PSZ2 cluster has a match in WHL with $\Delta z > 0.1$ and no redMaPPer counterpart. From Fig. \ref{fig:28}, which shows the SZ map for these clusters, it is not obvious which optical cluster would be a better counterpart to the PSZ2 cluster. The Northern Optical Survey \citep{ref:21} finds a cluster in the same region of the sky with a z=0.226. 

\begin{figure}[H]
\centering
\begin{minipage}{.75\textwidth}
  \centering
  \includegraphics[width=.5\textwidth, height=.5\textwidth]{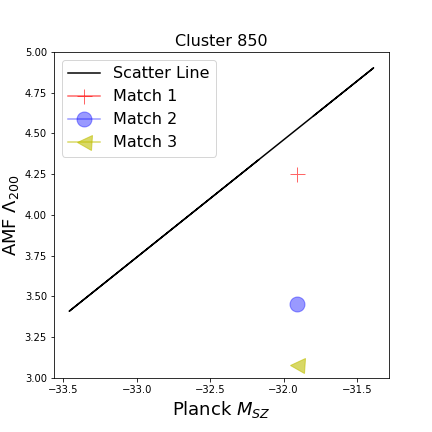}
  \caption{Cluster positions with respect to the scaling line for PSZ2 850}
  \label{fig:29}
\end{minipage}%
\begin{minipage}{.75\textwidth}
  \centering
  \includegraphics[width=.6\textwidth, height=.6\textwidth]{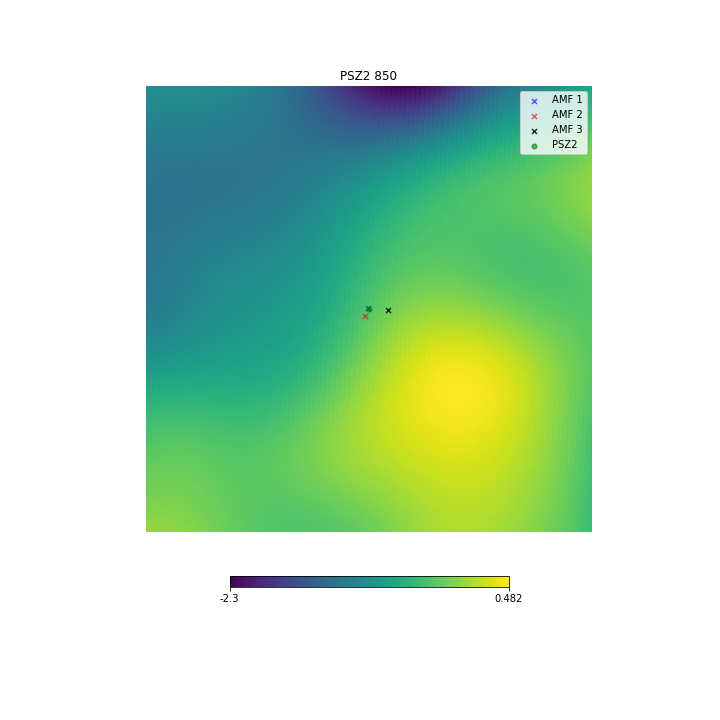}
  \caption{SZ map for PSZ2 850 and its AMF counterparts}
  \label{fig:30}
\end{minipage}
\end{figure}

\textbf{13. PSZ2 G183.90+42.99 (Cluster 850)}:  Three AMF counterparts (of richness $\Lambda_{200}$) 70, 32 and 22 and $\Delta z$ 0.0445, 0.0048, 0.1072 from the PSZ2 redshift of z=0.561). The angular separations $\Delta_{\theta}$ of the 3 clusters are 0.32, 1.31 and 3.2 respectively. All 3 optical clusters lie below the scaling relation (Fig. \ref{fig:29}), while, from the SZ map, it is evident that the Rank 1 and Rank 2 matches might correspond to clusters that contribute to the SZ signal (Fig. \ref{fig:30}). The CAMIRA survey \citep{ref:15} finds a cluster in this region of the sky at z $\sim$0.56. The PSZ2 cluster redshift is validated by an optical cluster from the WHL catalog and optical finders are typically unreliable at such high z values. 

\begin{figure}[H]
\centering
\begin{minipage}{.75\textwidth}
  \centering
  \includegraphics[width=.5\textwidth, height=.5\textwidth]{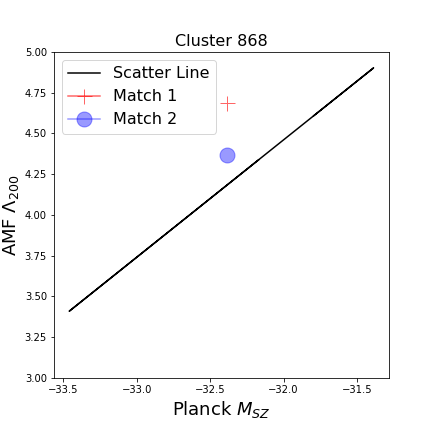}
  \caption{Cluster positions with respect to the scaling line for PSZ2 868}
  \label{fig:31}
\end{minipage}%
\begin{minipage}{.75\textwidth}
  \centering
  \includegraphics[width=.6\textwidth, height=.6\textwidth]{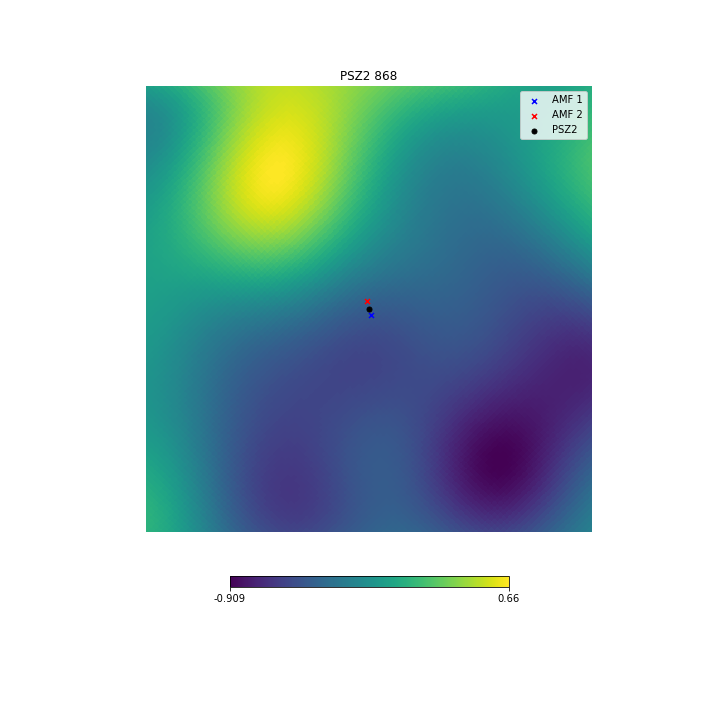}
  \caption{SZ map for PSZ2 868 and its AMF counterparts}
  \label{fig:32}
\end{minipage}
\end{figure}

\begin{figure}[H]
\centering
\begin{minipage}{.75\textwidth}
  \centering
  \includegraphics[width=.6\textwidth, height=.6\textwidth]{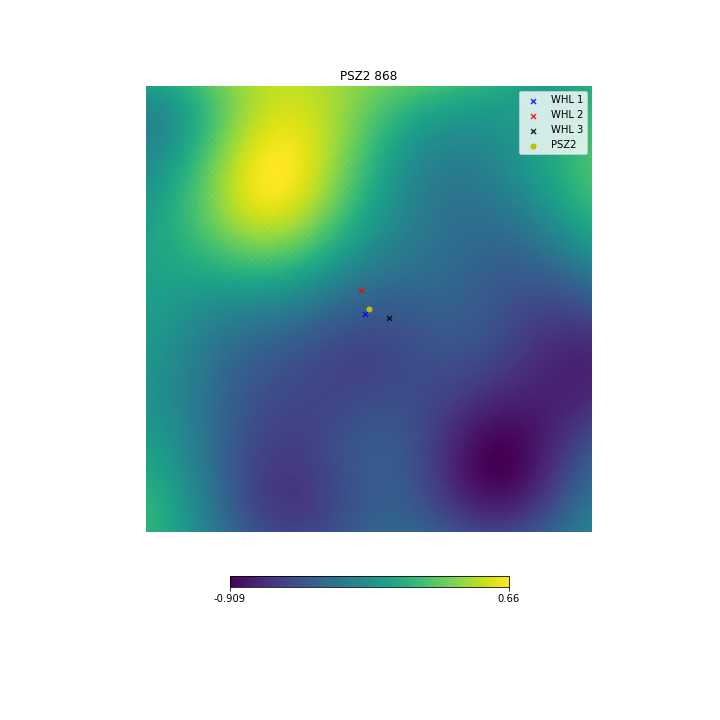}
  \caption{SZ map for PSZ2 868 and its WHL counterparts}
  \label{fig:33}
\end{minipage}
\end{figure}

\textbf{14. PSZ2 G187.74+20.66 (Cluster 868)}:  Two AMF counterparts (of richness $\Lambda_{200}$) 108 and 79 and $\Delta z$ 0.0066 and 0.0046 (from the PSZ2 redshift z=0.1929). The angular separations $\Delta_{\theta}$ of the 2 clusters are 1.05 and 1.49 respectively. Both AMF clusters lie above the scaling relation (Fig. \ref{fig:31}), while, from the SZ map (Fig \ref{fig:32}), the optical clusters lie on opposite sides of the PSZ2 cluster center, but very close to each other. From Fig. \ref{fig:33}, we see the SZ map for this Planck cluster and its multiple WHL matches (using our matching technique). The redshift of this PSZ2 cluster is drawn from the rich WHL cluster ($R_{L_{*}}\sim$104) whose center coincides with the center of the Planck cluster in Fig. \ref{fig:33}. 

\begin{figure}[H]
\centering
\begin{minipage}{.8\textwidth}
  \centering
  \includegraphics[width=.5\textwidth, height=.5\textwidth]{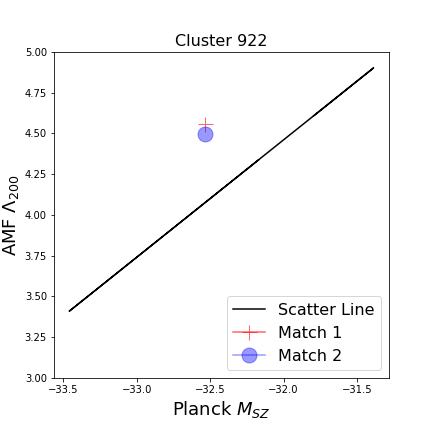}
  \caption{Cluster positions with respect to the scaling line for PSZ2 922}
  \label{fig:34}
\end{minipage}%
\begin{minipage}{.8\textwidth}
  \centering
  \includegraphics[width=.7\textwidth, height=.7\textwidth]{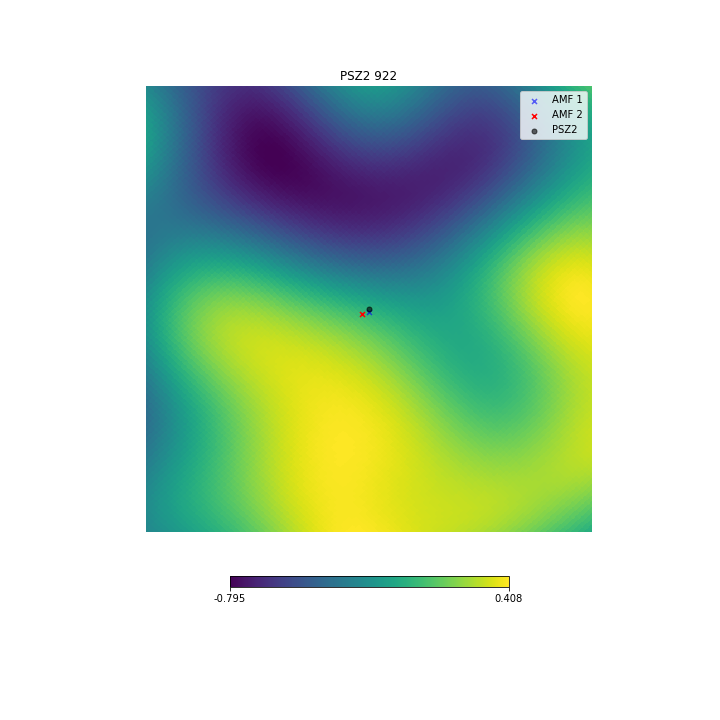}
  \caption{SZ map for PSZ2 922 and its AMF counterparts}
  \label{fig:35}
\end{minipage}
\end{figure}

\textbf{15. PSZ2 G204.10+16.51 (Cluster 922)}:  Two AMF counterparts (of richness $\Lambda_{200}$) 95 and 90 and $\Delta z$ 0.0193 and 0.0294 (from the PSZ2 redshift z=0.122). The angular separations $\Delta_{\theta}$ of the 2 clusters are 0.33 and 1.13 respectively. Both AMF clusters lie above the scaling relation (Fig. \ref{fig:34}). From the SZ map (Fig. \ref{fig:35}) we see that both AMF counterparts could possibly be clusters that contribute to the SZ signal, however, the Rank 1 match cluster center coincides better with the PSZ2 cluster center. In the region of the sky, the CAMIRA survey \citep{ref:15} finds a cluster with z $\sim$ 0.12. The PSZ2 cluster redshift is sourced from an Abell cluster with a redshift follow-up ( ZwCl 0733.1+1514).

\begin{figure}[H]
\centering
\begin{minipage}{.8\textwidth}
  \centering
  \includegraphics[width=.5\textwidth, height=.5\textwidth]{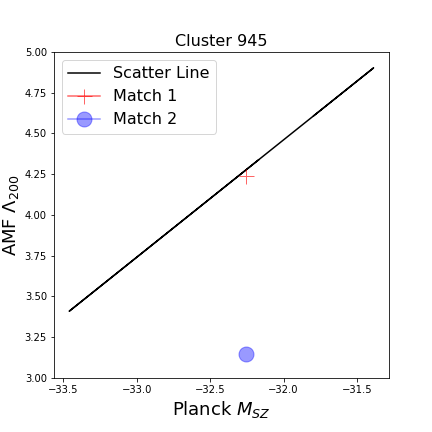}
  \caption{Cluster positions with respect to the scaling line for PSZ2 945}
  \label{fig:36}
\end{minipage}%
\begin{minipage}{.8\textwidth}
  \centering
  \includegraphics[width=.7\textwidth, height=.7\textwidth]{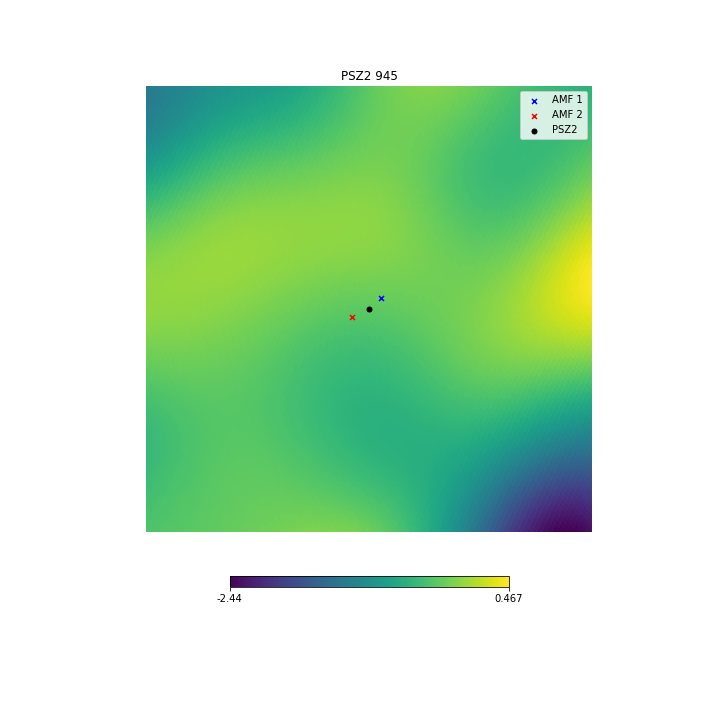}
  \caption{SZ map for PSZ2 945 and its AMF counterparts}
  \label{fig:37}
\end{minipage}
\end{figure}

\textbf{16. PSZ2 G210.01+50.85 (Cluster 945)}: Two AMF counterparts (of richness $\Lambda_{200}$) 69 and 23 and $\Delta z$ 0.074 and 0.036 (from the PSZ2 redshift z=0.319). The angular separations $\Delta_{\theta}$ of the 2 clusters are 1.36 and 1.55 respectively. The Rank 1 match lies on the scaling relation, while the Rank 2 match lies below the regression line (Fig. \ref{fig:36}). From the SZ map (Fig. \ref{fig:37}), it does seem more likely that the bulk of the contribution to the SZ signal comes from the location of the richer AMF cluster center. The PSZ2 cluster corresponds to an Abell cluster (A2634) which has had redshift follow-ups in the past \citep{ref:22} with a z$\sim$0.03. 

\begin{figure}[H]
\centering
\begin{minipage}{.8\textwidth}
  \centering
  \includegraphics[width=.5\textwidth, height=.5\textwidth]{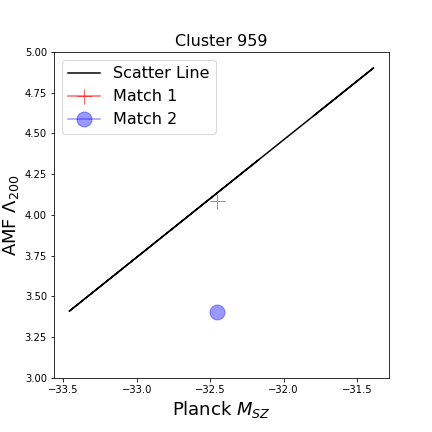}
  \caption{Cluster positions with respect to the scaling line for PSZ2 959}
  \label{fig:38}
\end{minipage}%
\begin{minipage}{.8\textwidth}
  \centering
  \includegraphics[width=.7\textwidth, height=.7\textwidth]{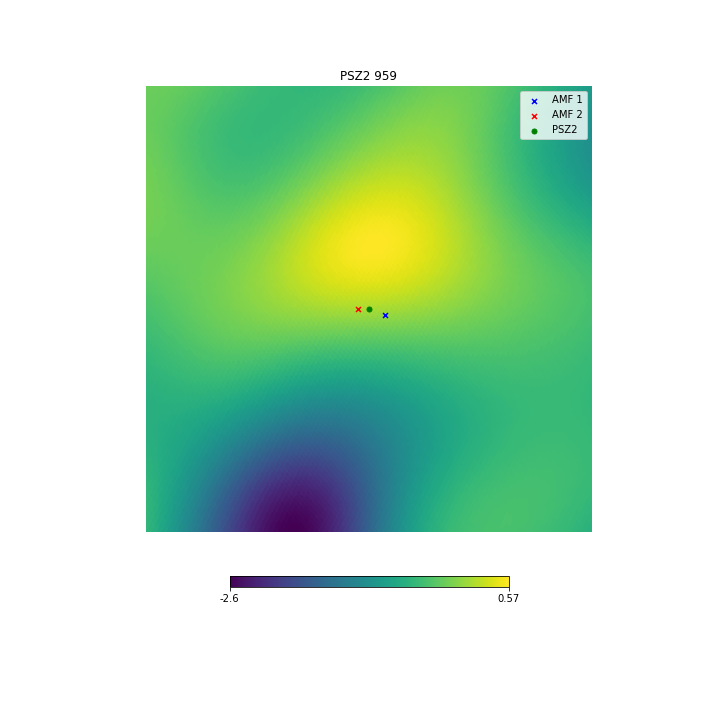}
  \caption{SZ map for PSZ2 959 and its AMF counterparts}
  \label{fig:39}
\end{minipage}
\end{figure}

\textbf{17. PSZ2 G212.80+50.63 (Cluster 959)}: Two AMF counterparts (of richness $\Lambda_{200}$) 59 and 30 and $\Delta z$ 0.123 and -0.007 (from the PSZ2 redshift z=0.2187). The angular separations $\Delta_{\theta}$ of the 2 clusters are 2.81 and 1.86 respectively. Both optical richnesses lie below the scaling relation (Fig. \ref{fig:38}), implying that the SZ signal could potentially have characterized two lower richness clusters as one cluster, while the SZ map shows optical clusters lying on either side of the PSZ2 cluster center (Fig. \ref{fig:39}). redMaPPer, similar to AMF DR9, has 2 possible optical counterparts, one of which is of lower richness and lies at a redshift value (z=0.21) close to that of the PSZ2 cluster. 

\begin{figure}[H]
\centering
\begin{minipage}{.75\textwidth}
  \centering
  \includegraphics[width=.5\textwidth, height=.5\textwidth]{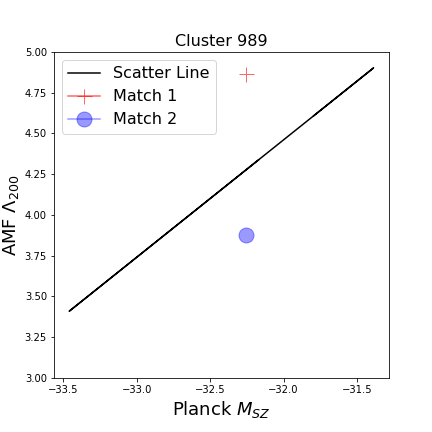}
  \caption{Cluster positions with respect to the scaling line for PSZ2 989}
  \label{fig:40}
\end{minipage}%
\begin{minipage}{.75\textwidth}
  \centering
  \includegraphics[width=.6\textwidth, height=.6\textwidth]{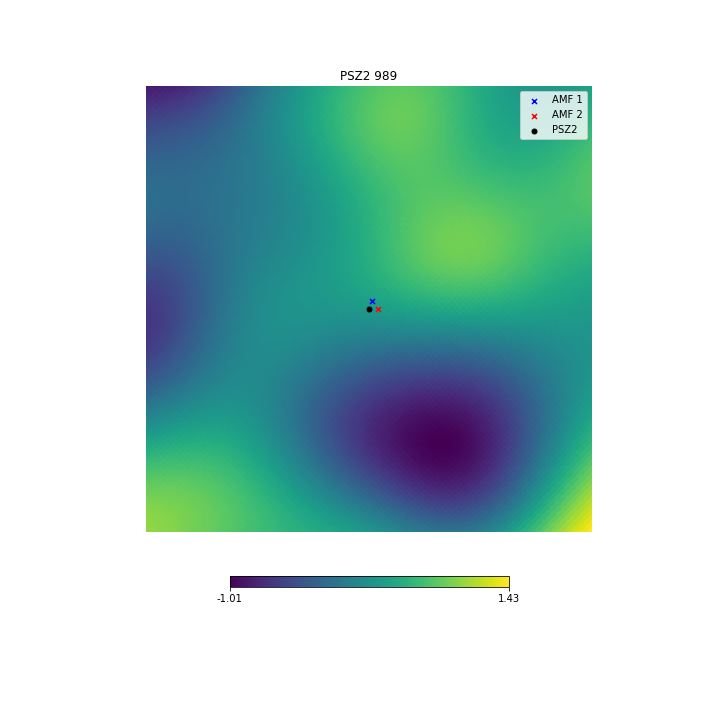}
  \caption{SZ map for PSZ2 989 and its AMF counterparts}
  \label{fig:41}
\end{minipage}
\end{figure}

\textbf{18. PSZ2 G220.11+22.91 (Cluster 989)}: Two AMF counterparts (of richness $\Lambda_{200}$) 129 and 48 and $\Delta z$ 0.2499 and 0.0005 (from the PSZ2 redshift z=0.2248). The angular separations $\Delta_{\theta}$ of the 2 clusters are 0.34 and 0.36 respectively. The optical matches lie on either side of the scaling relation (Fig. \ref{fig:40}). The SZ signal (Fig. \ref{fig:41}) could be from sources near either of the two optical matches. Redmapper has a near identical set of matching clusters (of richness $\Lambda$ 179 and 62) and $\Delta z$ 0.2503 and 0.0101 (from the PSZ2 redshift  z=0.2248) and at angular separations $\Delta_{\theta}$ 0.38 and 0.41 respectively. In the same region of the sky, CoMaLit \citep{ref:23} finds a cluster (\textbf{ZwCl 0823.2+0425N}) at z=0.47, which is the same redshift as the richer optical match for both AMF and redMaPPer. 

\begin{figure}[H]
\centering
\begin{minipage}{.8\textwidth}
  \centering
  \includegraphics[width=.5\textwidth, height=.5\textwidth]{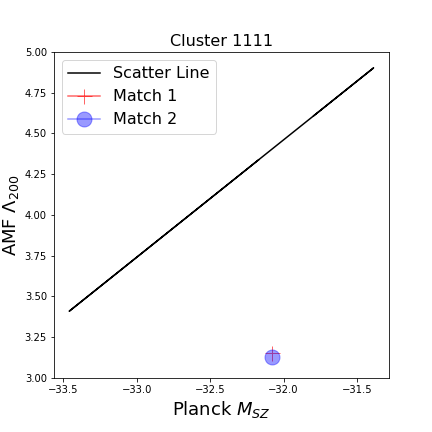}
  \caption{Cluster positions with respect to the scaling line for PSZ2 1111}
  \label{fig:42}
\end{minipage}%
\begin{minipage}{.8\textwidth}
  \centering
  \includegraphics[width=.7\textwidth, height=.7\textwidth]{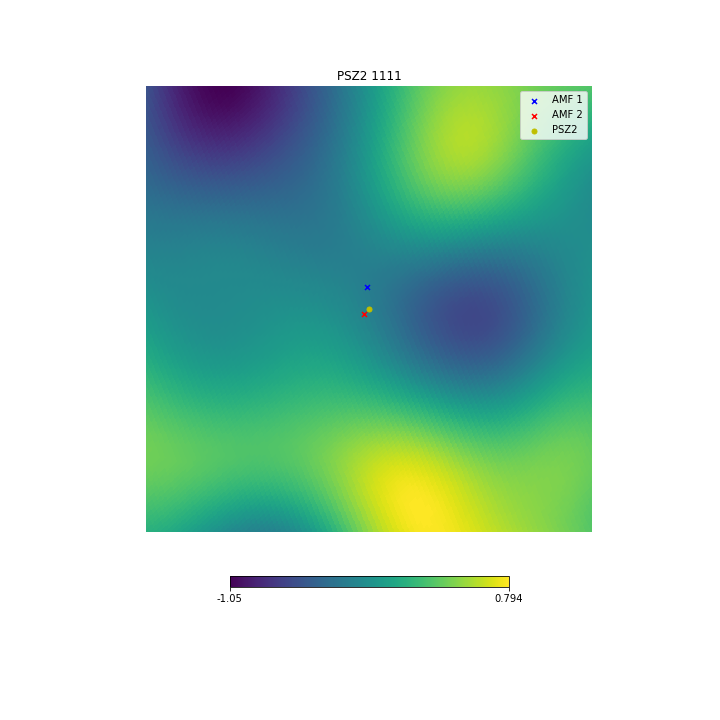}
  \caption{SZ map for PSZ2 1111 and its AMF counterparts}
  \label{fig:43}
\end{minipage}
\end{figure}

\textbf{19. PSZ2 G244.77+59.80 (Cluster 1111)}: Two AMF counterparts (of richness $\Lambda_{200}$) 23.4 and 22.9 and $\Delta z$ -0.0066 and -0.0016 (from the PSZ2 redshift z=0.4659). The angular separations $\Delta_{\theta}$ of the 2 clusters are 3.78 and 1.15 respectively. Both optical richnesses lie below the scaling relation (Fig. \ref{fig:42}), implying that the SZ signal could potentially have characterized two lower richness clusters as one cluster, while the SZ map shows optical clusters lying on either side of the PSZ2 cluster center. RedMaPPer also matches two optical counterparts of richness $\Lambda$ 74 and 38 lying at redshifts of 0.47 and 0.45 respectively. Fig. \ref{fig:43} shows the SZ map for the AMF matches for this Planck cluster. 

\begin{figure}[H]
\centering
\begin{minipage}{.75\textwidth}
  \centering
  \includegraphics[width=.5\textwidth, height=.5\textwidth]{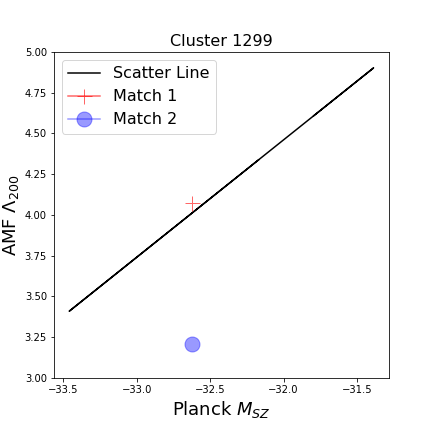}
  \caption{Cluster positions with respect to the scaling line for PSZ2 1299}
  \label{fig:45}
\end{minipage}%
\begin{minipage}{.75\textwidth}
  \centering
  \includegraphics[width=.6\textwidth, height=.6\textwidth]{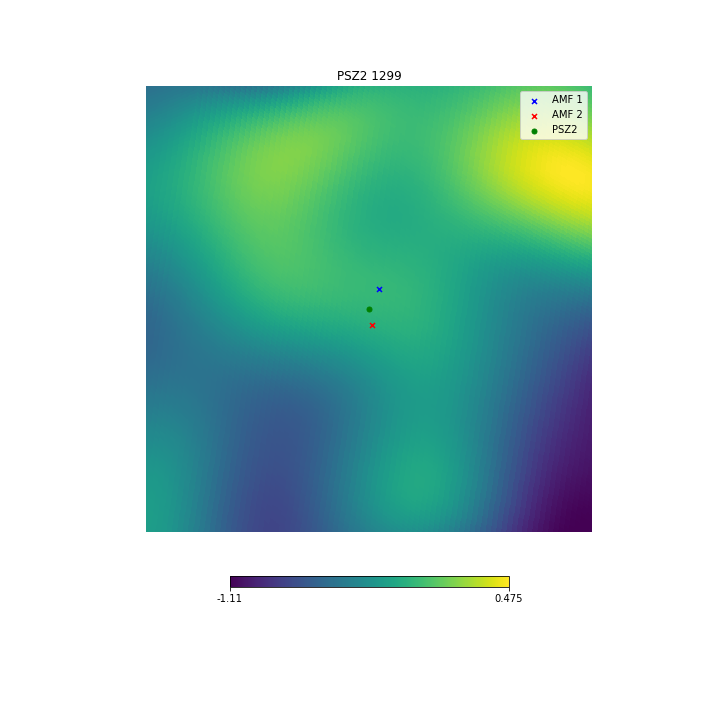}
  \caption{SZ map for PSZ2 1299 and its AMF counterparts}
  \label{fig:46}
\end{minipage}
\end{figure}

\textbf{20. PSZ2 G279.02+81.24 (Cluster 1299)}: Two AMF counterparts (of richness $\Lambda_{200}$) 59 and 25 and $\Delta z$ 0.0493 and 0.0905 (from the PSZ2 redshift z=0.18). The angular separations $\Delta_{\theta}$ of the 2 clusters are 3.84 and 2.61 respectively. The AMF clusters lie on either side of the scaling relation, with the richer cluster lying almost on the line (Fig. \ref{fig:45}) while from the SZ map (Fig. \ref{fig:46}) we see that the SZ source could be from either of the two locations (of the AMF cluster centers). Both the redMaPPer and the WHL have a mid-range richness cluster (redMaPPer $\Lambda$=49 and WHL $R_{L_{*}}$)=51 respectively) around z$\sim$ 0.22 at about a separation of $\Delta_{\theta} \sim$ 4. This might imply that the Planck cluster redshift might be better characterized. In the PSZ2 catalog itself, this cluster is described as being 'X--ray under-luminous'. 

\subsection{PSZ2 clusters (with external validation) matched by AMF DR9, with a redMaPPer counterpart} \label{appA_sub2}

In the following analyses, the AMF counterparts are ranked by richness, that is, if there are two possible AMF counterparts for a given Planck cluster, then the richer optical cluster is Ranked 1, and the AMF cluster of lower richness is Ranked 2. To look at the relative positions of the matching cluster centers in the SZ map, we utilize the MILCA full mission y maps (Compton parameter maps) as well as the standard deviation maps provided by the Planck collaboration. We obtained a signal-to-noise map from the y-map and the standard deviation map, and smoothed the result by 1$^\circ$. 

\begin{figure}[H]
\centering
\begin{minipage}{.8\textwidth}
  \centering
  \includegraphics[width=.6\textwidth, height=.6\textwidth]{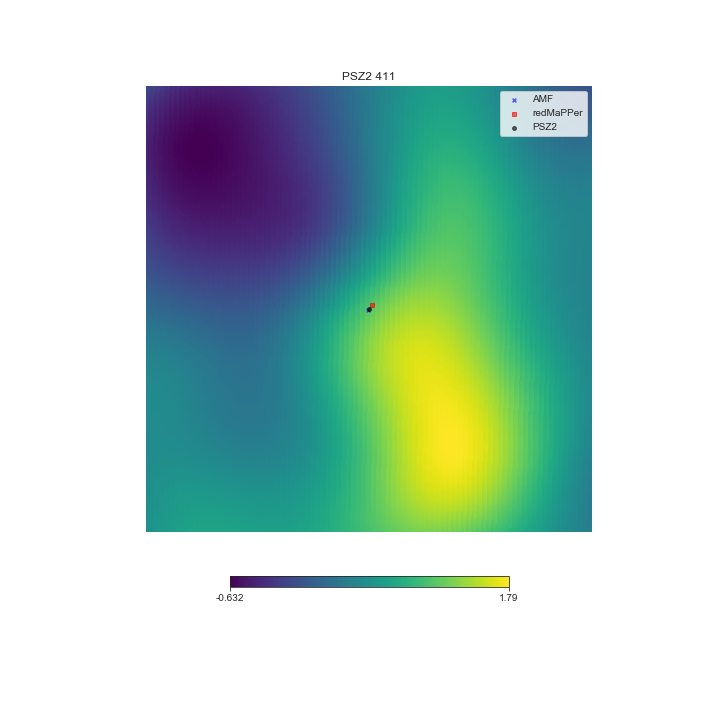}
  \caption{SZ map for PSZ2 411 and its AMF and redMaPPer counterparts}
  \label{fig:65}
\end{minipage}
\end{figure}

\textbf{1. PSZ2 G091.40-51.01 (Cluster 411)}: One AMF counterpart with a $\Delta z$=0.2188 (from the PSZ2 redshift z=0.099) with a richness $\Lambda_{200}$=138 and with a $\Delta_{\theta}$=0.17. The redMaPPer counterpart has a $\Delta z$=0.2015 with a richness $\Lambda$=144 and with a $\Delta_{\theta}$=1.03. There is a WHL counterpart which has a $\Delta z$=0.196 with a richness $R_{L_{*}}$=128 and with a $\Delta_{\theta}$=1.02. From Fig. \ref{fig:65} we see that the AMF cluster lies closer to the PSZ2 cluster center, however, the SZ signal contribution could have come sources at either location. Calculating the $\Lambda_{200}$ with the AMF value of redshift, the scatter $\Delta_{\Lambda_{200}}$ decreases from 3.2 to 2.5. 
The optical clusters identified are also extremely high in richness, and would be likely to be characterized properly. 

\begin{figure}[H]
\centering
\begin{minipage}{.8\textwidth}
  \centering
  \includegraphics[width=.5\textwidth, height=.5\textwidth]{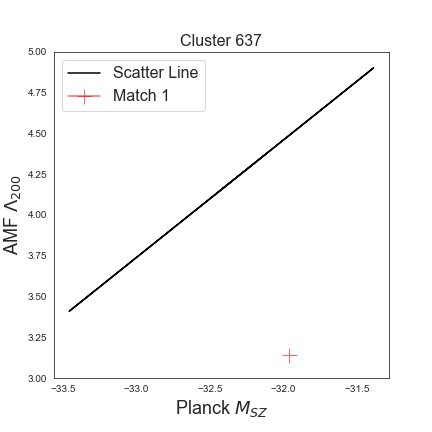}
  \caption{Cluster positions with respect to the scaling line for PSZ2 637}
  \label{fig:59}
\end{minipage}%
\begin{minipage}{.8\textwidth}
  \centering
  \includegraphics[width=.7\textwidth, height=.7\textwidth]{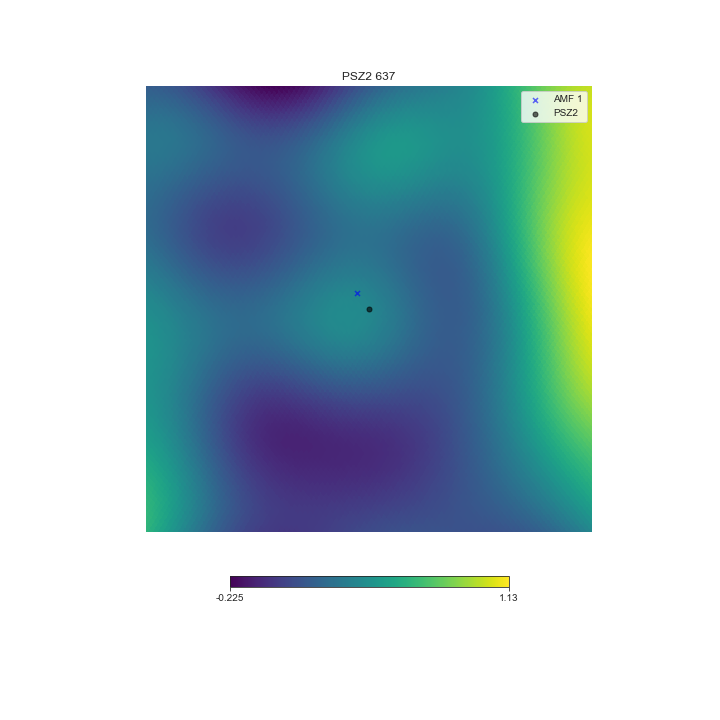}
  \caption{SZ map for PSZ2  637 and its AMF counterparts}
  \label{fig:60}
\end{minipage}
\end{figure}

\textbf{2. PSZ2 G128.18-51.08 (Cluster 637)}: One AMF counterpart with a $\Delta z$=-0.38 (from the PSZ2 redshift z=0.55) with a richness $\Lambda_{200}$=23 and with a $\Delta_{\theta}$=3.55. The redMaPPer counterpart has a richness $\Lambda$=79 and with a $\Delta_{\theta}$=0.24. There is a WHL counterpart which has a $\Delta z$=-0.0055 with a richness $R_{L_{*}}$=36 and with a $\Delta_{\theta}$=0.24. The AMF cluster lies below the scaling relation (Fig. \ref{fig:59}), while the SZ signal does seem to arise from the region of the AMF cluster center, though the two cluster centers do not coincide(Fig. \ref{fig:60}). The AMF finder is unreliable at such high values of redshift and the optical richness $\Lambda_{200}$ as well as the AMF redshift of the cluster has likely been underestimated.

\begin{figure}[H]
\centering
\begin{minipage}{.8\textwidth}
  \centering
  \includegraphics[width=.8\textwidth, height=.8\textwidth]{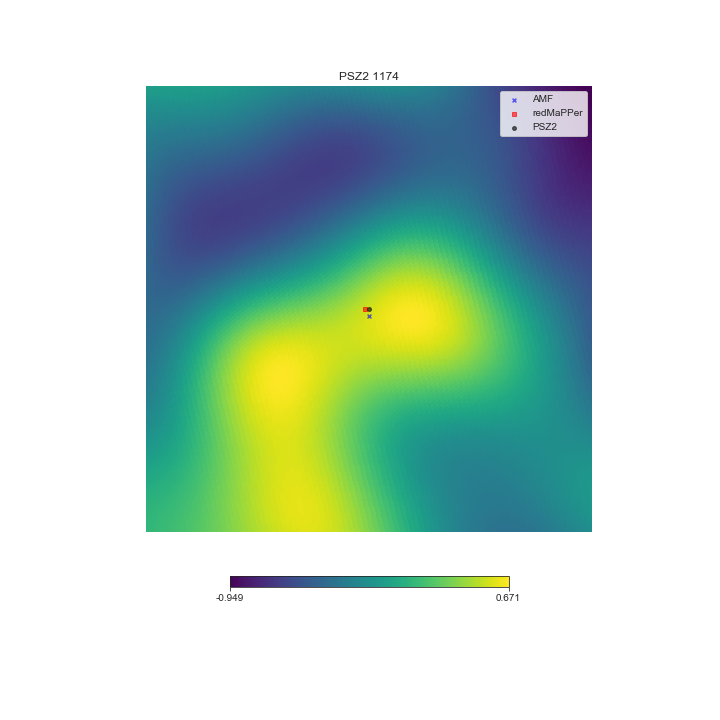}
  \caption{SZ map for PSZ2 1174 and its AMF and redMaPPer counterparts}
  \label{fig:66}
\end{minipage}
\end{figure}

\textbf{3. PSZ2 G254.96+55.88 (Cluster 1174)}: One AMF counterpart with a $\Delta z$=0.1871 (from the PSZ2 redshift z=0.078) with a richness $\Lambda_{200}$=68 and with a $\Delta_{\theta}$=0.69. The redMaPPer counterpart has a $\Delta z$=0.1942 with a richness $\Lambda$=78 and with a $\Delta_{\theta}$=0.45. There is a WHL counterpart which has a $\Delta z$=0.1865 with a richness $R_{L_{*}}$=61 and with a $\Delta_{\theta}$=0.45. From Fig. \ref{fig:66} we see that the redMaPPer cluster lies closer to the PSZ2 cluster center, however, the SZ signal contribution could have come sources at either location. Calculating the $\Lambda_{200}$ with the AMF value of redshift, the scatter $\Delta_{\Lambda_{200}}$ decreases from 1.2 to 0.2. Since all three optical counterparts have very similar properties, it is possible that the PSZ2 cluster can be better characterized. 

\begin{figure}[H]
\centering
\begin{minipage}{.8\textwidth}
  \centering
  \includegraphics[width=.6\textwidth, height=.6\textwidth]{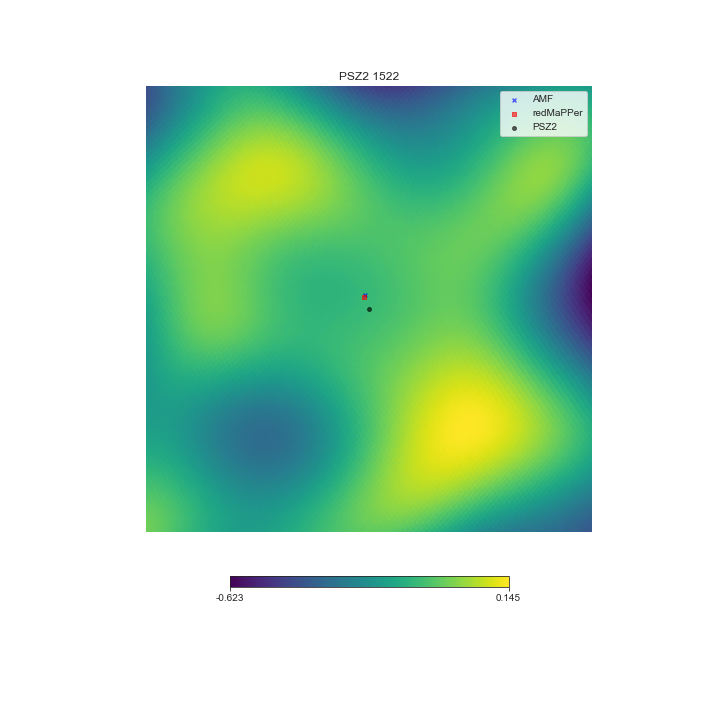}
  \caption{SZ map for PSZ2 1522 and its AMF and redMaPPer counterparts}
  \label{fig:67}
\end{minipage}
\end{figure}

\textbf{4. PSZ2 G323.39+81.61 (Cluster 1522)}: One AMF counterpart with a $\Delta z$=0.344 (from the PSZ2 redshift z=0.064) with a richness $\Lambda_{200}$=36 and with a $\Delta_{\theta}$=0.94. The redMaPPer counterpart has a $\Delta z$=0.341 with a richness $\Lambda$=43 and with a $\Delta_{\theta}$=0.91. There is a WHL counterpart which has a $\Delta z$=0.3474 with a richness $R_{L_{*}}$=28 and with a $\Delta_{\theta}$=0.91. From Fig. \ref{fig:67} we see that the SZ signal contribution could have come sources at either the location of the AMF or the redMaPPer counterpart. Calculating the $\Lambda_{200}$ with the AMF value of redshift, the scatter $\Delta_{\Lambda_{200}}$ increases from 0.4 to 2.5. It is possible that the optical and SZ clusters correspond to very different sources, or that the PSZ2 cluster could be blended. 

\subsection{PSZ2 clusters (with external validation) matched by the AMF DR9, without a redMaPPer counterpart} \label{appA_sub3}

\begin{figure}[H]
\centering
\begin{minipage}{.8\textwidth}
  \centering
  \includegraphics[width=.6\textwidth, height=.6\textwidth]{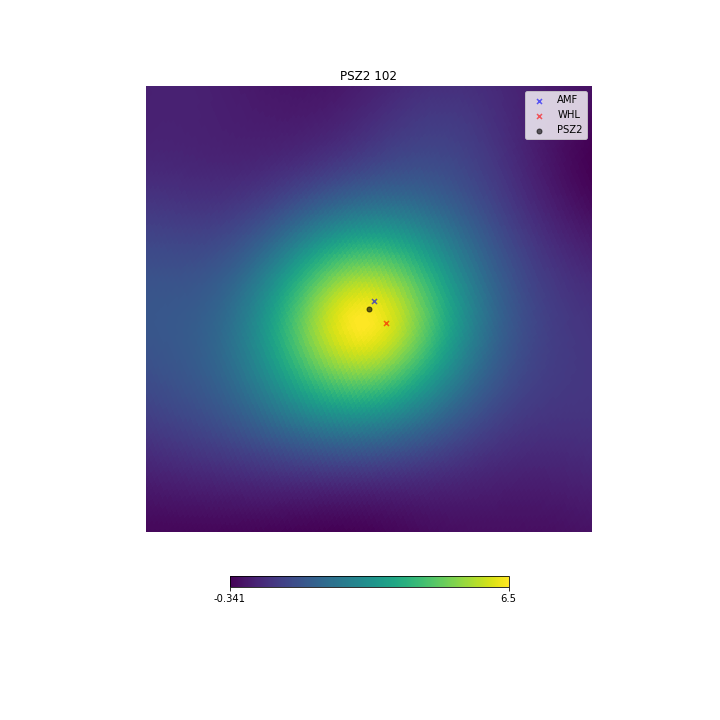}
  \caption{SZ map for PSZ2 102 and its AMF and WHL counterparts}
  \label{fig:73}
\end{minipage}
\end{figure}

\textbf{1. PSZ2 G029.06+44.55 (Cluster 102)} One AMF counterpart with a $\Delta z$=0.2446 (from the PSZ2 redshift z=0.0353) with a richness $\Lambda_{200}$=24 and with a $\Delta_{\theta}$=1.32. There is a WHL counterpart which has a $\Delta z$=0.3547 with a richness $R_{L_{*}}$=16 and with a $\Delta_{\theta}$=3.01. From the SZ map (Fig. \ref{fig:73}), the signal is more likely to have come from the location of the AMF cluster center. Both the AMF and WHL clusters have very low values of optical richness, however, the PSZ2 cluster has a high S/N (15.8) so it could be an instance of a blended cluster. The PSZ2 cluster also has its redshift sourced from an Abell cluster (A2147) with a well-established redshift \citep{ref:24}. 

\begin{figure}[H]
\centering
\begin{minipage}{.8\textwidth}
  \centering
  \includegraphics[width=.6\textwidth, height=.6\textwidth]{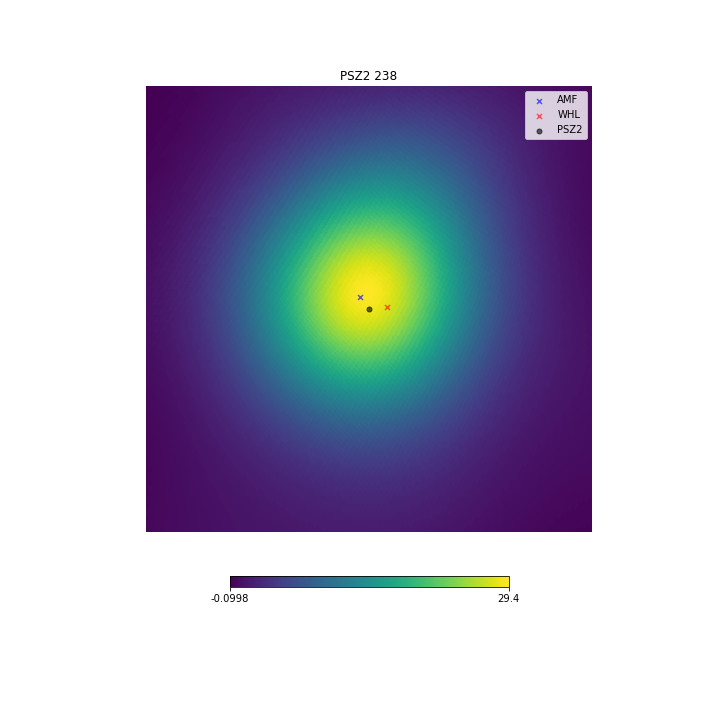}
  \caption{SZ map for PSZ2 238 and its AMF and WHL counterparts}
  \label{fig:70}
\end{minipage}
\end{figure}

\textbf{2. PSZ2 G057.80+88.00 (Cluster 238)} One AMF counterpart with a $\Delta z$=0.1352 (from the PSZ2 redshift z=0.0231) with a richness $\Lambda_{200}$=42 and with a $\Delta_{\theta}$=4.25. There is a WHL counterpart which has a $\Delta z$=0.5051 with a richness $R_{L_{*}}$=23 and with a $\Delta_{\theta}$=6.37. From the SZ map (Fig. \ref{fig:70}), the AMF DR9 cluster center lies closer to the PSZ2 cluster signal source than the WHL cluster center, but given the 5 arc-minute width of the Planck beam, the signal could have come from the location of either optical center. The Planck Cluster has an extremely high Signal-to-Noise of 35 so it could be a blended cluster. Occurs with a z$\sim$0.027 and Ngals=819 in the DR8 catalog compiled by \cite{ref:19}. 

\begin{figure}[H]
\centering
\begin{minipage}{.8\textwidth}
  \centering
  \includegraphics[width=.6\textwidth, height=.6\textwidth]{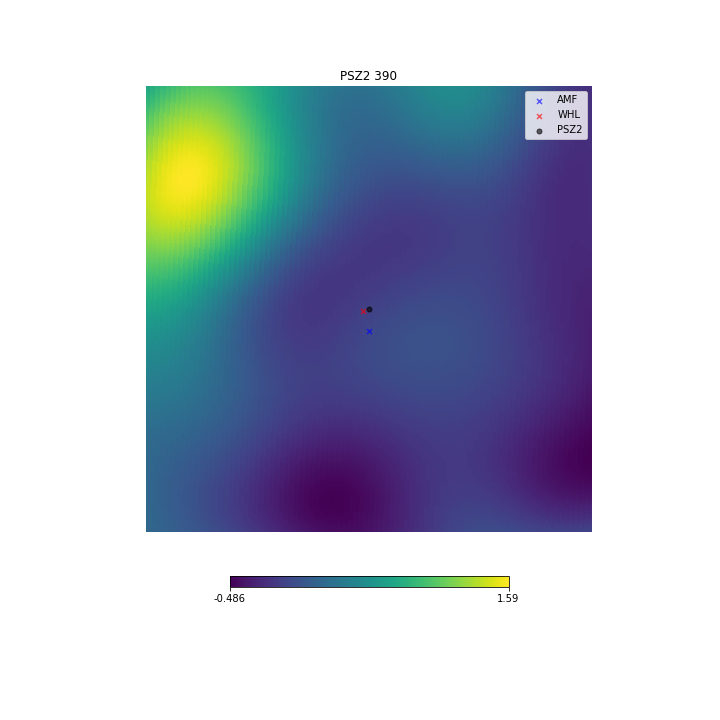}
  \caption{SZ map for PSZ2 390 and its AMF and WHL counterparts}
  \label{fig:74}
\end{minipage}
\end{figure}

\textbf{3. PSZ2 G086.93+53.18 (Cluster 390)} One AMF counterpart with a $\Delta z$=-0.2322 (from the PSZ2 redshift z=0.6752) with a richness $\Lambda_{200}$=28 and with a $\Delta_{\theta}$=3.92. The PSZ2 cluster sources its redshift from a WHL cluster (WHL J228.466+52.83) with a richness $R_{L_{*}}$=12 and with a $\Delta_{\theta}$=0.98. From the SZ map (Fig. \ref{fig:74}), the WHL cluster center lies closer to the Planck cluster than the AMF center, however the PSZ2 cluster center also does not seem to be at the location of a likely SZ source. The redshift of the PSZ2 cluster (0.67) is at a range beyond which the AMF cluster finder does not possess reliable characterization. The WHL counterpart is also of extremely low optical richness. 

\begin{figure}[H]
\centering
\begin{minipage}{.8\textwidth}
  \centering
  \includegraphics[width=.6\textwidth, height=.6\textwidth]{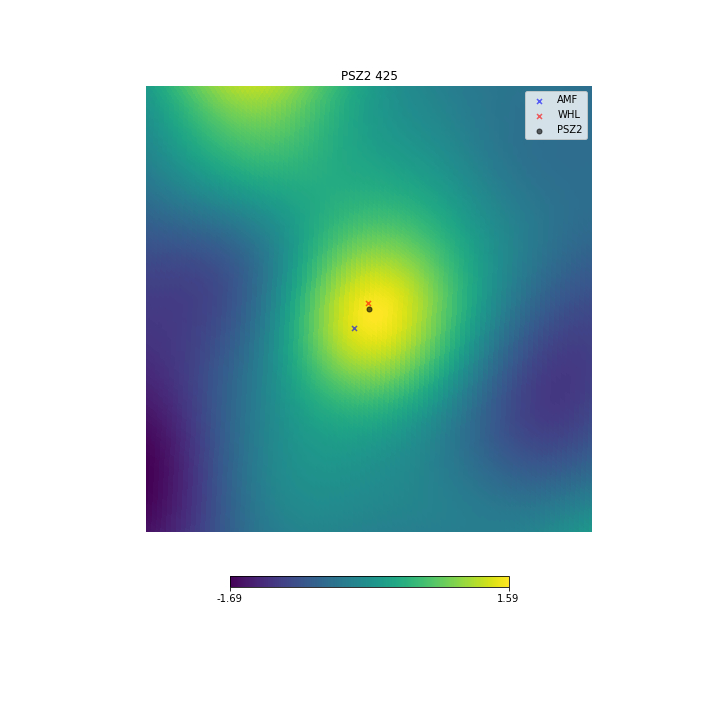}
  \caption{SZ map for PSZ2 425 and its AMF and WHL counterparts}
  \label{fig:75}
\end{minipage}
\end{figure}

\textbf{4. PSZ2 G093.42-43.21 (Cluster 425)} One AMF counterpart with a $\Delta z$=0.1082 (from the PSZ2 redshift z=0.0428) with a richness $\Lambda_{200}$=25 and with a $\Delta_{\theta}$=2.08. There is a WHL counterpart which has a $\Delta z$=0.0082 with a richness $R_{L_{*}}$=73 and with a $\Delta_{\theta}$=0.51. From the SZ map (Fig. \ref{fig:75}), the WHL cluster center lies closer to the PSZ2 cluster signal source than the AMF cluster center, but given the 5 arc-minute width of the Planck beam, the signal could have come from the location of either optical center. The Planck redshift is sourced from an Abell cluster (A2593), and the PSZ2 cluster also possesses an external MCXC X--ray validation (J2324.3+1439), as well as a Chandra follow-up \citep{ref:25}. 

\begin{figure}[H]
\centering
\begin{minipage}{.8\textwidth}
  \centering
  \includegraphics[width=.6\textwidth, height=.6\textwidth]{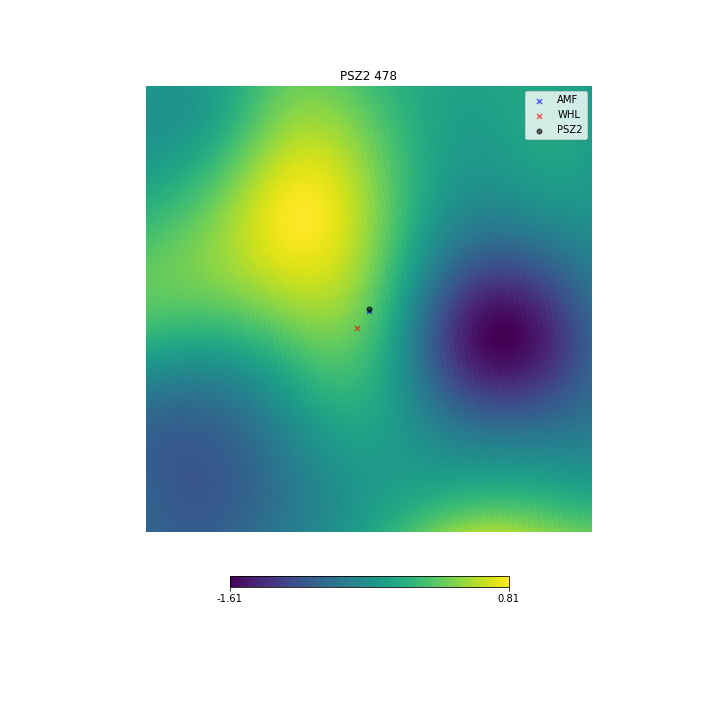}
  \caption{SZ map for PSZ2 478 and its AMF and WHL counterparts}
  \label{fig:76}
\end{minipage}
\end{figure}

\textbf{5. PSZ2 G099.86+58.45 (Cluster 478)} One AMF counterpart with a $\Delta z$=-0.1016 (from the PSZ2 redshift z=0.6305) with a richness $\Lambda_{200}$=42 and with a $\Delta_{\theta}$=0.34. The PSZ2 redshift is sourced from a WHL cluster (WHL J213.697+54.78) with a richness $R_{L_{*}}$=63 and with a $\Delta_{\theta}$=0.43. From the SZ map (Fig. \ref{fig:76}), the AMF cluster center lies closer to the PSZ2 cluster signal source than the WHL cluster center, but given the 5 arc-minute width of the Planck beam, the signal could have come from the location of either optical center. The redshift of the PSZ2 cluster (0.63) is at a range beyond which the AMF cluster finder does not possess reliable characterization. 

\section{Appendix B} \label{app:B}

This section lists the full table data for PSZ2 clusters without a previous external validation, which have an AMF DR9 counterpart (See Section 
\ref{sec:comp_nomsz}). 

For the reliability flag: 
\begin{itemize}
\item A: The PSZ2 cluster has matches in both redMaPPer and WHL, as well as a follow-up. The redshift difference between the AMF counterpart and the 3 individual matches are all $\le0.05$. Henceforth dz $\le0.05$ is defined as a matching redshift
\item B: The PSZ2 cluster has at least two matching redshifts
\item C: The PSZ2 cluster has at least one matching redshift
\item D: The PSZ2 cluster (matched by AMF) has at least one other counterpart, either redMaPPer, WHL or a followup
\item E: The PSZ2 cluster has no counterpart
\end{itemize}

The table displays the clusters with a counterpart in the extended AMF catalog, along with information about whether they occur in the AMF main, WHL and redMaPPer catalogs, the redshift of their AMF counterpart, and whether they have a redshift estimate in a follow-up study. In addition the table also lists the redshift differences between the matched optical counterparts and whether the PSZ2 clusters have had other follow-up studies. 
If there are no counterparts to calculate the redshift difference from then the column entry is an 'x'. If a particular cluster does not occur in AMF DR9 main then the redshift difference is calculated from the corresponding entry in the AMF DR9 extended catalog. Column 5 lists the reliability estimate of the cluster, the rubric of which is outlined above. Column 6 lists the calculated $M_{SZ}$ values for these PSZ2 clusters, according to the redshift of their AMF DR9 counterpart. Multiple entries in the columns indicate multiple possible AMF counterparts for the given PSZ2 cluster. A $^*$ indicates a possible blended cluster, while a $^+$ indicates an ambiguous match, in which we could not resolve the reason for the multiple optical counterparts. $M_{SZ}$ is given in units of $10^{14}$ $M_{\odot}$.

\begin{longtable*}{|c|c|c|c|c|c|c|}
\caption{Optical Counterparts for PSZ2 clusters without external validation \label{table:two_full}}\\
\hline
\textbf{PSZ2 Cluster}  & \textbf{$z_{AMF}$}  & \textbf{$z_{AMF}$ - $z_{Follow-up}$}
& \textbf{$z_{AMF}$ - $z_{redMaPPer}$} & \textbf{$z_{AMF}$ - $z_{WHL}$} &\textbf{Reliability}
& \textbf{$M_{SZ}$, Calc.} \\ 
\hline
161  & 0.38 & 0.0 & 0.02 & 0.04 & A & 6.0\\ 
176  & 0.26 & 0.06 & x & 0.04 & C & 4.6\\
194  & 0.37 & x & x & x & E & 5.9\\
243  & 0.37 & -0.04 & x & x & C & 5.1\\
279  & 0.40 & x & -0.11/-0.12 & -0.11/-0.12 & $D^*$ & 5.4/5.3\\
284 & 0.33 & x & x & -0.02/-0.1 & $C^+$ & 4.6/3.9 \\
295 & 0.18 & -0.18 & x & 0.10 & D & 3.6 \\
303 & 0.41 & -0.36 & x & x & D & 7.1\\
327 & 0.29 & -0.01 & -0.01 & -0.004 & A & 4.1 \\
328 & 0.41 & x & x & 0.02 & C & 5.1 \\
371 & 0.21 & 0.02 & 0.02 & 0.01 & A & 3.5 \\
380 & 0.28 & x & x & x & E & 4.65 \\
381 & 0.25 & -0.44/-0.45 & -0.01/-0.01 & 0.01/0.01 & $B^*$ & 4.0/3.9 \\
389 & 0.25 &-0.001 & x & 0.003 & B & 4.5\\
394 & 0.13 & -0.63 & x & 0.03 & C & 2.8 \\
410 & 0.31& x & x & x & $E^+$ & 5.8/6.4\\
415 & 0.18 & x & x & 0.02/0.16 & $C^+$ & 4.1/5.7\\
421 & 0.52 & 0.06 & 0.05 & x & C & 5.0\\
423 & 0.43 & x & -0.08 & -0.10 & D & 6.2 \\
426 & 0.36 & x & x & x & $E^*$ & 5.2/5.2 \\
438 & 0.46 & x & x & -0.05/-0.12 & $C^+$ & 6.5/\\
462 & 0.23 & -0.55 & -0.20 &  0.01 & C & 4.0 \\
465 & 0.29 & -0.02 & -0.01 & 0.28 & B & 3.3\\
468 & 0.25 & x & x & x & $E^+$ & 4.3/5.1\\
483 & 0.52 & -0.09 & x & -0.12 & D & 4.5 \\
510 & 0.31 & x & x & -0.09 & D & 8.6 \\
542  & 0.26 & x & x & 0.0 & C & 4.9 \\
546  & 0.47 & -0.04 & x & -0.02 & B & 4.6 \\
597  & 0.32 & x & x & x & E & 5.1\\
654  & 0.47 & x & x & 0.02 & C & 6.7 \\
666  & 0.07 & x & x & -0.10 & D & 2.6 \\
667  & 0.47 & 0.01 & -0.0 & -0.02 & A & 6.7 \\
668  & 0.47 & x & 0.02 & 0.15 & C & 6.0 \\
669  & 0.18 & x & x & x & E & 4.0\\
673  & 0.15 & -0.33 & -0.33 & 0.0 & C & 3.8 \\
681  & 0.15 & -0.45 & x & -0.24 & D & 2.8 \\
708  & 0.39 & x & x & x & E & 4.8\\
714  & 0.49 & 0.0 & x & - 0.01 & B & 6.7 \\
732  & 0.24 & 0.02 & x &  0.02 & B & 5.0 \\
739  & 0.46 & 0.01 & 0.01 &  0.01 & A & 5.2 \\
740  & 0.22 & x & x & x & E & 3.8 \\
769  &  0.19 & x & x & -0.01 & C & 3.3 \\
799  & 0.40 & x & x & -0.26 & D & 5.6 \\
831 & 0.14 & -0.02 & -0.15 &  -0.14 & C & 2.9 \\
836 & 0.38 & 0.04 & x & -0.01 & B & 5.8 \\
875 & 0.40 & x & x & x & $E^*$ & 6.9/7.0\\
882 & 0.35 & x & x & x & E & 6.9\\
906 & 0.27 & x & 0.02/-0.01 & -0.0/-0.03 & $B^*$ & 4.2/3.9 \\
920 & 0.26 & -0.01/-0.02 & x & 0.0/-0.01 & $B^*$ & 5.0/4.9 \\
921 & 0.55 & x & 0.18 &  -0.0 & C & 6.1 \\
1070 & 0.46 & x & -0.01 & -0.0 & B & 8.7 \\
1151 & 0.46 & x & x & -0.03 & C & 4.0\\
1510 & 0.46 & 0.0 & x &  0.0 & B & 9.3 \\
1513 & 0.18 & x & x & x & E & 3.4 \\
1548 & 0.09 & 0.0 & x & -0.51 & C & 2.3 \\
\hline
\end{longtable*}

\section{Appendix C} \label{app:C}

This section lists the tables for possible PSZ2 detections which were not included in the original catalog. (See Section \ref{sec:ext}). Table \ref{table:app} lists the clusters with S/N $>$ 4.5, while Table \ref{table:sn445} lists the clusters with 4 $<$ S/N $<$ 4.5. 

The Optical Detections flag is constructed as follows:
\begin{itemize}
\item \textbf{A}: Cluster occurs in all 3 optical catalogs (AMF, WHL and redMaPPer)
\item \textbf{B}: Cluster occurs in AMF and any one other optical catalog
\item \textbf{C}: Cluster occurs only in AMF DR9 catalog
\end{itemize}

Table \ref{table:app} also lists the Signal to Noise (SNR) calculated for these clusters after applying a Multi-frequency Matched Filter to the Planck HFI Maps (See Section \ref{sec:ext}). The table also provides information on whether or not there are any point sources in the vicinity of the cluster candidate, where the vicinity has been defined as a disk with of radius $5\sigma_{\mathrm{beam}}$ with the sitioned at the center.

\vspace{2cm}

\begin{center}
\begin{longtable*}{|c|c|c|c|c|c|c|} 
\caption{Possible PSZ2 Detections with S/N $>$ 4.5. Candidates with MMF S/N $>4.5$ are highlighted with an $\ast$ next to their AMF Cluster column\label{table:app}} \\ 
\hline
\textbf{AMF Cluster}  & \textbf{$z_{AMF}$} & \textbf{$\Lambda_{200}$} & \textbf{Planck S/N} & \textbf{Optical Detections}
& \textbf{SNR} & \textbf{Point Source} \\
\hline
69  & 0.38 & 129 & 4.6 & A & 2.7 & N\\
127 & 0.24 & 112 & 5.2 & A & 4.2 & N\\
146 & 0.39 & 109 & 4.7 & A & 3.0 & N\\
352 & 0.51 & 89 & 4.6 & A & 3.5 & Y\\
408 & 0.23 & 86 & 4.9 & B & 3.3 & Y\\
524 & 0.37 & 80 & 4.5 & B & 2.9 & Y\\
548 & 0.33 & 80 & 7.1 & A & 2.9 & N\\
600 & 0.52 & 78 & 4.8 & A & 4.2 & N\\
632 & 0.42 & 76 & 5.0 & A & 3.3 & N\\
676 & 0.23 & 75 & 4.8 & A & 4.0 & N\\
731 & 0.51 & 73 & 8.3 & B & 3.7 & Y\\
739 & 0.11 & 73 & 4.5 & A & 3.5 & N\\
768 & 0.27 & 72 & 4.6 & A & 4.2 & N\\
820* & 0.23 & 71 & 7.3 & C & 9.2 & N\\
843 & 0.19 & 70 & 5.6 & A & 4.5 & N\\
844 & 0.39 & 70 & 5.2 & A & 3.4 & N\\
924 & 0.40 & 69 & 5.6 & A & 7.4 & N\\
943 & 0.22 & 69 & 4.8 & A & 3.6 & N\\
1008* & 0.21 & 68 & 5.4 & A & 4.9 & Y\\
1049 & 0.32 & 67 & 4.7 & A & 2.9 & N\\
1075 & 0.14 & 67 & 4.9 & A & 2.5 & N\\
1246* & 0.11 & 65 & 4.8 & A & 4.6 & N\\
1263* & 0.12 & 64 & 5.3 & C & 9.7 & N\\
1412 & 0.32 & 62 & 5.5 & A & 3.0 & N\\
1424 & 0.36 & 62 & 5.5 & A & 3.7 & N\\
1517 & 0.12 & 61 & 4.8 & A & 3.8 & N\\
1644 & 0.43 & 60 & 4.7 & A & 3.7 & N\\
1685 & 0.16 & 59 & 4.6 & A & 3.0 & N\\
1713* & 0.36 & 59 & 4.9 & A & 4.7 & N\\
2106* & 0.17 & 56 & 4.8 & A & 7.5 & Y\\
2155 & 0.09 & 56 & 4.8 & B & 4.1 & N\\
2164* & 0.13 & 56 & 4.9 & A & 4.7 & Y\\
2243 & 0.48 & 55 & 4.6 & C & 2.8 & N\\
2262 & 0.28 & 55 & 4.8 & A & 3.6 & N\\
2287 & 0.41 & 55 & 4.7 & A & 3.9 & N\\
2306 & 0.1 & 54 & 5.2 & B & 3.0 & N\\
2383 & 0.35 & 54 & 4.8 & A & 2.7 & N\\
2391 & 0.26 & 54 & 4.7 & A & 3.3 & N\\
2512 & 0.29 & 53 & 4.6 & A & 3.3 & N\\
2592 & 0.28 & 53 & 4.6 & A & 4.5 & N\\
2662 & 0.17 & 52 & 4.7 & A & 2.6 & N\\
2697 & 0.39 & 52 & 4.6 & A & 3.5 & N\\
2788 & 0.40 & 52 & 4.8 & A & 2.6 & N\\
2815 & 0.20 & 51 & 4.6 & A & 3.4 & N\\
2890 & 0.27 & 51 & 4.8 & A & 3.3 & N\\
2940* & 0.41 & 51 & 5.1 & A & 9.0 & Y\\
3088 & 0.31 & 50 & 4.7 & A & 4.1 & N\\
3107* & 0.17 & 50 & 5.5 & A & 8.8 & Y\\
3147 & 0.33 & 50 & 4.7 & B & 3.6 & Y\\
3337 & 0.28 & 49 & 5.1 & A & 4.2 & N\\
3678 & 0.42 & 48 & 4.6 & A & 2.6 & N\\
3888* & 0.37 & 47 & 6.4 & C & 7.0 & Y\\
3899 & 0.35 & 47 & 4.8 & A & 3.6 & N\\
3906 & 0.24 & 47 & 5.4 & A & 4.0 & N\\
4053* & 0.43 & 47 & 4.6 & A & 4.5 & N\\
4483 & 0.43 & 45 & 4.7 & A & 3.4 & N\\
4792 & 0.47 & 45 & 4.7 & C & 3.9 & N\\
4975 & 0.1 & 44 & 5.0 & A & 3.9 & N\\
5233 & 0.21 & 43 & 4.9 & B & 3.5 & N\\
5278 & 0.47 & 43 & 4.8 & A & 3.2 & N\\
5418* & 0.35 & 43 & 5.5 & C & 5.1 & Y\\
5468* & 0.20 & 43 & 9.3 & C & 13.2 & N\\
6007 & 0.45 & 42 & 5.1 & A & 2.8 & N\\
6237 & 0.34 & 41 & 4.9 & A & 3.3 & N\\
6854* & 0.08 & 40 & 10.8 & B & 11.1 & N\\
7097 & 0.27 & 40 & 4.6 & A & 3.9 & N\\
7742* & 0.38 & 38 & 6.1 & B & 6.6 & Y\\
8969 & 0.43 & 37 & 5.4 & B & 3.4 & Y\\
9038 & 0.19 & 37 & 4.6 & A & 1.9 & N\\
9339 & 0.22 & 36 & 5.9 & A & 3.0 & Y\\
9391 & 0.41 & 36 & 4.6 & B & 2.8 & N\\
9701 & 0.55 & 36 & 5.0 & B & 3.7 & N\\
11869* & 0.21 & 34 & 8.0 & C & 7.2 & N\\
13788 & 0.36 & 32 & 4.8 & B & 2.3 & N\\
14140 & 0.24 & 32 & 4.7 & A & 3.2 & N\\
14951* & 0.18 & 31 & 4.7 & B & 13.3 & N\\
15913* & 0.27 & 31 & 7.3 & C & 9.2 & Y\\
22833 & 0.32 & 27 & 5.4 & C & 5.3 & Y\\
24357 & 0.30 & 27 & 4.9 & A & 3.3 & N\\
30722 & 0.57 & 25 & 4.6 & B & 4.0 & N\\
31242 & 0.31 & 25 & 4.8 & C & 4.0 & N\\
32253 & 0.27 & 24 & 5.2 & B & 2.7 & N\\
33395* & 0.37 & 24 & 4.5 & B & 5.5 & Y\\
35342 & 0.46 & 23 & 4.7 & A & 2.7 & N\\
36776 & 0.13 & 23 & 4.5 & C & 2.3 & N\\
51665 & 0.28 & 20 & 5.0 & B & 1.7 & N\\
\hline
\end{longtable*}
\end{center}

\vspace{1cm}

 \begin{longtable*}{|c|c|c|c|c|c|c|}
\caption{Possible PSZ2 Detections with 4 $<$  S/N $<$ 4.5 \label{table:sn445}}\\
\hline
\textbf{AMF Cluster}  & \textbf{$z_{AMF}$} & \textbf{$\Lambda_{200}$} & \textbf{Planck S/N} & \textbf{Optical Detections Flag} \\
\hline
113  & 0.38 & 116 & 4.1 & A \\
164 & 0.21 & 105 & 4.4 & A \\
178 & 0.29 & 104 & 4.0 & A \\
211 & 0.36 & 100 & 4.2 & A \\
245 & 0.42 & 97 & 4.2 & B \\
315 & 0.49 & 92 & 4.1 & B \\
355 & 0.24 & 89 & 4.5 & A \\
400 & 0.25 & 86 & 4.1 & C \\
407 & 0.40 & 86 & 4.1 & A \\
428 & 0.32 & 85 & 4.4 & A \\
431 & 0.34 & 85 & 4.4 & A \\
492 & 0.28 & 82 & 4.3 & A \\
545 & 0.29 & 80 & 4.1 & A \\
549 & 0.38 & 80 & 4.5 & A \\
595 & 0.39 & 78 & 4.2 & A \\
637 & 0.45 & 76 & 4.4 & A \\
643 & 0.18 & 76 & 4.1 & A \\
645 & 0.38 & 76 & 4.5 & B \\
691 & 0.11 & 75 & 4.4 & A \\
703 & 0.41 & 74 & 4.2 & A \\
753 & 0.21 & 73 & 4.1 & A \\
832 & 0.25 & 71 & 4.5 & A \\
858 & 0.32 & 70 & 4.4 & A \\
860 & 0.14 & 70 & 4.2 & A \\
926 & 0.37 & 69 & 4.2 & A \\
973 & 0.28 & 68 & 4.1 & A \\
1003 & 0.36 & 68 & 4.5 & A \\
1024 & 0.28 & 68 & 4.5 & A \\
1105 & 0.14 & 67 & 4.1 & A \\
1237 & 0.25 & 65 & 4.3 & A \\
1380 & 0.15 & 63 & 4.1 & A \\
1395 & 0.19 & 63 & 4.4 & B \\
1417 & 0.40 & 62 & 4.3 & A \\
1508 & 0.1 & 61 & 4.0 & B \\
1521 & 0.43 & 61 & 4.2 & A \\
1628 & 0.49 & 60 & 4.1 & A \\
1769 & 0.33 & 59 & 4.2 & A \\
1778 & 0.19 & 59 & 4.2 & A \\
1862 & 0.21 & 58 & 4.2 & A \\
1919 & 0.32 & 58 & 4.0 & A \\
1999 & 0.15 & 57 & 4.5 & A \\
2076 & 0.29 & 56 & 4.1 & A \\
2080 & 0.10 & 56 & 4.1 & A \\
2253 & 0.39 & 55 & 4.3 & A \\
2274 & 0.40 & 55 & 4.3 & A \\
2304 & 0.51 & 5.5 & 4.4 & A \\
2385 & 0.26 & 54 & 4.2 & A \\
2412 & 0.36 & 54 & 4.1 & A \\
2457 & 0.49 & 54 & 4.1 & A \\
2557 & 0.33 & 53 & 4.2 & A \\
2679 & 0.19 & 51 & 4.1 & A \\
2829 & 0.48 & 51 & 4.1 & A \\
2937 & 0.20 & 51 & 4.1 & A \\
2979 & 0.19 & 51 & 4.1 & B \\
3056 & 0.21 & 50 & 4.1 & A \\
3199 & 0.17 & 50 & 4.3 & A \\
3243 & 0.28 & 50 & 4.1 & A \\
3367 & 0.46 & 49 & 4.1 & A \\
3535 & 0.34 & 48 & 4.2 & A \\
3580 & 0.11 & 48 & 4.4 & A \\
3655 & 0.30 & 48 & 4.4 & A \\
3675 & 0.12 & 48 & 4.4 & B \\
4484 & 0.36 & 45 & 4.3 & A \\
4533 & 0.51 & 45 & 4.3 & B \\
4558 & 0.32 & 45 & 4.3 & A \\
4610 & 0.12 & 45 & 4.3 & A \\
4820 & 0.14 & 44 & 4.0 & A \\
5195 & 0.54 & 43 & 4.3 & A \\
5222 & 0.36 & 43 & 4.3 & B \\
5248 & 0.37 & 43 & 4.1 & A \\
5517 & 0.37 & 43 & 4.1 & A \\
5777 & 0.35 & 42 & 4.3 & A \\
6228 & 0.39 & 41 & 4.2 & A \\
6426 & 0.26 & 41 & 4.1 & A \\
6805 & 0.53 & 40 & 4.2 & A \\
6879 & 0.15 & 40.0 & 4.4 & A \\
6935 & 0.20 & 40 & 4.2 & A \\
7176 & 0.26 & 39 & 4.1 & A \\
7302 & 0.38 & 39 & 4.2 & A\\
7491 & 0.46 & 39 & 4.3 & A \\
7852 & 0.51 & 38 & 4.2 & A \\
7990 & 0.15 & 38 & 4.1 & A \\
8396 & 0.50 & 37 & 4.1 & A \\
8523 & 0.35 & 37 & 4.3 & A \\
8536 & 0.15 & 37 & 4.2 & B \\
8923 & 0.39 & 37 & 4.0 & B \\
9388 & 0.41 & 36 & 4.1 & A \\
9520 & 0.29 & 36 & 4.2 & A \\
9683 & 0.53 & 36 & 4.0 & A \\
10060 & 0.23 & 35 & 4.1 & B \\
10202 & 0.23 & 35 & 4.3 & A \\
10433 & 0.14 & 35 & 4.2 & A \\
10523 & 0.14 & 35 & 4.2 & A \\
11014 & 0.09 & 34 & 4.1 & B \\
11129 & 0.40 & 34 & 4.2 & C \\
11354 & 0.15 & 34 & 4.2 & A \\
11475 & 0.23 & 34 & 4.3 & C \\
11518 & 0.47 & 34 & 4.1 & A \\
12830 & 0.44 & 33 & 4.1 & A \\
12928 & 0.21 & 33 & 4.4 & A \\
13423 & 0.46 & 32 & 4.3 & A \\
13564 & 0.44 & 32 & 4.3 & A \\
13934 & 0.40 & 32 & 4.3 & A \\
14169 & 0.08 & 32 & 4.4 & B \\
15374 & 0.30 & 31 & 4.1 & A \\
15484 & 0.23 & 31 & 4.0 & A \\
16339 & 0.30 & 30 & 4.3 & B \\
16349 & 0.21 & 30 & 4.3 & A \\
17010 & 0.27 & 30 & 4.4 & A \\
17057 & 0.20 & 30 & 4.3 & C \\
17106 & 0.26 & 30 & 4.1 & A \\
17224 & 0.26 & 30 & 4.1 & A \\
19208 & 0.53 & 29 & 4.0 & C \\
19713 & 0.47 & 29 & 4.2 & A \\
22616 & 0.17 & 27 & 4.5 & B \\
23146 & 0.27 & 27 & 4.1 & C \\
23620 & 0.30 & 27 & 4.1 & B \\
24337 & 0.41 & 27 & 4.2 & A \\
24722 & 0.31 & 27 & 4.5 & A \\
26709 & 0.36 & 26 & 4.1 & C \\
27335 & 0.12 & 26 & 4.2 & A \\
31485 & 0.43 &  24 & 4.1 & C \\
34269 & 0.47 & 24 & 4.1 & C \\
34430 & 0.33 & 24 & 4.1 & A \\
34568 & 0.29 & 24 & 4.4 & C \\
34777 & 0.30 & 24 & 4.1 & B \\
36952 & 0.46 & 23 & 4.2 & B \\
37074 & 0.34 & 23 & 4.0 & C \\
38203 & 0.38 & 23 & 4.1 & A\\
38490 & 0.48 & 23 & 4.0 & A \\
40357 & 0.47 & 22 & 4.1 & A\\
40483 & 0.30 & 22 & 4.2 & C \\
42589 & 0.20 & 22 & 4.3 & C\\
44537 & 0.51 & 22 & 4.5 & C \\
46750 & 0.42 & 21 & 4.1 & A \\
47036 & 0.12 & 21 & 4.0 & A \\
47291 & 0.45 & 21 & 4.3 & B \\
48943 & 0.37 & 21 & 4.5 & C \\
49067 & 0.20 & 21 & 4.2 & C \\
49091 & 0.18 & 21 & 4.0 & C \\
49121 & 0.51 & 21 & 4.3 & C \\
51823 & 0.47 & 20 & 4.2 & B \\
54088 & 0.41 & 20 & 4.4 & C \\
\hline
\end{longtable*}

\newpage

\bibliographystyle{plainnat}
\bibliography{Paper_refs.bib}

\begin{thebibliography}{35}
\providecommand{\natexlab}[1]{#1}
\providecommand{\url}[1]{\texttt{#1}}
\expandafter\ifx\csname urlstyle\endcsname\relax
  \providecommand{\doi}[1]{doi: #1}\else
  \providecommand{\doi}{doi: \begingroup \urlstyle{rm}\Url}\fi

\bibitem[Adam et~al.(2016{\natexlab{a}})Adam, Ade, Aghanim, Akrami, Alves,
  Arg{\"u}eso, Arnaud, Arroja, Ashdown, Aumont, et~al.]{ref:35}
R~Adam, PAR Ade, N~Aghanim, Y~Akrami, MIR Alves, F~Arg{\"u}eso, M~Arnaud,
  F~Arroja, Mark Ashdown, J~Aumont, et~al.
\newblock Planck 2015 results-i. overview of products and scientific results.
\newblock \emph{Astronomy \& Astrophysics}, 594:\penalty0 A1,
  2016{\natexlab{a}}.

\bibitem[Adam et~al.(2016{\natexlab{b}})Adam, Ade, Aghanim, Alves, Arnaud,
  Ashdown, Aumont, Baccigalupi, Banday, Barreiro, et~al.]{ref:34}
R~Adam, PAR Ade, N~Aghanim, MIR Alves, M~Arnaud, M~Ashdown, J~Aumont,
  C~Baccigalupi, AJ~Banday, RB~Barreiro, et~al.
\newblock Planck 2015 results-x. diffuse component separation: Foreground maps.
\newblock \emph{Astronomy \& Astrophysics}, 594:\penalty0 A10,
  2016{\natexlab{b}}.

\bibitem[Ade et~al.(2016)Ade, Aghanim, Arnaud, Ashdown, Aumont, Baccigalupi,
  Banday, Barreiro, Barrena, Bartlett, et~al.]{ref:8}
PAR Ade, N~Aghanim, M~Arnaud, M~Ashdown, J~Aumont, C~Baccigalupi, AJ~Banday,
  RB~Barreiro, R~Barrena, JG~Bartlett, et~al.
\newblock Planck 2015 results-xxvii. the second planck catalogue of
  sunyaev-zeldovich sources.
\newblock \emph{Astronomy \& Astrophysics}, 594:\penalty0 A27, 2016.

\bibitem[Allen et~al.(2008)Allen, Rapetti, Schmidt, Ebeling, Morris, and
  Fabian]{ref:5}
SW~Allen, DA~Rapetti, RW~Schmidt, H~Ebeling, RG~Morris, and AC~Fabian.
\newblock Improved constraints on dark energy from chandra x-ray observations
  of the largest relaxed galaxy clusters.
\newblock \emph{Monthly Notices of the Royal Astronomical Society},
  383\penalty0 (3):\penalty0 879--896, 2008.

\bibitem[Andernach et~al.(1995)Andernach, Tago, and Stengler-Larrea]{ref:24}
H~Andernach, E~Tago, and E~Stengler-Larrea.
\newblock A compilation of measured redshifts of aco clusters.
\newblock \emph{Astrophysical Letters and Communications}, 31:\penalty0 27--30,
  1995.

\bibitem[Andrade-Santos et~al.(2017)Andrade-Santos, Jones, Forman, Lovisari,
  Vikhlinin, van Weeren, Murray, Arnaud, Pratt, D{\'e}mocl{\`e}s,
  et~al.]{ref:25}
Felipe Andrade-Santos, Christine Jones, William~R Forman, Lorenzo Lovisari,
  Alexey Vikhlinin, Reinout~J van Weeren, Stephen~S Murray, Monique Arnaud,
  Gabriel~W Pratt, Jessica D{\'e}mocl{\`e}s, et~al.
\newblock The fraction of cool-core clusters in x-ray versus sz samples using
  chandra observations.
\newblock \emph{The Astrophysical Journal}, 843\penalty0 (1):\penalty0 76,
  2017.

\bibitem[Banerjee et~al.(2018)Banerjee, Szabo, Pierpaoli, Franco, Ortiz,
  Oramas, and Tornello]{ref:1}
Panchajanya Banerjee, Thad Szabo, Elena Pierpaoli, Gerardo Franco, Maria Ortiz,
  Aris Oramas, and Brianna Tornello.
\newblock An optical catalog of galaxy clusters obtained from an adaptive
  matched filter finder applied to sdss dr9 data.
\newblock \emph{New Astronomy}, 58:\penalty0 61--71, 2018.

\bibitem[Burenin(2017)]{ref:27}
RA~Burenin.
\newblock An extension of the planck galaxy cluster catalogue.
\newblock \emph{Astronomy Letters}, 43\penalty0 (8):\penalty0 507--515, 2017.

\bibitem[Dong et~al.(2008)Dong, Pierpaoli, Gunn, and Wechsler]{ref:11}
Feng Dong, Elena Pierpaoli, James~E Gunn, and Risa~H Wechsler.
\newblock Optical cluster finding with an adaptive matched-filter technique:
  algorithm and comparison with simulations.
\newblock \emph{The Astrophysical Journal}, 676\penalty0 (2):\penalty0 868,
  2008.

\bibitem[Erler et~al.(2019)Erler, Ramos-Ceja, Basu, and Bertoldi]{ref:39}
Jens Erler, Miriam~E Ramos-Ceja, Kaustuv Basu, and Frank Bertoldi.
\newblock Introducing constrained matched filters for improved separation of
  point sources from galaxy clusters.
\newblock \emph{Monthly Notices of the Royal Astronomical Society},
  484\penalty0 (2):\penalty0 1988--1999, 2019.

\bibitem[Fujita et~al.(2006)Fujita, Sarazin, and Sivakoff]{ref:18}
Yutaka Fujita, Craig~L Sarazin, and Gregory~R Sivakoff.
\newblock Chandra observations of a 2670 and a 2107: A comet galaxy and cds
  with large peculiar velocities.
\newblock \emph{Publications of the Astronomical Society of Japan}, 58\penalty0
  (1):\penalty0 131--141, 2006.

\bibitem[Gal et~al.(2009)Gal, Lopes, De~Carvalho, Kohl-Moreira, Capelato, and
  Djorgovski]{ref:21}
RR~Gal, PAA Lopes, RR~De~Carvalho, JL~Kohl-Moreira, HV~Capelato, and
  SG~Djorgovski.
\newblock The northern sky optical cluster survey. iii. a cluster catalog
  covering pi steradians.
\newblock \emph{The Astronomical Journal}, 137\penalty0 (2):\penalty0 2981,
  2009.

\bibitem[Haehnelt and Tegmark(1996)]{ref:38}
Martin~G Haehnelt and Max Tegmark.
\newblock Using the kinematic sunyaev-zeldovich effect to determine the
  peculiar velocities of clusters of galaxies.
\newblock \emph{Monthly Notices of the Royal Astronomical Society},
  279\penalty0 (2):\penalty0 545--556, 1996.

\bibitem[Haslam et~al.(1978)Haslam, Kronberg, Waldthausen, Wielebinski, and
  Schallwich]{ref:22}
CGT Haslam, PP~Kronberg, H~Waldthausen, R~Wielebinski, and D~Schallwich.
\newblock A radio survey of clusters of galaxies. i. 11.1 cm observations of
  a591 a754, a1066, a1314, a1517, a2094, a2142, a2255, a2256, a2319 and a2462.
\newblock \emph{Astronomy and Astrophysics Supplement Series}, 31, 1978.

\bibitem[Hogan et~al.(2015)Hogan, Edge, Hlavacek-Larrondo, Grainge, Hamer,
  Mahony, Russell, Fabian, McNamara, and Wilman]{ref:17}
MT~Hogan, AC~Edge, J~Hlavacek-Larrondo, KJB Grainge, SL~Hamer, EK~Mahony,
  HR~Russell, AC~Fabian, BR~McNamara, and RJ~Wilman.
\newblock A comprehensive study of the radio properties of brightest cluster
  galaxies.
\newblock \emph{Monthly Notices of the Royal Astronomical Society},
  453\penalty0 (2):\penalty0 1201--1222, 2015.

\bibitem[Hurier et~al.(2013)Hurier, Macias-Perez, and Hildebrandt]{ref:33}
G~Hurier, JF~Macias-Perez, and S~Hildebrandt.
\newblock Milca, a modified internal linear combination algorithm to extract
  astrophysical emissions from multifrequency sky maps.
\newblock \emph{Astronomy \& Astrophysics}, 558:\penalty0 A118, 2013.

\bibitem[Kaiser(2013)]{ref:6}
Nick Kaiser.
\newblock Measuring gravitational redshifts in galaxy clusters.
\newblock \emph{Monthly Notices of the Royal Astronomical Society},
  435\penalty0 (2):\penalty0 1278--1286, 2013.

\bibitem[Kepner et~al.(1999)Kepner, Fan, Bahcall, Gunn, Lupton, and Xu]{ref:10}
Jeremy Kepner, Xiaohui Fan, Neta Bahcall, James Gunn, Robert Lupton, and
  Guohong Xu.
\newblock An automated cluster finder: the adaptive matched filter.
\newblock \emph{The Astrophysical Journal}, 517\penalty0 (1):\penalty0 78,
  1999.

\bibitem[Khatri(2016)]{ref:31}
Rishi Khatri.
\newblock An alternative validation strategy for the planck cluster catalogue
  and y-distortion maps.
\newblock \emph{Astronomy \& Astrophysics}, 592:\penalty0 A48, 2016.

\bibitem[Lopes(2007)]{ref:16}
PAA Lopes.
\newblock Empirical photometric redshifts of luminous red galaxies and clusters
  in the sloan digital sky survey.
\newblock \emph{Monthly Notices of the Royal Astronomical Society},
  380\penalty0 (4):\penalty0 1608--1620, 2007.

\bibitem[Melin et~al.(2005)Melin, Bartlett, and Delabrouille]{ref:36}
J-B Melin, JG~Bartlett, and J~Delabrouille.
\newblock The selection function of sz cluster surveys.
\newblock \emph{Astronomy \& Astrophysics}, 429\penalty0 (2):\penalty0
  417--426, 2005.

\bibitem[Melin et~al.(2006)Melin, Bartlett, and Delabrouille]{ref:37}
J-B Melin, James~G Bartlett, and Jacques Delabrouille.
\newblock Catalog extraction in sz cluster surveys: a matched filter approach.
\newblock \emph{Astronomy \& Astrophysics}, 459\penalty0 (2):\penalty0
  341--352, 2006.

\bibitem[Oguri(2014)]{ref:15}
Masamune Oguri.
\newblock A cluster finding algorithm based on the multiband identification of
  red sequence galaxies.
\newblock \emph{Monthly Notices of the Royal Astronomical Society},
  444\penalty0 (1):\penalty0 147--161, 2014.

\bibitem[Pierpaoli et~al.(2001)Pierpaoli, Scott, and White]{ref:3}
Elena Pierpaoli, Douglas Scott, and Martin White.
\newblock Power-spectrum normalization from the local abundance of rich
  clusters of galaxies.
\newblock \emph{Monthly Notices of the Royal Astronomical Society},
  325\penalty0 (1):\penalty0 77--88, 2001.

\bibitem[Piffaretti et~al.(2011)Piffaretti, Arnaud, Pratt, Pointecouteau, and
  Melin]{ref:14}
R~Piffaretti, M~Arnaud, GW~Pratt, E~Pointecouteau, and J-B Melin.
\newblock The mcxc: a meta-catalogue of x-ray detected clusters of galaxies.
\newblock \emph{Astronomy \& Astrophysics}, 534:\penalty0 A109, 2011.

\bibitem[Rines et~al.(2007)Rines, Diaferio, and Natarajan]{ref:4}
Kenneth Rines, Antonaldo Diaferio, and Priyamvada Natarajan.
\newblock The virial mass function of nearby sdss galaxy clusters.
\newblock \emph{The Astrophysical Journal}, 657\penalty0 (1):\penalty0 183,
  2007.

\bibitem[Rykoff et~al.(2014)Rykoff, Rozo, Busha, Cunha, Finoguenov, Evrard,
  Hao, Koester, Leauthaud, Nord, et~al.]{ref:9}
ES~Rykoff, E~Rozo, MT~Busha, CE~Cunha, A~Finoguenov, A~Evrard, J~Hao,
  BP~Koester, A~Leauthaud, B~Nord, et~al.
\newblock redmapper. i. algorithm and sdss dr8 catalog.
\newblock \emph{The Astrophysical Journal}, 785\penalty0 (2):\penalty0 104,
  2014.

\bibitem[Seljak(2000)]{ref:2}
Uro{\v{s}} Seljak.
\newblock Analytic model for galaxy and dark matter clustering.
\newblock \emph{Monthly Notices of the Royal Astronomical Society},
  318\penalty0 (1):\penalty0 203--213, 2000.

\bibitem[Sereno(2015)]{ref:23}
Mauro Sereno.
\newblock Comalit--iii. literature catalogues of weak lensing clusters of
  galaxies (lc2).
\newblock \emph{Monthly Notices of the Royal Astronomical Society},
  450\penalty0 (4):\penalty0 3665--3674, 2015.

\bibitem[Szabo et~al.(2011)Szabo, Pierpaoli, Dong, Pipino, and Gunn]{ref:12}
Thad Szabo, Elena Pierpaoli, Feng Dong, Antonio Pipino, and J~Gunn.
\newblock An optical catalog of galaxy clusters obtained from an adaptive
  matched filter finder applied to sloan digital sky survey data release 6.
\newblock \emph{The Astrophysical Journal}, 736\penalty0 (1):\penalty0 21,
  2011.

\bibitem[Tarr{\'\i}o et~al.(2019)Tarr{\'\i}o, Melin, and Arnaud]{ref:32}
Paula Tarr{\'\i}o, Jean-Baptiste Melin, and Monique Arnaud.
\newblock Comprass: a combined planck-rass catalogue of x-ray-sz clusters.
\newblock \emph{arXiv preprint arXiv:1901.00873}, 2019.

\bibitem[Tempel et~al.(2012)Tempel, Tago, and Liivam{\"a}gi]{ref:19}
E~Tempel, E~Tago, and LJ~Liivam{\"a}gi.
\newblock Groups and clusters of galaxies in the sdss dr8-value-added
  catalogues.
\newblock \emph{Astronomy \& Astrophysics}, 540:\penalty0 A106, 2012.

\bibitem[Toffolatti et~al.(2014)Toffolatti, Collaboration, et~al.]{ref:29}
Luigi Toffolatti, Planck Collaboration, et~al.
\newblock Planck 2013 results. xx. cosmology from sunyaev--zeldovich cluster
  counts.
\newblock \emph{Astronomy and Astrophysics, 571, A20}, 2014.

\bibitem[Umetsu et~al.(2016)Umetsu, Zitrin, Gruen, Merten, Donahue, and
  Postman]{ref:7}
Keiichi Umetsu, Adi Zitrin, Daniel Gruen, Julian Merten, Megan Donahue, and
  Marc Postman.
\newblock Clash: Joint analysis of strong-lensing, weak-lensing shear, and
  magnification data for 20 galaxy clusters.
\newblock \emph{The Astrophysical Journal}, 821\penalty0 (2):\penalty0 116,
  2016.

\bibitem[Wen et~al.(2012)Wen, Han, and Liu]{ref:13}
ZL~Wen, JL~Han, and FS~Liu.
\newblock A catalog of 132,684 clusters of galaxies identified from sloan
  digital sky survey iii.
\newblock \emph{The Astrophysical Journal Supplement Series}, 199\penalty0
  (2):\penalty0 34, 2012.

\end{thebibliography}

\end{document}